\AtBeginDocument{\RenewCommandCopy\qty\SI}
\documentclass[preprint, 3p]{elsarticle}

\journal{Science of Computer Programming}
\biboptions{sort&compress}

\usepackage{amsmath}
\usepackage{amssymb}
\usepackage{derivative}
\usepackage{xfrac} 
\usepackage{physics} 
\usepackage{siunitx}
\usepackage{pifont}
\newcommand{\cmark}{\text{\ding{51}}}%
\newcommand{\xmark}{\text{\ding{55}}}%

\usepackage[utf8]{inputenc}
\usepackage{mlmodern}
\usepackage{microtype}

\usepackage{tikz}
\usepackage{pgfplots}
\usepackage{pgfplotstable}
\usepgfplotslibrary{colormaps, colorbrewer, groupplots}

\usepackage{amsmath}

\pgfplotsset{
  compat=1.18,
  cycle list/Set1,
  cycle multiindex* list={
        mark list*\nextlist
        Set1\nextlist
    },
  results/.style={
    width=5.3cm,
    height=4.3cm,
    ytick distance=0.2,
    xtick={0,10,...,50},
    tick label style={font=\tiny},
    title style={font=\tiny},
    xlabel style={font=\tiny},
    ylabel style={font=\tiny},
    ytick={0,0.2,...,1},
    ymin=0,
    ymax=1,
    yticklabels={0\%,20\%,40\%,60\%,80\%,100\%}, 
    legend cell align={left},
    legend style={
      legend columns=3,
      anchor=center,
      align=center,
      font=\tiny
    },
    legend image post style={mark indices={}},
    enlargelimits=true,
    xlabel=Epoch,
    every axis plot post/.append style={
      thick,
      table/col sep=comma,
      table/x=Epoch,
    },
  },
  group/results/.style={
    group style={
      group size=3 by 1, 
      ylabels at=edge left,
      x descriptions at=edge bottom,
      horizontal sep=0.25cm,
    }, 
    results,
  },
}

\usepackage[bottom]{footmisc}

\usepackage{tabularray}
\UseTblrLibrary{booktabs}
\UseTblrLibrary{siunitx}
\SetTblrStyle{note}{font=\footnotesize}

\usepackage{tblr-extras}
\UseTblrLibrary{caption}

\usepackage{xparse}
\NewDocumentCommand{\rot}{O{45} O{2em} m}{\makebox[#2][l]{\rotatebox{#1}{#3}}}%

\usepackage[shortcuts,acronym]{glossaries}
\usepackage{subcaption}

\usepackage{paralist}

\newacronym{ml}{ML}{machine learning}
\newacronym{vnncomp}{VNN-COMP}{Neural Network Verification Competition}
\newacronym{ldl}{LDL}{Logic of Differentiable Logics}
\newacronym{stl}{STL}{Signal Temporal Logic}
\newacronym{ltl}{LTL}{Linear-time Temporal Logic}
\newacronym{gtsrb}{GTSRB}{German Traffic Sign Recognition Benchmark}
\newacronym{cnn}{CNN}{convolutional neural network}
\newacronym{smt}{SMT}{Satisfiability Modulo Theories}
\newacronym{milp}{MILP}{Mixed-Integer Linear Programming}

\usepackage[breaklinks=true]{hyperref}
\usepackage[capitalize]{cleveref}

\newcommand{\dl}[2]{[\![#1]\!]_{\text{#2}}}
\newcommand{\epsball}[2]{\mathbb{B}(#1;#2)}

\renewcommand{\vec}[1]{\boldsymbol{#1}}

\DeclareMathOperator{\argmax}{arg\,max}
\DeclareMathOperator{\argmin}{arg\,min}

\renewcommand{\implies}{\:\longrightarrow\:}
\renewcommand{\iff}{\:\longleftrightarrow\:}

\newcommand{\ldotpp}{.\ \ }

\journal{\textcopyright\ 2025. This manuscript version is made available under the CC-BY-NC-ND 4.0 license (\url{https://creativecommons.org/licenses/by-nc-nd/4.0/}).}

\makeatletter
\def\ps@pprintTitle{%
     \let\@oddhead\@empty
     \let\@evenhead\@empty
     \def\@oddfoot
       {\hbox to \textwidth%
        {\ifnopreprintline\relax\else
        \@myfooterfont%
         \ifx\@elsarticlemyfooteralign\@elsarticlemyfooteraligncenter%
           \hfil\@elsarticlemyfooter\hfil%
         \else%
         \ifx\@elsarticlemyfooteralign\@elsarticlemyfooteralignleft%
           \@elsarticlemyfooter\hfill{}%
         \else%
         \ifx\@elsarticlemyfooteralign\@elsarticlemyfooteralignright%
           {}\hfill\@elsarticlemyfooter%
         \else%
            \parbox[t]{\textwidth}{\@journal}\fi%
         \fi%
         \fi%
         \fi%
         }%
       }%
     \let\@evenfoot\@oddfoot}
\makeatother

\begin{document}
  \begin{frontmatter}
    \title{Comparing differentiable logics for learning with logical constraints\tnoteref{t1}}
    
    \tnotetext[t1]{This paper is an extended version of our paper~\cite{flinkowComparingDifferentiableLogics2023} published in FMAS 2023, containing an extended and revised experimental setup that leads to a more fair and meaningful comparison. \Cref{sec:contributions} explains our extensions and contributions in more detail.}


    \author[1]{Thomas Flinkow\corref{cor1}} \ead{thomas.flinkow@mu.ie} \ead[url]{https://www.cs.nuim.ie/~tflinkow/}
    \author[1,2]{Barak A. Pearlmutter} \ead{barak@pearlmutter.net} \ead[url]{https://www.bcl.hamilton.ie/~barak/}
    \author[1,2]{Rosemary Monahan} \ead{rosemary.monahan@mu.ie} \ead[url]{http://rosemarymonahan.com/}
    
    \cortext[cor1]{Corresponding author}

    \affiliation[1]{organization={Department of Computer Science}, addressline={Maynooth University}, city={Maynooth, Co. Kildare}, country={Ireland}}
     \affiliation[2]{organization={Hamilton Institute}, addressline={Maynooth University}, city={Maynooth, Co. Kildare}, country={Ireland}}

    \begin{abstract}
      Extensive research on formal verification of machine learning systems indicates that learning from data alone often fails to capture underlying background knowledge, such as specifications implicitly available in the data.
      Various neural network verifiers have been developed to ensure that a machine-learnt model satisfies correctness and safety properties; however, they typically assume a trained network with fixed weights.
      A promising approach for creating machine learning models that inherently satisfy constraints after training is to encode background knowledge as explicit logical constraints that guide the learning process via so-called differentiable logics.
      In this paper, we experimentally compare and evaluate various logics from the literature, present our findings, and highlight open problems for future work.
      We evaluate differentiable logics with respect to their suitability in training, and use a neural network verifier to check their ability to establish formal guarantees.
      The complete source code for our experiments is available as an easy-to-use framework for training with differentiable logics at \url{https://github.com/tflinkow/comparing-differentiable-logics}.
    \end{abstract}

    \begin{keyword}
      machine learning \sep neuro-symbolic \sep differentiable logic \sep verification
    \end{keyword}
  \end{frontmatter}
  
  \section{Introduction}
  \label{sec:introduction}
  Advancements in \ac{ml} in the past few years indicate great potential for applying \ac{ml} to various domains.
  Autonomous systems are one such application domain, but the use of \ac{ml} components in such a safety-critical domain presents unique new challenges for formal verification.
  These include
  \begin{inparaenum}[(i)]
      \item \ac{ml} failing to learn background knowledge from data alone \cite{wangEfficientFormalSafety2018},
      \item neural networks being susceptible to adversarial inputs \cite{szegedyIntriguingPropertiesNeural2014,goodfellowExplainingHarnessingAdversarial2015}, and
      \item a lack of specifications, particularly when continuous learning is permitted \cite{seshiaFormalSpecificationDeep2018,leuckerFormalVerificationNeural2020,farrellExploringRequirementsSoftware2023}.
  \end{inparaenum}    
  
  Addressing these challenges is even more important and more difficult when the \ac{ml}-enabled system is permitted to continue to learn after deployment, either to adapt to changing environments or to correct and improve itself when errors are detected \cite{chengContinuousSafetyVerification2021}.

  \subsection{Formal verification of neural networks}
  A multitude of neural network verifiers have been presented in the past few years.
  We refer the reader to the \ac{vnncomp} reports \cite{bakSecondInternationalVerification2021,mullerThirdInternationalVerification2022,brixFourthInternationalVerification2023,brixFirstThreeYears2023b} for an overview of state-of-the-art neural network verifiers, and to \citet{huangSurveySafetyTrustworthiness2020,liuAlgorithmsVerifyingDeep2021,urbanReviewFormalMethods2021,albarghouthiIntroductionNeuralNetwork2021,mengAdversarialRobustnessDeep2022} for in-depth surveys on neural network verification.
  
  Reluplex, one of the first verifiers for neural networks, was provided by \citet{katzReluplexEfficientSMT2017a}.
  State-of-the-art tools are its successor Marabou \cite{katzMarabouFrameworkVerification2019,wuMarabou20Versatile2024}, along with NNV \cite{tranNNVNeuralNetwork2020}, MN-BaB \cite{ferrariCompleteVerificationMultiNeuron2021}, and $\alpha,\beta$\nobreakdash-CROWN \cite{zhangEfficientNeuralNetwork2018a,xuAutomaticPerturbationAnalysis2020a,wangBetaCROWNEfficientBound2021b,xuFastCompleteEnabling2021a,zhangGeneralCuttingPlanes2022a,shiNeuralNetworkVerification2024} (the winner of the 2021--2023 \ac{vnncomp} competitions~\cite{bakSecondInternationalVerification2021,mullerThirdInternationalVerification2022,brixFourthInternationalVerification2023}).
  However, as noted by \citet{kwiatkowskaSafetyVerificationDeep2019} they typically assume trained networks with fixed weights and do not target the learning process itself.

  \subsection{Loss-based methods for guiding network training}
  There are multiple reasons for injecting background knowledge into neural networks: improved performance, learning with noisy or sparse data, or to guarantee compliance of the network predictions with the background knowledge \cite{giunchigliaDeepLearningLogical2022a}.
  One step in the direction of correct-by-construction neural networks are so-called \emph{differentiable logics}, which transform a logical constraint $\phi$ into an additional constraint loss term $\mathcal{L}_{\text{C}}$ that measures how close the network is to satisfying the constraint.
  This is in addition to standard prediction loss\footnote{Without loss of generality, standard cross-entropy loss (a measure of the difference between the predicted and true probabilities) will be used in the following.} $\mathcal{L}_{\text{P}}$.

  The total loss $\mathcal{L}$ is then a weighted sum of constraint and cross-entropy loss, as shown in \cref{eq:loss_total}, where $\lambda_{\text{P}}$ and $\lambda_{\text{C}}$ balance the different loss terms, $\vec{x}$ is training data, and $y$ the true label.
  \begin{equation}\label{eq:loss_total}
    \mathcal{L}(\vec{x}, y, \phi):=\lambda_{\text{P}}\mathcal{L}_{\text{P}}(\vec{x},y)+\lambda_{\text{C}}\mathcal{L}_{\text{C}}(\vec{x},y,\phi).
  \end{equation}

  The optimal network weights $\vec{\theta}^+$, needed to map the input to a desired output, are then obtained by minimising the total loss using standard gradient descent, where $\vec{\theta}$ denotes the network weights, formally shown in~\cref{eq:optimal_weights}.
  \begin{equation}\label{eq:optimal_weights}
    \vec{\theta}^+=\argmin_{\vec{\theta}}\; \mathcal{L}(\vec{x}, y, \phi).
  \end{equation}
  
  In order to translate logical constraints into this constraint loss, a mapping must be defined that allows for real-valued truth values and is differentiable (almost everywhere) for use with standard gradient-based methods.
  In Section \ref{sec:background} we provide an overview of two of these mappings (so-called \emph{differentiable logics}) from popular literature, namely DL2 and fuzzy logics.

  \subsection{Our contributions}
  \label{sec:contributions}
  In this section, we outline our contributions and improvements over our work in~\cite{flinkowComparingDifferentiableLogics2023}.
  Our original presentation provided a preliminary experimental comparison of differentiable logics.
  We now extend that work, by contributing a complete revision of the experimental setup providing for a more measurable experimental comparison of differentiable logics as explained in~\cref{sec:experimental}.

  \begin{description}
    \item[Training with counterexamples.] Learning to satisfy logical constraints means learning from examples that do not satisfy the constraint.
    The experiments in our earlier work were limited to constraints on data already in the training set.
    As a result, the constraints did not have a major impact on the learning process, as the datasets do not necessarily contain sufficiently many counterexamples for the constraints.
    In this work, we now find counterexamples outside of the dataset that violate the logical constraint using Projected Gradient Descent (PGD) \cite{madryDeepLearningModels2018} and incorporate them into the learning process, thus ensuring that differentiable logics have a positive impact on the learning process.
    \item[Adaptive loss-balancing.] Our earlier comparison of differentiable logics also suffered from a lack of fairness, due to a difficult-to-tune hyperparameter that prevented the logics to perform at their best.
    Our extended work addresses this through use of an adaptive loss-balancing approach called GradNorm~\cite{chenGradNormGradientNormalization2018a}, which allows each logic to perform close to optimally within our experiments, facilitating a fairer comparison of the logics in~\cref{sec:experimental}.
    \item[Differentiable logics and formal verification.] A further contribution is provided via a new experiment, described in~\cref{subsec:verification}, which investigates whether training with differentiable logic also has benefits for the verification of neural networks.
    \item[Improved presentation and theoretical comparison metrics.] We also improve our presentation of differentiable logics in~\cref{sec:background} to ensure that differences between logics are more easily identified.
    Further, we apply theoretical comparison metrics from the literature, such as derivatives and consistency, to more differentiable logics in~\cref{sec:theory}.
    \item[Improved Python implementation.] Finally, our implementation of various differentiable logics in Python has been vastly improved: we provide a reusable, general-purpose framework for training with differentiable logics on arbitrary datasets with neural networks with arbitrary constraints.
    Additionally, the logics are now implemented not only for use in training with PyTorch~\cite{paszkePyTorchImperativeStyle2019}, but also to allow the investigation of mathematical properties (such as derivatives and integrals) in a automated manner with SymPy~\cite{meurerSymPySymbolicComputing2017} and Numpy~\cite{harrisArrayProgrammingNumPy2020}.
  \end{description}
  
  \subsection{Notation \& definitions}
  In the following, vectors will be denoted using bold face, e.g. $\vec{x}$.

  Without loss of generality, we consider only classifier neural networks that output probabilities.
  The function approximated by the neural network will be denoted by $\mathcal{N}$, and for our purposes will be a function $\mathcal{N}:\mathbb{R}^m\to[0,1]^n$, mapping images to a vector containing the predicted probabilities of the input belonging to any of the $n$ possible classes.
  Each vector element $\mathcal{N}(\vec{x})_k$ represents the prediction for class $k$ (where $0\le k<n$). We further define $\epsball{\vec{x}}{\epsilon}$ to be the $\ell_{\infty}$ ball with radius $\epsilon$ around $\vec{x}$, i.e.
  \begin{equation}\label{eq:epsilon_ball}
    \epsball{\vec{x}}{\epsilon}:=\{ \vec{x}'\in\mathbb{R}^m : \lVert\vec{x}'-\vec{x}\rVert_{\infty}\le\epsilon \}.
  \end{equation}
  This gives us a shorthand to refer to the set of all points close to point $\vec{x}$ with respect to the $\ell_{\infty}$ norm.
  This will be convenient later when discussing local robustness, where a property is supposed to hold not just for one particular input, but also for all points that are similar to that input.
  Lastly, given logical formula $\phi$, we let $\dl{\phi}{$L$}$ denote its many-valued relaxation under a specific logic $L$, based on notation from \citet{slusarzLogicDifferentiableLogics2023a}.

  \section{Background: DL2 and Fuzzy Logics}
  \label{sec:background}

  \emph{Deep Learning with Differentiable Logics (DL2)} is a system built by \citet{fischerDL2TrainingQuerying2019} for querying and training neural networks with logical constraints that are mapped into $[0, \infty)$, where $0$ represents absolute truth, and any positive value represents a degree of falsehood that is directly used as a penalty.
  Whereas DL2 was designed specifically for deep learning contexts, fuzzy logics are well-studied logical systems that also happen to be suitable for use as a differentiable logic due to their many-valued nature, with operators that are often differentiable almost everywhere.
  Fuzzy logics express degrees of truth in the unit interval $[0,1]$, with absolute falsehood mapped to $0$, and absolute truth mapped to $1$.

  The first in-depth study investigating the use of fuzzy logics in loss functions (called \emph{Differentiable Fuzzy Logics}) was provided by \citet{vankriekenAnalyzingDifferentiableFuzzy2022} and 
  presented insights into the learning characteristics of differentiable logic operators.
  The use of fuzzy-logic-based loss was also investigated by~\citet{marraTNormsDrivenLoss2023}, who focused on keeping logical relations intact, instead of combining arbitrary fuzzy logic operators.
  
  As a follow up, \citet{slusarzLogicDifferentiableLogics2023a} provide a common presentation of differentiable logics including DL2, fuzzy logics, and \ac{stl}, in a unified, extensible framework called \emph{\ac{ldl}}.
  This framework allows the investigation of properties of differentiable logics in general, without having to consider each logic on its own.
  
  The \ac{ldl} framework additionally provides mechanised proofs of theoretical properties of differentiable logics, thus providing a stepping stone for the development of programming language support for verification of machine learning \cite{affeldtTamingDifferentiableLogics2024b}.

  In this section, we present the logics that we use in our experimental comparison when  mapping logical constraints $\phi\in\Phi$ into real-valued loss.
  These logics are the DL2 logic relaxation $\dl{\cdot}{DL2}:\Phi\to[0,\infty)$, and the fuzzy-logic-based relaxation $\dl{\cdot}{FL}:\Phi\to[0,1]$.
  Here, we provide details of their atomic terms, their operators and their quantifiers.
  
  \subsection{Atomic terms}
  Atomic terms in DL2 are comparisons that are translated as
  \begin{equation}\label{eq:dl2_comparisons}
    \dl{x\le y}{DL2}:=\max\{x-y, 0\}\qquad\text{and}\qquad\dl{x\neq y}{DL2}:=\xi[x=y],
  \end{equation}
  where $\xi>0$ is a constant (found not to have significant influence~\cite{fischerDL2TrainingQuerying2019}), and $[x=y]$ is a standard indicator function.

  Fuzzy logics typically do not define fuzzy comparison operators, which is why \citet{slusarzLogicDifferentiableLogics2023a} introduce a fuzzy comparison mapping shown in~\cref{eq:slusarz_oracle} below:
  \begin{equation}\label{eq:slusarz_oracle}
      \dl{\le}{FL}:[0,1]^2\to [0,1], \qquad
      \dl{\le}{FL}= 1-\max\left\{\frac{x-y}{x+y}, 0\right\}.
  \end{equation}

  As with all fuzzy logic operators, the truth values of the terms $x,y$ must be mapped into $[0,1]$ by some oracle.
  In our constraints,  atomic terms are comparisons, so we take the liberty of changing this mapping to the one shown in~\cref{eq:oracle_modified}, allowing us to forgo the need for an external oracle.
  Note that although this is not technically a pure fuzzy logic operator anymore, fuzzy logics usually do not have fuzzy comparison operators at all. 
  \begin{equation}\label{eq:oracle_modified}
    \dl{\le}{FL}:\mathbb{R}^2\to [0,1],\qquad
    \dl{\le}{FL}= 1-\frac{\max\{x-y, 0\}}{|x|+|y|}
  \end{equation}

  Note also that this mapping has a property that we intuitively might wish to hold:
  for example, we might want $21\leq 20$ to be as much of a violation as $\num{21000}\leq\num{20000}$.
  This cannot be achieved in DL2, where the violation depends only on the absolute difference of values.

  \begin{table}
    \caption{Common t-norms, t-conorms and fuzzy implications.}
    \label{tab:fuzzy_logics}
    \centerline{%
      \small
\begin{tblr}{
  colspec={lQ[c, mode=dmath]Q[c, mode=dmath]Q[c, mode=dmath]},
  row{1}={font=\bfseries, mode=text},
  cell{2}{2}={r=2}{c},
  cell{2}{3}={r=2}{c},
  cell{3}{4}={r=3}{c},
  cell{5}{2}={r=2}{c},
  cell{5}{3}={r=2}{c},
  vline{2-4} = {2-Z}{dotted},
  hline{3-Y} = {dotted},
}
  \toprule
  Logic $L$           & T-norm $T_L(x,y)$                                               & S-norm $S_L(x,y)$                                      & Implication $I_L(x,y)$                                                     \\
  \midrule
  Gödel          & \min\{x,y\}                                 & \max\{x,y\}                        & \begin{cases}1,&\text{if }x<y,\\y,&\text{else.}\end{cases}             \\
  Kleene-Dienes   &                                                               &                                                      & S_L(N(x), y)                                             \\ \addlinespace
  \L ukasiewicz   & \max\{0,x+y-1\}                             & \min\{1, x+y\}                     &                                                                          \\ \addlinespace
  Reichenbach     & xy                                          & x+y-xy                             &                                                                          \\
  Goguen          &                                                               &                                                      & \begin{cases}1,&\text{if }x<y,\\y^x,&\text{else.}\end{cases}           \\
  Yager ($p\ge 1$) & \max\{1-((1-x)^p+(1-y)^p)^{\sfrac{1}{p}},0\} & \min\{(x^p+y^p)^{\sfrac{1}{p}}, 1\} & \begin{cases}1,&\text{if }x=y=0,\\\sfrac{y}{x},&\text{else.}\end{cases} \\
  \bottomrule
\end{tblr}
    }
  \end{table}
  
  \subsection{Conjunction and disjunction}
  In DL2, conjunction and disjunction are mapped to addition and multiplication, thus being associative and commutative:
  \begin{equation}\label{eq:dl2_conjunction_disjunction}
    \dl{x\wedge y}{DL2}:=\dl{x}{DL2}+\dl{y}{DL2}\qquad\text{and}\qquad\dl{x\vee y}{DL2}:=\dl{x}{DL2}\cdot\dl{y}{DL2}.
  \end{equation}

  Fuzzy logics are based on functions $T:[0,1]^2\to[0,1]$ that are commutative, associative, monotonic, and satisfy $T(1,y)=y$.
  These are called triangular norms~\cite{klementTriangularNorms2000} (abbreviated as \emph{t-norms}) and generalise conjunction.
  A t-conorm (also called \emph{s-norm}) generalises disjunction and can be obtained from a t-norm using $S(x,y)=1-T(1-x, 1-y)$; thus, for our fuzzy logic mapping we have:
  \begin{equation}\label{eq:fuzzy_conjunction_disjunction}
    \dl{x\wedge y}{$L$}:=T_L(x,y)\qquad\text{and}\qquad\dl{x\vee y}{$L$}:=S_L(x,y),
  \end{equation}
  where $L$ stands for a chosen fuzzy logic.
  
  \subsection{Negation}
  Because DL2 maps into $[0,\infty)$, a separate translation of negation does not exist.
  Instead, negation is handled by pushing it inwards to the level of comparison, e.g. $\dl{\lnot(x\le y)}{DL2}=\dl{y<x}{DL2}$.

  In fuzzy logics, negation is a function $N:[0,1]\to[0,1]$ such that $N(1)=0$ and for all $x$, $N(N(x))\ge x$.
  We will only use the standard strong negation $N(x)=1-x$.
  
  \subsection{Implication}
  DL2 does not provide a separate translation for implication, instead, material implication is used:
  \begin{equation}\label{eq:dl2_implication}
    \dl{x \implies y}{DL2}:=\dl{\lnot x\vee y}{DL2}.
  \end{equation}

  In fuzzy logics, there are multiple ways to obtain fuzzy implications.
  \Cref{tab:fuzzy_logics} lists the definitions of the mentioned t-norms, t-conorms, as well as implications. 
  \citet{baczynskiFuzzyImplications2008} give a detailed overview of fuzzy implications.
  \begin{itemize}
    \item From a t-conorm $S$ and fuzzy negation $N$, one obtains a so-called $(S,N)$-implication (which generalises material implication) as $I(x,y)= S(N(x), y)$).
    Example $(S,N)$-implications are the Kleene-Dienes implication $I_\text{Kleene-Dienes}(x,y):=\max\{1-x,y\}$ and Reichenbach implication $I_\text{Reichenbach}(x,y):=1-x+xy$, both with the standard negation $N(x)=1-x$.
    \item Other implications generalise the intuitionistic implication and are called $R$-implications, because they use the t-norm residuum $R(x,y):=\sup\{t\in[0, 1] \mid T(x, t) \le y\}$.
    Example $R$-implications are the Gödel and Goguen implications.
    \item The \L ukasiewicz implication $I_\text{\L ukasiewicz}(x,y):=\min\{1-x+y,1\}$ is both an $(S,N)$-implication and an $R$-implication.
  \end{itemize}
  Additionally, \citet{vankriekenAnalyzingDifferentiableFuzzy2022} propose \emph{sigmoidal implications} in order to improve the derivatives of the original implication, while preserving its characteristics.
  \Cref{eq:implication_sigmoidal} shows the sigmoidal implication $I^{(s)}$ for a given fuzzy implication $I$ and parameter $s$ for controlling the steepness.
  In our experiments, we use the sigmoidal Reichenbach implication with $s=9$, as suggested by \citet{vankriekenAnalyzingDifferentiableFuzzy2022}.
  \begin{equation}\label{eq:implication_sigmoidal}
    I^{(s)}(x,y)=\dfrac{(1+\exp(\sfrac{s}{2})) \sigma(sI(x,y)-\sfrac{s}{2})-1}{\exp(\sfrac{s}{2})-1},\qquad\text{with}\quad\sigma(x)=\frac{1}{(1+\exp(-x)).}
  \end{equation}
  
  \subsection{Universal quantification}
  In DL2, \citet{fischerDL2TrainingQuerying2019} provide an important insight: training to satisfy a universally quantified constraint can be approximated by finding a counterexample that violates the constraint, and using that counterexample in training.
  This counterexample can in theory be provided by any suitable oracle---DL2 uses PGD to find the worst possible perturbation around particular inputs.
  In DL2, universal quantification thus is not general, and is treated outside of the logic.

  With t-norms being associative and commutative, universal fuzzy quantification is given by repeated application of conjunction.
  
  The LDL framework can handle arbitrarily nested, general universal (and existential) quantifiers.

  \subsection{Other operators}
  Based on the comparison, negation, conjunction, disjunction, and implication operators shown above, other operators can be derived (such as $\dl{x\iff y}{$L$}=\dl{(x\implies y)\wedge(y\implies x)}{$L$}$).
  Note that as shown in \cref{tab:fuzzy_logics}, we will only use \emph{symmetric configurations} of fuzzy logic operators in our investigations and experiments, meaning a logic will consist of some t-norm for conjunction and its dual t-conorm for disjunction, and its implication will be either the $(S,N)$-implication based on its t-conorm, or the R-implication based on its t-norm.
  Symmetric configurations usually retain logical relations, but they might not perform as well as arbitrarily combined fuzzy operators.
  An analysis of both symmetric as well as arbitrary combinations is provided by \citet{vankriekenAnalyzingDifferentiableFuzzy2022}.
  
  \section{Theoretical Comparison}
  \label{sec:theory}
  With this wide range of differentiable logics to choose from, we first focus our comparison on investigating which logical operators satisfy theoretical properties such as shadow-lifting and derivatives, Modus Ponens and Modus Tollens reasoning, as well as consistency, as these properties have implications for the logical operators' suitability for reasoning.

  \subsection{Shadow-lifting (derivatives of conjunction)}
  \label{sec:shadow-lifting}

  \begin{table}
      \centerline{%
        \small
\begin{talltblr}[
  note{a} = {For brevity, we let $\mathcal{S}=(1-x)^p+(1-y)^p$.},
  label = {tab:derivatives_conjunctions},
  caption = {The derivatives of the DL2 and fuzzy conjunctions and their properties. Note that all conjunctions shown are commutative, so we only show the derivatives with respect to $x$.}
]{
  colspec={lQ[c, m, mode=dmath]Q[c, m, mode=dmath]Q[l, m, mode=text]Q[c, m, mode=dmath]},  
  row{1}={font=\bfseries, mode=text},
}
  \toprule
  Logic $L$ & Conjunction $\dl{x\wedge y}{$L$}$ & $\displaystyle\pdv{\dl{x\wedge y}{$L$}}{x}$ & Behaviour & {Shadow-\\lifting} \\
  \midrule
  DL2 & x+y & 1 & strong derivative & \cmark \\
  Gödel & \min\{x,y\} & \begin{cases}1,&\text{if }x\le y\\0\end{cases} & strong derivative & \xmark \\
  \L ukasiewicz & \max\{0,x+y-1\} & \begin{cases}1,&\text{if }x+y>1\\0\end{cases} & {strong derivative;\\but vanishing on half\\ of the domain} & \xmark \\
  Reichenbach & xy & y & {derivative low, if $y$ low;\\(not suitable for learning)} & \cmark \\
  Yager\TblrNote{a} ($p\ge 1$) & \max\{1-\mathcal{S}^{\sfrac{1}{p}},0\} & \begin{cases}(1-x)^{p-1} \mathcal{S}^{\sfrac{1}{p-1}},&\text{if }\mathcal{S}^{\sfrac{1}{p}}>1\\0 \end{cases} & {derivative vanishes on\\considerable part of\\the domain; derivative\\high, if $x$ or $y$ low\\(suitable for learning)} & \xmark \\
  \bottomrule
\end{talltblr}
      }
  \end{table}

  \citet{varnaiRobustnessMetricsLearning2020a} introduce the \emph{shadow-lifting} property for conjunction, which requires the truth value of a conjunction to increase when the truth value of a conjunct increases; formally expressed\footnote{Note that the original version from \citet{varnaiRobustnessMetricsLearning2020a} is defined for conjunctions with $M$ terms, as \ac{stl} has conjunctions with $M$ terms by design.} in~\cref{eq:shadow-lifting}: the shadow-lifting property is satisfied, if for any $\rho\neq 0$,
  \begin{equation}\label{eq:shadow-lifting}
      \pdv{\dl{x_1\wedge x_2}{$L$}}{x_i}\bigg\vert_{x_1=x_2=\rho} >0 \quad\text{for all }i\in\{1,2\}.
  \end{equation}

  This property is highly desirable for learning, as it allows for gradual improvement.
  For example, the formula $0.1\wedge 0.9$ should be more true than $0.1\wedge 0.2$, but the Gödel t-norm $T_\text{G}(x,y)=\min\{x,y\}$ yields the same truth value in both cases.
  Note that \citet{varnaiRobustnessMetricsLearning2020a} have proven that no conjunction operator can be associative and shadow-lifting at the same time.

  DL2 uses addition for conjunction, trivially satisfying shadow-lifting.
  The only t-norm to satisfy the shadow-lifting property is the product t-norm $T(x,y)=xy$.
  The Gödel, \L ukasiewicz, and Yager t-norms are not differentiable everywhere due to their use of the $\min$ and $\max$ operators, and do not satisfy shadow-lifting~\cite{affeldtTamingDifferentiableLogics2024b}.
  However, a conjunction that enjoys shadow-lifting might not be favourable for other reasons; as noted by \citet{vankriekenAnalyzingDifferentiableFuzzy2022}, the derivatives of the product t-norm will be low if $x$ and $y$ are both low, making it hard for the learning process to make progress at all.
  In contrast to that, the DL2 conjunction is not just shadow-lifting, but also has strong derivatives regardless of the values of $x$ and $y$.
  \Cref{tab:derivatives_conjunctions} shows the derivatives of the conjunctions presented earlier.

  \subsection{Modus Ponens and Modus Tollens reasoning (derivatives of implication)}
  \label{sec:modus-ponens-modus-tollens}

  \begin{table}
      \centerline{%
        \small
\begin{talltblr}[
  note{a}={Derivatives vanish on half of the domain.},
  note{b}={Modus Ponens reasoning is only possible for $x>1$. However, when $x<1$ (corresponding to being very confident in $x$), it is only possible to decrease confidence in $y$, which is the opposite of what should happen with Modus Ponens.},
  caption={The derivatives of the DL2 and fuzzy implications and their properties. A \cmark{} symbol in the Modus Ponens (or, respectively, the Modus Tollens) column indicates that the particular implication closely follows Modus Ponens (or, respectively, Modus Tollens) reasoning, while a \xmark{} indicates that this particular implication does not follow Modus Ponens (or, respectively, Modus Tollens) reasoning.},
  label={tab:derivatives_implications}
]{
  colspec={lQ[c, m, mode=dmath]Q[c, m, mode=dmath]Q[c, m, mode=dmath]Q[c, m, mode=text]Q[c, m, mode=text]},  
  row{1}={font=\bfseries, mode=text},
}
  \toprule
  Logic $L$ & Implication $\dl{x\wedge y}{$L$}$ & $\displaystyle\pdv{\dl{x\implies y}{$L$}}{x}$ & $\displaystyle\pdv{\dl{x\implies y}{$L$}}{y}$ & {Modus\\Ponens} &  {Modus\\Tollens} \\
  \midrule
  DL2 & y(1-x) & -y & 1-x & (\cmark)\TblrNote{b} & \xmark \\
  Gödel & \begin{cases}1,&\text{if }x<y,\\y&\text{else.}\end{cases} & 0 & \begin{cases}0,&\text{if }x<y\TblrNote{a},\\1,&\text{if }x>y.\end{cases} & \xmark & \xmark \\
  Kleene-Dienes & \max\{1-x,y\} & \begin{cases}0,&\text{if }x+y>1\\-1,&\text{if }x+y<1.\end{cases} & \begin{cases}0,&\text{if }x+y<1\\1,&\text{if }x+y>1.\end{cases} & \xmark & \xmark \\
  \L ukasiewicz & \min\{1,1-x+y\} & \begin{cases}0,&\text{if }x<y\TblrNote{a}\\-1,&\text{if }x>y.\end{cases} & \begin{cases}0,&\text{if }x<y\TblrNote{a}\\1,&\text{if }x>y.\end{cases} & \xmark & \xmark \\
  Reichenbach & 1-x+xy & y-1 & x & \cmark & \cmark \\
  Goguen & \begin{cases}1,&\text{if }x\le y,\\\sfrac{y}{x},&\text{else}.\end{cases} & \begin{cases}0,&\text{if }x\le y\TblrNote{a},\\-\sfrac{y}{x^2},&\text{else}.\end{cases} & \begin{cases}0,&\text{if }x\le y\TblrNote{a},\\\sfrac{1}{x},&\text{else}.\end{cases} & \xmark & \xmark \\
  Yager ($p\ge 1$) & \begin{cases}1,&\text{if }x,y=0,\\\sfrac{y}{x},&\text{else.}\end{cases} & \begin{cases}0,&\text{if }x,y=0,\\ y^x\log y,&\text{else.}\end{cases} & \begin{cases}0,&\text{if }x,y=0,\\ xy^{x-y},&\text{else.}\end{cases} & \xmark & \cmark \\
  \bottomrule
\end{talltblr}
      }
  \end{table}

  Given an implication $x\implies y$, $x$ is called the \emph{antecedent}, and $y$ is called the \emph{consequent}.

  \emph{Modus Ponens} reasoning (affirming the antecedent) allows to infer $y$ from $x$.
  On the other hand, \emph{Modus Tollens} reasoning (denying the consequent) is used to infer $\lnot x$ from $\lnot y$.

  Due to the different notions of truth in DL2 and fuzzy logics, in the following we will not refer to truth values directly, but rather use the concept of \emph{confidence}.

  Modus Ponens reasoning needs to be able to increase the confidence in the consequent $y$, if the confidence in the antecedent $x$ is high.
  Modus Tollens reasoning needs to be able to decrease the confidence in the antecedent $x$, if the confidence in the consequent $y$ is low.
  Thus, due to their different interpretations of truth, in DL2, increasing the confidence in $\phi$ means decreasing the value of $\dl{\phi}{DL2}$, whereas in fuzzy logics, increasing the confidence in $\phi$ corresponds to increasing the value of $\dl{\phi}{FL}$.
  
  \Cref{tab:derivatives_implications} lists derivatives of the DL2 and fuzzy implications, along with a description of their behaviour, and whether they permit Modus Ponens and Modus Tollens reasoning.

  As noted by \citet{vankriekenAnalyzingDifferentiableFuzzy2022}, not only is most background knowledge phrased as implications, but also there are often far more negative examples than they are positive ones in \ac{ml} contexts, thus requiring Modus Tollens reasoning more frequently.
  Choosing a suitable implication that performs well in the presence of this Modus Ponens / Modus Tollens imbalance is thus an important task to guarantee best learning.

  Next, we examine derivatives of $\dl{x\implies y}{$L$}$ and investigate whether they permit Modus Ponens reasoning, and, more importantly, Modus Tollens reasoning.
  For the Gödel, Kleene-Dienes, \L ukasiewicz, Reichenbach, and Goguen implication, we briefly summarise known results from the literature~\cite{vankriekenAnalyzingDifferentiableFuzzy2022}, and contribute investigations into the derivatives and Modus Ponens and Modus Tollens behaviour for DL2 and the Yager implication.

  \begin{itemize}
      \item In DL2, Modus Tollens reasoning is not possible; since the derivative with respect to the antecedent $x$ is $-y$, it is never possible to increase $x$ (i.e. to decrease the confidence in $x$).
      Modus Ponens reasoning is possible only for values of $x>1$, which means when we are very confident in $x$, it is not possible to increase the confidence in $y$---even worse, the confidence in $y$ will be \emph{decreased}.
      \item For the Gödel implication, derivatives with respect to the antecedent $x$ do not exist; it is thus impossible to perform Modus Tollens reasoning.
      Whenever $x>y$, the confidence in $y$ is arbitrarily increased, even when the confidence in $x$ is low; thus not exactly following Modus Ponens.
      Further, when $x<y$, derivatives vanish.
      \item The Kleene-Dienes implication does not closely follow Modus Ponens or Modus Tollens; instead, the confidence in the antecedent $x$ is decreased when the confidence in $x$ and $y$ is low, and the confidence in the consequent $y$ is increased, if the confidence in $x$ and $y$ is high.
      \item The \L ukasiewicz logic has another issue: the confidence in the antecedent can never be decreased; it therefore does not follow Modus Tollens reasoning.
      Whenever we are more confident in antecedent $x$ than the in consequent $y$, the confidence in $y$ will be increased, which does not closely follow Modus Ponens.
      Further, when $x<y$, derivatives vanish.
      \item The Reichenbach implication closely follows Modus Ponens (whenever the confidence in the antecedent $x$ is high, the confidence in the consequent $y$ will be increased) and Modus Tollens (if the confidence in the consequent $y$ is low, the confidence in the antecedent $x$ will be decreased).
      \item The Goguen implication does not follow Modus Ponens: whenever the confidence in $x$ is very low, the confidence in $y$ will be increased strongly.
      Modus Tollens reasoning is possible, however, as $-\sfrac{y}{x^2}$ gets smaller for increasing values of $x$, it also behaves the opposite of what should be expected: when the confidence in $x$ is very high, it should be decreased faster than when it is low.
      Further, when $x\le y$, derivatives vanish.
      \item The Yager implication follows Modus Tollens reasoning but not Modus Ponens reasoning; when the confidence in $x$ is high but the confidence in $y$ is low, derivatives with respect to $y$ are low instead of high; but when $x$ is high, derivatives are high, too.
  \end{itemize}

  \subsection{Consistency (integrals of fuzzy logic operators)}
  \label{sec:consistency}

  \begin{table}
    \centering
    \caption{The consistency of different fuzzy logics over a representative set of axioms, showing that the \L ukasiewicz, sigmoidal Reichenbach, and Goguen logics are the most consistent, and the Gödel logic is the least consistent for this set of axioms.}
    \label{tab:consistency}
    \centerline{
      \small
\begin{tblr}
{
    colspec=lccccccc,
    cells={mode=dmath},
    row{1}={font=\bfseries, mode=text},
    row{Z}={font=\bfseries, mode=text},
}
\toprule
\textbf{Tautology} & \rot{Gödel} & \rot{Kleene-Dienes} & \rot{Łukasiewicz} & \rot{Reichenbach} & \rot{Goguen} & \rot{sig. Reichenbach} & \rot{Yager}\\
\midrule
{\footnotesize\text{\textbf{Axiom schemata}}} &  &  &  &  &  &  & \\
P \implies (Q \implies P) & 0.67 & 0.79 & 1 & 0.92 & 1 & 0.99 & 0.89\\
(P \implies (Q \implies R)) \implies ( (P \implies Q) \implies (P \implies R) ) & 0.75 & 0.75 & 0.96 & 0.87 & 0.93 & 0.97 & 0.85\\
(\lnot P \implies \lnot Q) \implies (Q \implies P) & 0.79 & 0.75 & 1 & 0.86 & 0.90 & 0.98 & 0.82\\
\midrule
{\footnotesize\text{\textbf{Primitive propositions}}} &  &  &  &  &  &  & \\
(P \vee P) \implies P & 0.50 & 0.75 & 0.75 & 0.75 & 0.69 & 0.88 & 0.72\\
Q \implies (P \vee Q) & 0.83 & 0.79 & 1 & 0.92 & 1 & 0.98 & 0.90\\
(P \vee Q) \implies (Q \vee P) & 0.67 & 0.75 & 1 & 0.86 & 1 & 0.96 & 0.85\\
(P \vee (Q \vee R)) \implies (Q \vee (P \vee R)) & 0.75 & 0.78 & 1 & 0.91 & 1 & 0.98 & 0.91\\
(Q \implies R) \implies ( (P \vee Q) \implies (P \vee R)) & 0.88 & 0.76 & 1 & 0.90 & 1 & 0.99 & 0.90\\
\midrule
{\footnotesize\text{\textbf{Law of excluded middle}}} &  &  &  &  &  &  & \\
P \vee \lnot P & 0.75 & 0.75 & 1 & 0.83 & 0.83 & 0.83 & 0.81\\
\midrule
{\footnotesize\text{\textbf{Law of contradiction}}} &  &  &  &  &  &  & \\
\lnot(P \wedge \lnot P) & 0.75 & 0.75 & 1 & 0.83 & 0.83 & 0.83 & 0.81\\
\midrule
{\footnotesize\text{\textbf{Law of double negation}}} &  &  &  &  &  &  & \\
P \iff \lnot (\lnot P)) & 0.50 & 0.75 & 1 & 0.70 & 1 & 0.91 & 0.69\\
\midrule
{\footnotesize\text{\textbf{Principles of transposition}}} &  &  &  &  &  &  & \\
(P \iff Q) \iff (\lnot P \iff \lnot Q) & 0.17 & 0.67 & 1 & 0.61 & 0.59 & 0.93 & 0.57\\
( (P \wedge Q) \implies R ) \iff ( (P \wedge \lnot R) \implies \lnot Q) & 0.51 & 0.64 & 0.67 & 0.67 & 0.65 & 0.73 & 0.64\\
\midrule
{\footnotesize\text{\textbf{Laws of tautology}}} &  &  &  &  &  &  & \\
P \iff (P \wedge P) & 0.50 & 0.75 & 0.75 & 0.69 & 0.50 & 0.87 & 0.44\\
P \iff (P \vee P) & 0.50 & 0.75 & 0.75 & 0.69 & 0.69 & 0.87 & 0.69\\
\midrule
{\footnotesize\text{\textbf{Laws of absorption}}} &  &  &  &  &  &  & \\
(P \implies Q) \iff (P \iff (P \wedge Q)) & 0.33 & 0.71 & 0.83 & 0.66 & 0.67 & 0.94 & 0.49\\
Q \implies (P \iff (P \wedge Q)) & 0.33 & 0.75 & 1 & 0.82 & 1 & 0.98 & 0.57\\
\midrule
{\footnotesize\text{\textbf{Assoc., comm., dist. laws}}} &  &  &  &  &  &  & \\
(P \wedge (Q \vee R)) \iff ( (P \wedge Q) \vee (P \wedge R) ) & 0.42 & 0.72 & 0.90 & 0.69 & 0.89 & 0.90 & 0.67\\
(P \vee (Q \wedge R)) \iff ( (P \vee Q) \wedge (P \vee R) ) & 0.58 & 0.72 & 0.90 & 0.69 & 0.90 & 0.90 & 0.69\\
\midrule
{\footnotesize\text{\textbf{De Morgan's laws}}} &  &  &  &  &  &  & \\
\lnot(P \wedge Q) \iff (\lnot P \vee \lnot Q) & 0.67 & 0.75 & 1 & 0.75 & 1 & 0.93 & 0.78\\
\lnot(P \vee Q) \iff (\lnot P \wedge \lnot Q) & 0.33 & 0.75 & 1 & 0.75 & 1 & 0.93 & 0.73\\
\midrule
{\footnotesize\text{\textbf{Material excluded middle}}} &  &  &  &  &  &  & \\
(P \implies Q) \vee (Q \implies P) & 1 & 0.83 & 1 & 0.97 & 1 & 1 & 1\\
\bottomrule
Average Consistency & 0.60 & 0.75 & 0.93 & 0.79 & 0.87 & 0.92 & 0.75\\
\end{tblr}
    }
  \end{table}
  
  Whereas derivatives of logical operators give insights into how they behave in the machine learning process, another view is from a logic perspective, looking at their integrals, as proposed by~\citet{grespanEvaluatingRelaxationsLogic2021}, who introduce a notion of \emph{consistency} (that is, a measure of how well truth is preserved) for fuzzy logics based on the intuition that a fuzzy logic tautology $\tau$ should be absolutely true for all possible truth values, i.e. its integral should evaluate to $1$, formally expressed in~\cref{eq:consistency} for a given fuzzy logic relaxation $L$.
  \begin{equation}\label{eq:consistency}
    \int \cdots \int_{[0,1]} \dl{\tau(x_1, \dots , x_n)}{$L$} \,\mathrm{d}x_n \cdots \mathrm{d}x_1
  \end{equation}

  We use the same representative set of axioms as~\cite{grespanEvaluatingRelaxationsLogic2021}, and evaluate the consistency of these for all fuzzy logics listed in~\cref{tab:fuzzy_logics}.
  The results are displayed in~\cref{tab:consistency} and indicate that the \L ukasiewicz logic, the Goguen logic, and the sigmoidal Reichenbach logic~ \cite{vankriekenAnalyzingDifferentiableFuzzy2022} are the most consistent, i.e. they preserve truth the most.
  Similarly, we observe that Gödel is generally the least consistent for the set of axioms chosen.

  This confirms the results presented in~\cite{grespanEvaluatingRelaxationsLogic2021}, who found the \L ukasiewicz and Goguen logics (logics with $R$-implications) to be more consistent than the Reichenbach and Kleene-Dienes logics (logics with $S$-implications).
  We found the Gödel logic (a logic with an $R$-implication) to preserve truth the least, so it cannot be concluded that logics with $R$-implications are generally preferable to those with $S$-implications.
  
  Note however, that consistency is not a definite metric to evaluate fuzzy logics with, as it depends on a specific set of axioms chosen and should only be used as an indicator.

  Further, it is not possible to evaluate DL2 with this measure. This is because the set of axioms on which we could evaluate it is severely restricted due to both DL2's lack of a general negation and the fact that atomic terms in DL2 are comparison, rather than atomic truth values.
  
  \section{Experimental Evaluation \& Results}
  \label{sec:experimental}
  In this section we revise the experimental set up in~\citet{flinkowComparingDifferentiableLogics2023} so that constraints are not limited to data that is already in the training set, and loss terms are fairly balanced.

  As differentiable logics translate constraint satisfaction into loss, the only time the network can learn to satisfy a constraint is when it encounters examples that do not satisfy the constraint.
  Instead of relying on the dataset to contain sufficiently many, and sufficiently suitable counterexamples, we now employ an oracle to find counterexamples \emph{outside} of the dataset that violate the constraint.
  This leads to an improved comparison of differentiable logics by ensuring that the logics have an optimal positive impact on the learning process.

  The idea of learning with constraints that can refer to data outside of the training set by finding a counterexample that violates the constraint was introduced in DL2~\cite{fischerDL2TrainingQuerying2019} and is one of its major strengths.
  Following DL2, we limit our experiments to training networks to satisfy constraints of the form
  \begin{equation}\label{eq:constraint_general}
    \forall\vec{x}'\in \epsball{\vec{x}}{\epsilon} \ldotpp \phi(\vec{x}, \vec{x}', y),
  \end{equation}
  given an input $\vec{x}$ and label $y$.

  This restriction allows PGD to approximate a counterexample $\vec{x}^*$ for constraint $\phi$ by maximising its loss.
  Formally,
  \begin{equation}
    \vec{x}^*=\argmax_{\vec{x}'\in \epsball{\vec{x}}{\epsilon}}\; \mathcal{L}_{\text{C}}(\vec{x}, \vec{x}', y, \phi).
  \end{equation}

  Learning to satisfy constraints of the form shown in \cref{eq:constraint_general} thus corresponds to a two-step process of first finding a counterexample $\vec{x}^*$ that violates the constraint, and then minimising the total loss
  \begin{equation}\label{eq:loss_total_counterexample}
    \mathcal{L}(\vec{x}, y, \phi)=\lambda_{\text{P}}\mathcal{L}_{\text{P}}(\vec{x},y)+\lambda_{\text{C}}\mathcal{L}_{\text{C}}(\vec{x},\vec{x}^*,y,\phi),
  \end{equation}
  where the constraint loss is given by
  \begin{equation}
      \mathcal{L}_{\text{C}}(\vec{x},\vec{x}^*,y,\phi)=\begin{cases}
          \dl{\phi(\vec{x},\vec{x}^*,y)}{DL2},&\text{for DL2},\\
          1-\dl{\phi(\vec{x},\vec{x}^*, y)}{$L$},&\text{for a given fuzzy logic $L$}.
      \end{cases}
  \end{equation}

  \subsection{Practical considerations}
  \subsubsection{Auto-PGD}
  In theory, we assume the oracle to always succeed in finding a counterexample for the constraint.
  In practice, when using PGD as the oracle as initially suggested by \citet{fischerDL2TrainingQuerying2019}, PGD attacks can fail for various reasons, which has implications for the comparison: when the PGD attack fails to find a counterexample, this does not necessarily mean the network is robust, but can also indicate unfavourable attack parameters.

  As noted by \citet{mosbachLogitPairingMethods2019,croceScalingRandomizedGradientFree2020}, PGD attacks in particular rely heavily on the attack parameters (iterations and step size) and can fail quite often.
  \citet{croceReliableEvaluationAdversarial2020a} identify the fixed step size as a particular issue, and propose Auto-PGD, which improves attack success by adapting the step size dynamically based on the loss surface.
  At the same time, their approach reduces the number of free parameters to only the number of iterations.

  We use Auto-PGD with \numrange{20}{40} iterations and \numrange{3}{70} random restarts in our experiments, and twice as many iterations during testing to evaluate constraint satisfaction on stronger counterexamples.
  
  \subsubsection{GradNorm}
  The additional constraint loss term $\mathcal{L}_{\text{C}}$ introduces hyperparameters $\lambda_{\text{P}}$ and $\lambda_{\text{C}}$ that are responsible for balancing the two loss terms in \cref{eq:loss_total_counterexample}.
  These hyperparameters need to be well fine-tuned to properly balance the different loss terms.
  This is a prerequisite for the logical loss to perform optimally and have the most impact.
  Our previous work used grid search to find approximate values; however, despite being time-consuming, it is also unlikely to yield optimal values, which had consequences for the fairness of our comparison.

  In this paper, we use GradNorm by \citet{chenGradNormGradientNormalization2018a} to provide a fairer comparison.
  GradNorm is an adaptive loss balancing algorithm in multi-task learning scenarios, where the total loss is a weighted sum of individual task losses.
  GradNorm treats these loss weights as learnable parameters that are allowed to vary for each epoch.
  The network is penalised when backpropagated gradients from any task are too large or too small, which should lead to the different tasks training at similar rates.
  GradNorm at least matches, and often outperforms, exhaustive grid search, not only reducing the time needed to train with close-to-optimal weights, but also leading to overall improved task performance.
  GradNorm provides only one hyperparameter, the so-called \emph{asymmetry parameter} $\alpha$, where higher values of $\alpha$ lead to stronger training rate balancing. 

  \subsection{Setup}
  Our training experiment is implemented in PyTorch~\cite{paszkePyTorchImperativeStyle2019} and based on the original DL2 experiments~\cite{fischerDL2TrainingQuerying2019,EthsriDl22024}.
  The code, together with the experimental data and our reported results, is publicly available at \url{https://github.com/tflinkow/comparing-differentiable-logics}.
  The goal of this experiment is to compare the differentiable logics introduced in~\cref{sec:background}, and investigate which logic performs most favourable during training, focusing specifically on implication and conjunction, as these have noticeable consequences for the learning process.
  
  All training experiments were conducted on a system with an NVIDIA RTX4090 GPU with \qty{24}{\giga\byte} of VRAM.
  All experiments were run for 50 epochs.
  Running all experiments took \qty{2}{\day} \qty{10}{\hour} \qty{51}{\minute} to complete.
  
  The baseline for each experiment was achieved by training without any logical loss translation.
  We train with the MNIST~\cite{lecunGradientbasedLearningApplied1998}, Fashion-MNIST~\cite{xiaoFashionMNISTNovelImage2017}, CIFAR-10~\cite{krizhevskyLearningMultipleLayers2009}, and the \ac{gtsrb}~\cite{stallkampGermanTrafficSign2011} datasets with various logical constraints.
  For MNIST, Fashion-MNIST, and CIFAR-10 we use simple \acp{cnn}, and for \ac{gtsrb} we use a \ac{cnn} with a spatial transformer\footnote{\url{https://github.com/poojahira/gtsrb-pytorch}}.
  
  We use the AdamW~\cite{loshchilovDecoupledWeightDecay2018} optimiser with learning rates of \numrange[scientific-notation=true]{0.00001}{0.001} and a weight decay of \num[scientific-notation=true]{0.0001}.
  To keep computation tractable while enabling strong PGD attacks during training, we use large batch sizes ranging from \numrange{1024}{4096}.

  In order to ensure that networks do not overfit despite this, data augmentation techniques (including random rotations, random crops, colour jitter, and random affine transformations) are used during training, and all metrics reported in the following are obtained from the networks' performance on the test set.
  
  \subsection{Evaluation metrics}  
\begin{figure}
    \centering
    \begin{subfigure}{.3\textwidth}
      \centering
      \includegraphics[width=.35\linewidth]{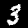}  
      \caption{Original image $\vec{x}$.}
      \label{fig:img_input}
    \end{subfigure}
    \begin{subfigure}{.3\textwidth}
      \centering
      \includegraphics[width=.35\linewidth]{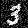}  
      \caption{Random sample from $\epsball{\vec{x}}{0.4}$.}
      \label{fig:img_random}
    \end{subfigure}
    \begin{subfigure}{.3\textwidth}
      \centering
      \includegraphics[width=.35\linewidth]{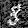}  
      \caption{Counterexample from $\epsball{\vec{x}}{0.4}$.}
      \label{fig:img_pgd}
    \end{subfigure}
    \caption{Example images from training with the $\mathsf{StandardRobustness}(\epsilon=0.4, \delta=0.1)$ constraint on MNIST showing an original image in~\cref{fig:img_input}, a random sample from $\epsball{\vec{x}}{0.4}$ in~\cref{fig:img_random}, and a constraint counterexample in~\cref{fig:img_pgd}.}
    \label{fig:metrics}
  \end{figure}

  In the following, let $X=\{(\vec{x}_i, y_i)\}$ be the test set consisting of pairs of input data $\vec{x}$ and true label $y$, and further, let $[\cdot]$ denote the standard indicator function.
  As shown on an example on MNIST in~\cref{fig:metrics}, we use $\mathsf{random}(\vec{x};\epsilon)$ to denote a randomly-uniformly-sampled element from $\epsball{\vec{x}}{\epsilon}$, and $\mathsf{counterexample}_{\phi}(\vec{x};\epsilon)$ to denote an approximate counterexample for constraint $\phi$ from $\epsball{\vec{x}}{\epsilon}$. 

  Our evaluation metrics are as formally defined by~\citet[Def.~5--7]{casadioNeuralNetworkRobustness2022}:
  \begin{itemize}
  \item \emph{Prediction Accuracy (PAcc)} is the fraction of correct predictions:
  \begin{equation}\label{eq:pacc}
      \text{PAcc}:=\frac{1}{|X|}\sum\limits_{(\vec{x},y)\in X}[\mathcal
      {N}(\vec{x})=y].
  \end{equation}
 \item \emph{Constraint Accuracy (CAcc)} is the fraction of random samples that satisfy the constraint $\phi$:
  \begin{equation}\label{eq:cacc}
      \text{CAcc}:=\frac{1}{|X|}\sum\limits_{(\vec{x},y)\in X}[\mathcal{N}(\vec{x})\vDash\phi(\vec{x}, \mathsf{random}(\vec{x};\epsilon), y)].
  \end{equation}
  \item \emph{Constraint Security (CSec)} is the fraction of adversarial samples that satisfy the constraint $\phi$:
  \begin{equation}\label{eq:csec}
      \text{CSec}:=\frac{1}{|X|}\sum\limits_{(\vec{x},y)\in X}[\mathcal{N}(\vec{x})\vDash\phi(\vec{x}, \mathsf{counterexample}_{\phi}(\vec{x};\epsilon), y)].
  \end{equation}
  \end{itemize}

  The results reported in~\cref{tab:robustness_results,tab:groups_gtsrb,tab:even-odd-mnist,tab:class-similarity_cifar10} show the scenarios with the highest combined performance  of the last \qty{10}{\percent} of experimental runs.
  The highest combined performance is the one that maximises the product of prediction accuracy and constraint security.
  
  The plots in~\cref{fig:robustness_results,fig:groups_gtsrb,fig:even-odd-mnist,fig:class-similarity_cifar10} show how prediction accuracy, constraint accuracy, and constraint security change over time.
  The best result (i.e. the values reported in the tables) is indicated in each plot.

  \subsection{Training neural networks with differentiable logics}\label{subsec:training_experiment}
  In the following, we explain the logical constraints used during training, specifically chosen to evaluate different logical operators.
  In practice, constraints are often large conjunctions of basic atomic propositions, such as in the ROAD-R dataset~\cite{giunchigliaROADRAutonomousDriving2023}: $(\lnot\mathsf{Pedestrian}\wedge\lnot\mathsf{Cyclist})\wedge(\lnot\mathsf{Car}\wedge\lnot\mathsf{Cyclist})\wedge\, \ldots\ $.
  These conjunctions are useful for checking and correcting network predictions, ensuring predictions align with the background knowledge.
  However, in this work, we want to compare operators other than conjunction.
  For this reason, we use constraints specifically selected to involve certain logical operators.

  \subsubsection{Comparing comparison operators}
  Classification robustness~\cite[Def.~1]{casadioNeuralNetworkRobustness2022} is expressed as
  \begin{equation}\label{eq:classification_robustness}
    \mathsf{ClassificationRobustness}(\epsilon):=\ \forall \vec{x}'\in \epsball{\vec{x}}{\epsilon} \ldotpp \argmax_i\mathcal{N}(\vec{x}')=y,
  \end{equation}
  where $y$ is the true label, requiring the prediction for the perturbed input to be correct.
  This involves the non-differentiable $\argmax$ operator, making it unsuitable for use in a differentiable loss function setting.
  
  A variant to circumvent this problem was proposed in DL2~\cite{fischerDL2TrainingQuerying2019}:
  \begin{equation}\label{eq:robustness_constraint_dl2}
      \mathsf{StrongClassificationRobustness}(\epsilon,\delta):=\ \forall \vec{x}'\in \epsball{\vec{x}}{\epsilon} \ldotpp \mathcal{N}(\vec{x}')_y\ge \delta.
  \end{equation}
  Note that this variant only compares the network's prediction for the true class against some threshold value (chosen to be $\delta=0.52$ in the original DL2 experiments).

  \citet{casadioNeuralNetworkRobustness2022} note that both of these robustness formulations fail to capture the intended semantics, because they refer to the prediction for the true class instead of the current network output.
  This is unfavourable, because the standard loss term already penalises incorrect predictions, and the standard robustness term should only focus on robustness, that is, the network's prediction should not change when slightly perturbing the input (even when the original prediction is incorrect).
  Therefore, we use the notion of \emph{standard robustness}~\cite[Def.~2]{casadioNeuralNetworkRobustness2022} and train with the \textsf{StandardRobustness} constraint shown in \cref{eq:constraint_standard_robustness}:
  \begin{equation}\label{eq:constraint_standard_robustness}
    \mathsf{StandardRobustness}(\epsilon, \delta):=\ \forall \vec{x}'\in \epsball{\vec{x}}{\epsilon} \ldotpp \lVert\mathcal{N}(\vec{x}')-\mathcal{N}(\vec{x})\rVert_{\infty}\le \delta.
  \end{equation}

  We use this constraint on MNIST and Fashion-MNIST.

  \paragraph{Results}
  
  \begin{figure}
    \begin{subfigure}{\textwidth}
      \centering
      \begin{tikzpicture}[font=\small]
  \begin{groupplot}[group/results]
    \nextgroupplot[title={Prediction Accuracy (PAcc)},]
\addplot+[mark indices=47, densely dotted] table [y=Test-P-Acc] {r_standard-robustness-mnist-Baseline.csv};
\addplot+[mark indices=49] table [y=Test-P-Acc] {r_standard-robustness-mnist-DL2.csv};
\addplot+[mark indices=47] table [y=Test-P-Acc] {r_standard-robustness-mnist-Goedel.csv};

\coordinate (c1) at (rel axis cs:0,1);
    \nextgroupplot[title={Constraint Accuracy (CAcc)},
      yticklabels={},
      xlabel={},
    ]
\addplot+[mark indices=47, densely dotted] table [y=Test-C-Acc] {r_standard-robustness-mnist-Baseline.csv};
\addplot+[mark indices=49] table [y=Test-C-Acc] {r_standard-robustness-mnist-DL2.csv};
\addplot+[mark indices=47] table [y=Test-C-Acc] {r_standard-robustness-mnist-Goedel.csv};

\coordinate (c2) at (rel axis cs:0,1);
    \nextgroupplot[title={Constraint Security (CSec)},
      yticklabel pos=right,
      yticklabel style={anchor=east,xshift=2.5em},
      legend to name=full-legend
    ]
\addplot+[mark indices=47, densely dotted] table [y=Test-C-Sec] {r_standard-robustness-mnist-Baseline.csv};
\addplot+[mark indices=49] table [y=Test-C-Sec] {r_standard-robustness-mnist-DL2.csv};
\addplot+[mark indices=47] table [y=Test-C-Sec] {r_standard-robustness-mnist-Goedel.csv};
\addlegendentry {Baseline};
\addlegendentry {DL2};
\addlegendentry {Fuzzy Logic};

\coordinate (c3) at (rel axis cs:1,1);
  \end{groupplot}
  \coordinate (c4) at ($(c1)!.5!(c3)$);
  \node[below, yshift=0.5cm] at (c4 |- current bounding box.south) {\pgfplotslegendfromname{full-legend}};
\end{tikzpicture}
      \caption{$\mathsf{StandardRobustness}(\epsilon=0.4, \delta=0.1)$ on MNIST.}
      \label{fig:robustness_mnist}
    \end{subfigure}
    
    \bigskip

    \begin{subfigure}{\textwidth}
      \centering
      \begin{tikzpicture}[font=\small]
  \begin{groupplot}[group/results]
    \nextgroupplot[title={Prediction Accuracy (PAcc)},]
\addplot+[mark indices=50, densely dotted] table [y=Test-P-Acc] {r_standard-robustness-fmnist-Baseline.csv};
\addplot+[mark indices=50] table [y=Test-P-Acc] {r_standard-robustness-fmnist-DL2.csv};
\addplot+[mark indices=51] table [y=Test-P-Acc] {r_standard-robustness-fmnist-Goedel.csv};

\coordinate (c1) at (rel axis cs:0,1);
    \nextgroupplot[title={Constraint Accuracy (CAcc)},
      yticklabels={},
      xlabel={}
    ]
\addplot+[mark indices=50, densely dotted] table [y=Test-C-Acc] {r_standard-robustness-fmnist-Baseline.csv};
\addplot+[mark indices=50] table [y=Test-C-Acc] {r_standard-robustness-fmnist-DL2.csv};
\addplot+[mark indices=51] table [y=Test-C-Acc] {r_standard-robustness-fmnist-Goedel.csv};

\coordinate (c2) at (rel axis cs:0,1);
    \nextgroupplot[title={Constraint Security (CSec)},
      yticklabel pos=right,
      yticklabel style={anchor=east,xshift=2.5em},
      legend to name=full-legend
    ]
\addplot+[mark indices=50, densely dotted] table [y=Test-C-Sec] {r_standard-robustness-fmnist-Baseline.csv};
\addplot+[mark indices=50] table [y=Test-C-Sec] {r_standard-robustness-fmnist-DL2.csv};
\addplot+[mark indices=51] table [y=Test-C-Sec] {r_standard-robustness-fmnist-Goedel.csv};
\addlegendentry {Baseline};
\addlegendentry {DL2};
\addlegendentry {Fuzzy Logic};

\coordinate (c3) at (rel axis cs:1,1);
  \end{groupplot}
  \coordinate (c4) at ($(c1)!.5!(c3)$);
  \node[below, yshift=0.5cm] at (c4 |- current bounding box.south) {\pgfplotslegendfromname{full-legend}};
\end{tikzpicture}
      \caption{$\mathsf{StandardRobustness}(\epsilon=0.2, \delta=0.1)$ on Fashion-MNIST.}
      \label{fig:robustness_fmnist}
    \end{subfigure}
    \caption{Results of training with the $\mathsf{StandardRobustness}$ constraint on MNIST in~\cref{fig:robustness_mnist} and on Fashion-MNIST in~\cref{fig:robustness_fmnist}.}
    \label{fig:robustness_results}
  \end{figure}
  
  \begin{table}
    \centering
    \caption{Results of training with the $\mathsf{StandardRobustness}$ constraint on MNIST in~\cref{tab:robustness_mnist} and on Fashion-MNIST in~\cref{tab:robustness_fmnist}.
    The best result is displayed in boldface in each table.}
    \label{tab:robustness_results}
    \begin{subtable}{.45\textwidth}
      \centering
      \caption{$\mathsf{StandardRobustness}(\epsilon=0.4, \delta=0.1)$ on MNIST.}
      \label{tab:robustness_mnist}
      \scriptsize
      \begin{tblr}
  {
    colspec={Q[l, mode=text]Q[c, mode=text]Q[c, mode=text]Q[c, mode=text]},
    row{1}={font=\bfseries, mode=text},
  }
    \toprule
      Logic & PAcc & CAcc & CSec \\
    \midrule
Baseline & \qty{99.17}{\percent} & \qty{79.41}{\percent} & \qty{0.02}{\percent} \\
DL2 & \qty{92.97}{\percent} & \qty{99.44}{\percent} & \qty{80.82}{\percent} \\
Fuzzy Logic & \textbf{\qty{91.02}{\percent}} & \textbf{\qty{100.00}{\percent}} & \textbf{\qty{93.47}{\percent}} \\

    \bottomrule
  \end{tblr}
    \end{subtable}%
    \hfil
    \begin{subtable}{.45\textwidth}
      \centering
      \caption{$\mathsf{StandardRobustness}(\epsilon=0.2, \delta=0.1)$ on Fashion-MNIST.}
      \label{tab:robustness_fmnist}
      \scriptsize
      \begin{tblr}
  {
    colspec={Q[l, mode=text]Q[c, mode=text]Q[c, mode=text]Q[c, mode=text]},
    row{1}={font=\bfseries, mode=text},
  }
    \toprule
      Logic & PAcc & CAcc & CSec \\
    \midrule
Baseline & \qty{89.79}{\percent} & \qty{55.81}{\percent} & \qty{0.02}{\percent} \\
DL2 & \qty{79.22}{\percent} & \qty{99.66}{\percent} & \qty{58.14}{\percent} \\
Fuzzy Logic & \textbf{\qty{75.23}{\percent}} & \textbf{\qty{99.55}{\percent}} & \textbf{\qty{86.89}{\percent}} \\

    \bottomrule
  \end{tblr}
    \end{subtable}
  \end{table}

  Results of training with the $\mathsf{StandardRobustness}$ constraint are shown in~\cref{fig:robustness_results,tab:robustness_results} and show that both DL2 and fuzzy logic lead to significantly improved constraint security.

  On MNIST, as shown in~\cref{fig:robustness_mnist,tab:robustness_mnist}, DL2 is able to improve constraint security to \qty{80.82}{\percent} and fuzzy logic to \qty{93.47}{\percent}, compared to the baseline with \qty{0.02}{\percent}.
  Prediction accuracy drops from \qty{99.17}{\percent} for the baseline to \qty{92.97}{\percent} for DL2, and to \qty{91.02}{\percent} for fuzzy logic.

  On Fashion-MNIST, shown in~\cref{fig:robustness_fmnist,tab:robustness_fmnist}, the baseline has a constraint security of \qty{0.02}{\percent}, whereas DL2 has a constraint security of \qty{58.14}{\percent}, and fuzzy logic \qty{86.89}{\percent}.
  Prediction accuracy drops from \qty{89.79}{\percent} of the baseline to \qty{79.22}{\percent} for DL2 and to \qty{75.23}{\percent} for fuzzy logic.

  The fuzzy logic comparison operator performs best in these experiments as it leads to higher constraint security compared to DL2, albeit at the expense of prediction accuracy.

  \subsubsection{Comparing conjunction and disjunction}

  \begin{figure}
  \centering
    \begin{subfigure}[c]{.25\textwidth}   
      \centering 
      \includegraphics[width=.1333\linewidth]{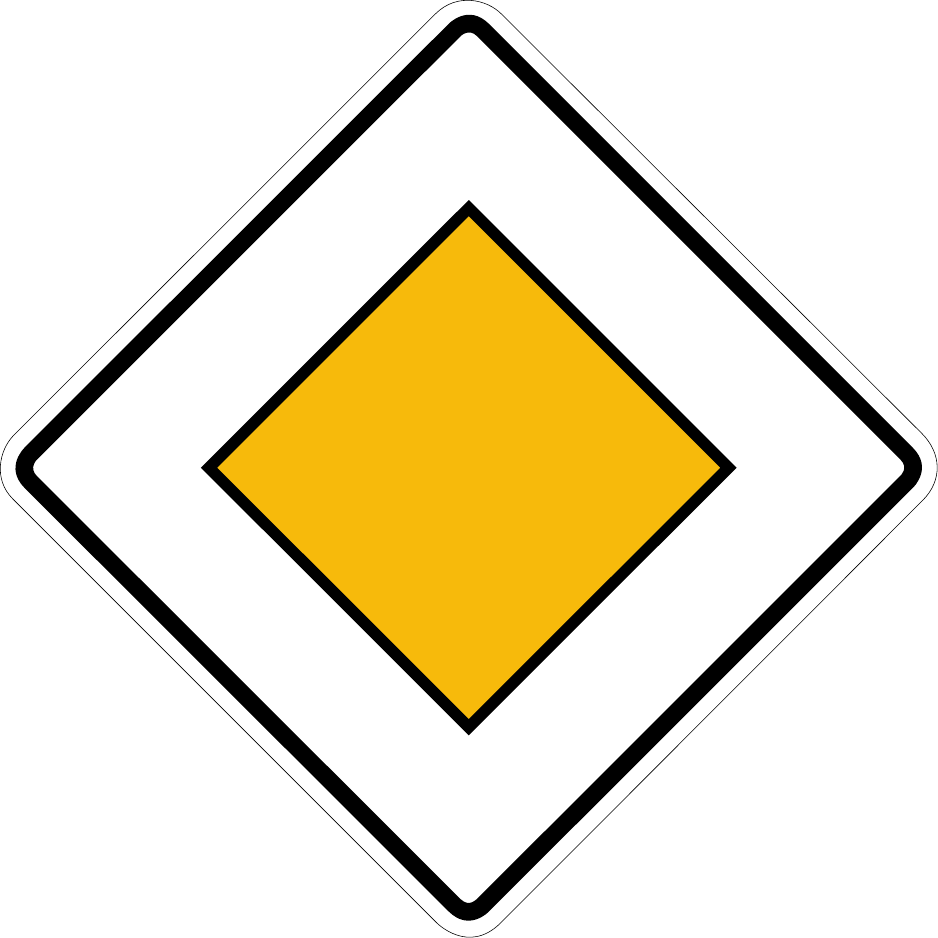}
      \includegraphics[width=.1333\linewidth]{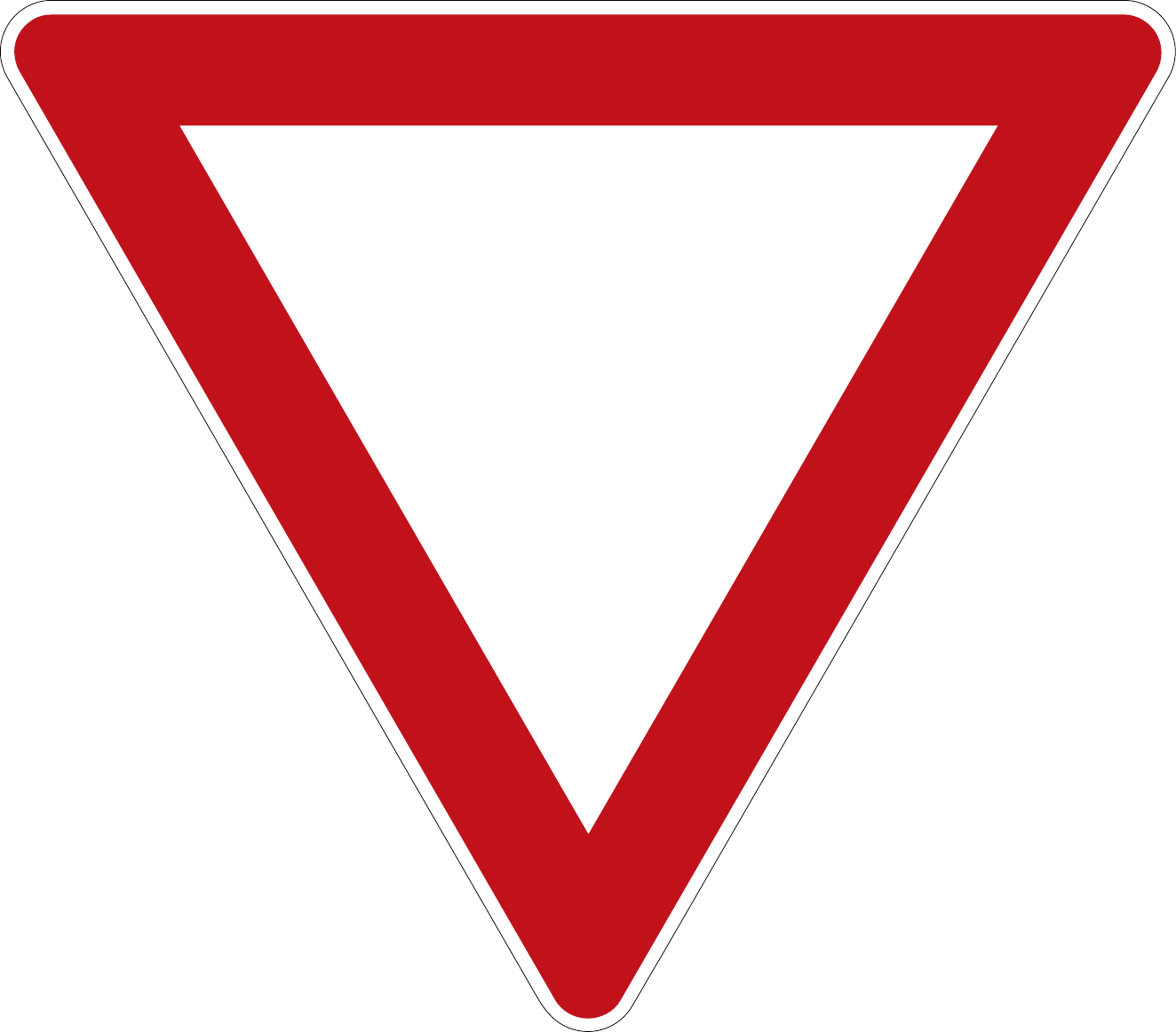}
      \includegraphics[width=.1333\linewidth]{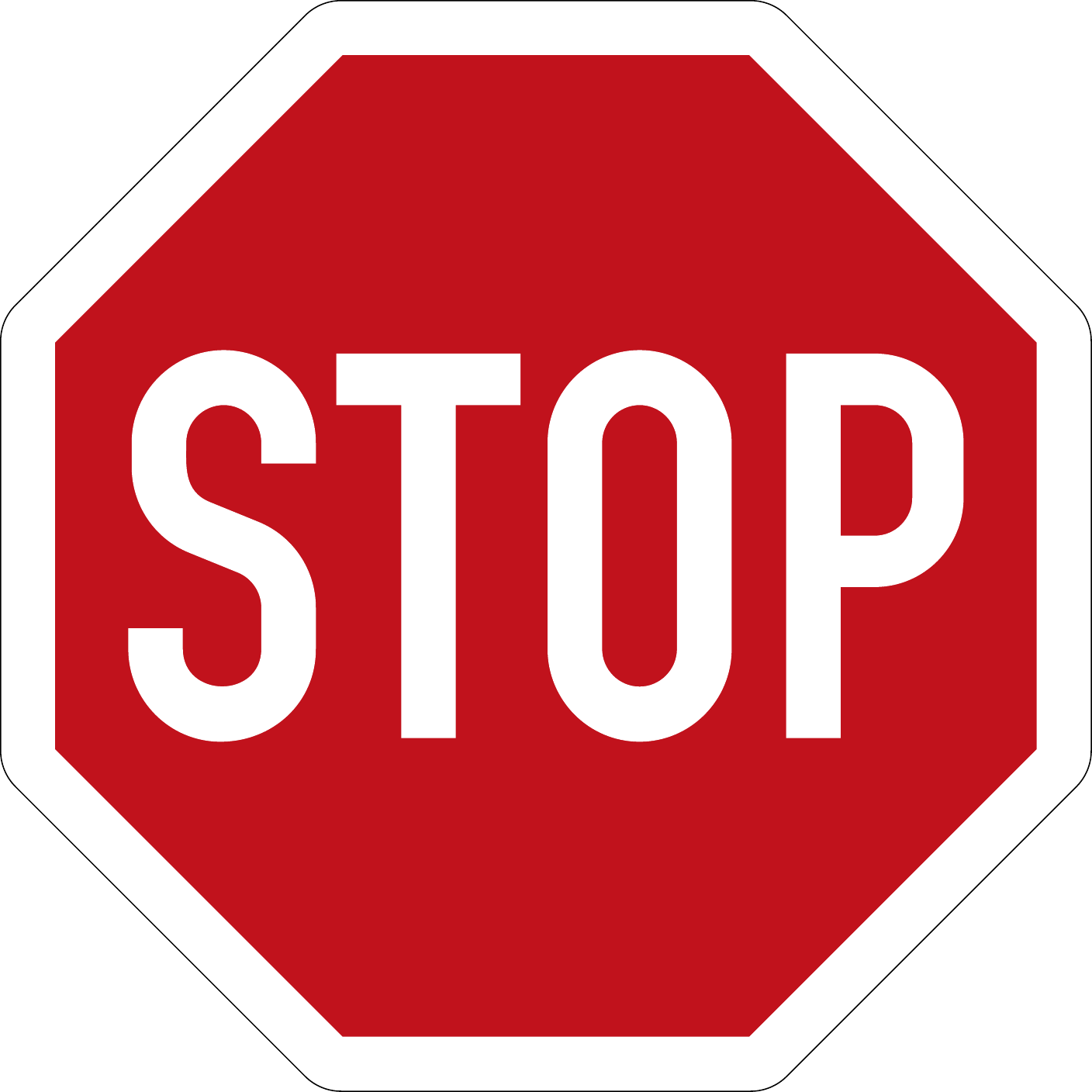}
      \includegraphics[width=.1333\linewidth]{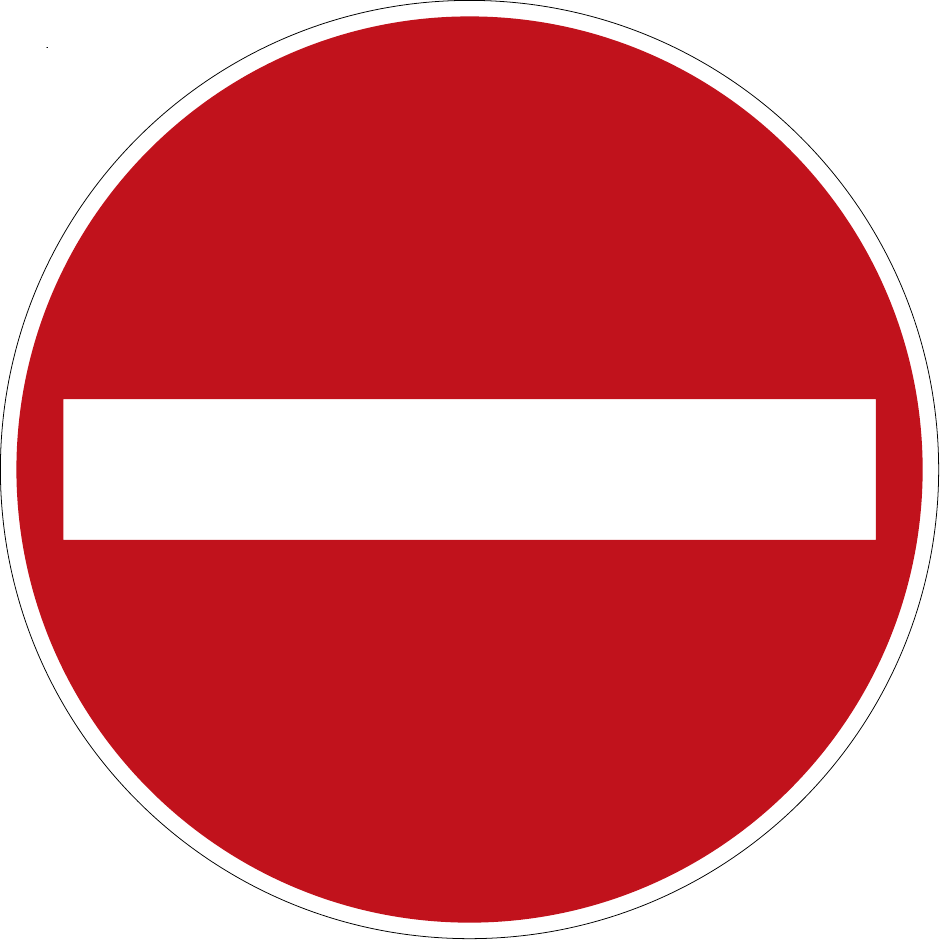}
      \caption{unique signs}
    \end{subfigure}%
    \hspace{.2cm}
    \begin{subfigure}[c]{.5\textwidth}
      \centering
      \includegraphics[width=.0665\linewidth]{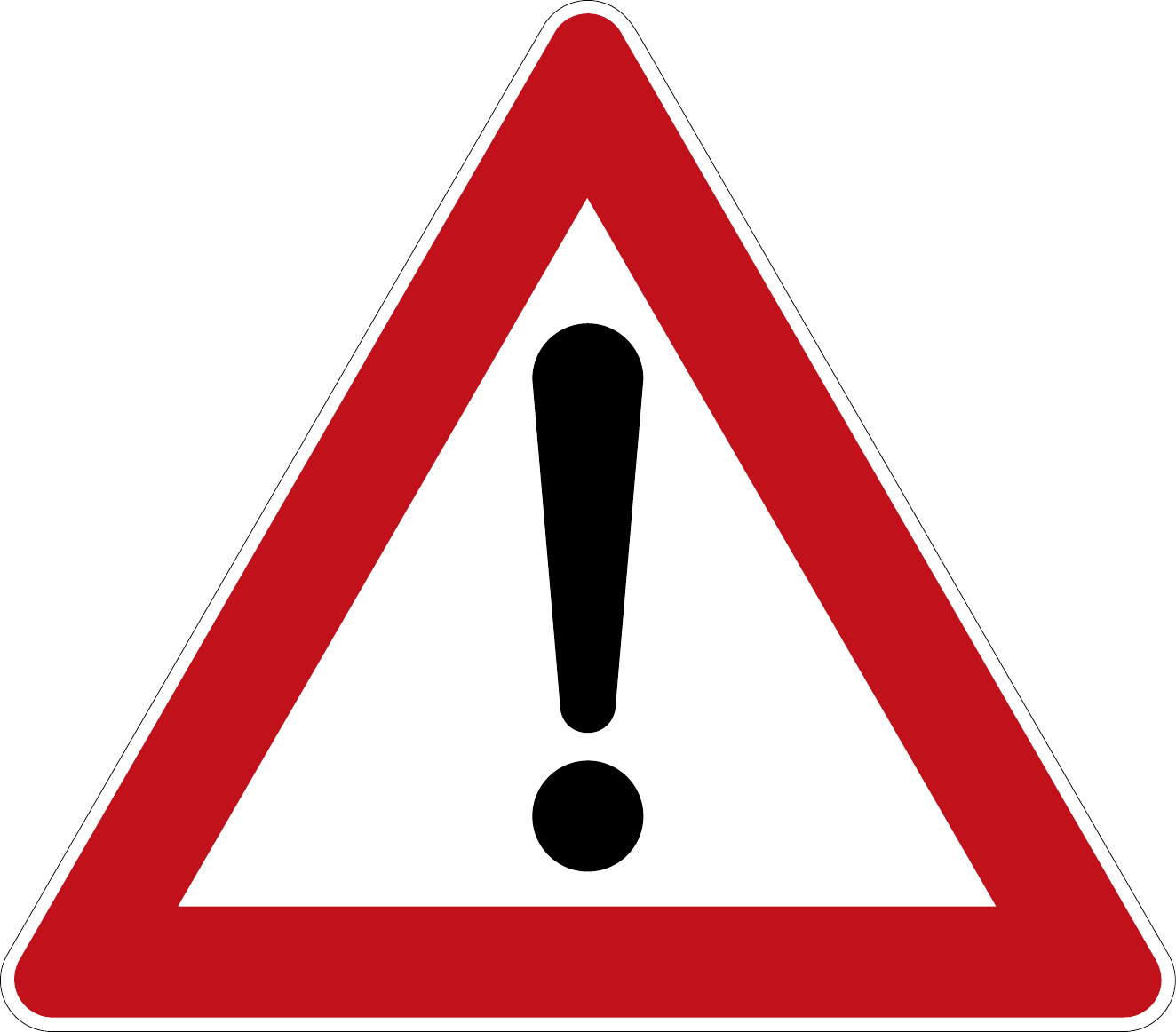}
      \includegraphics[width=.0665\linewidth]{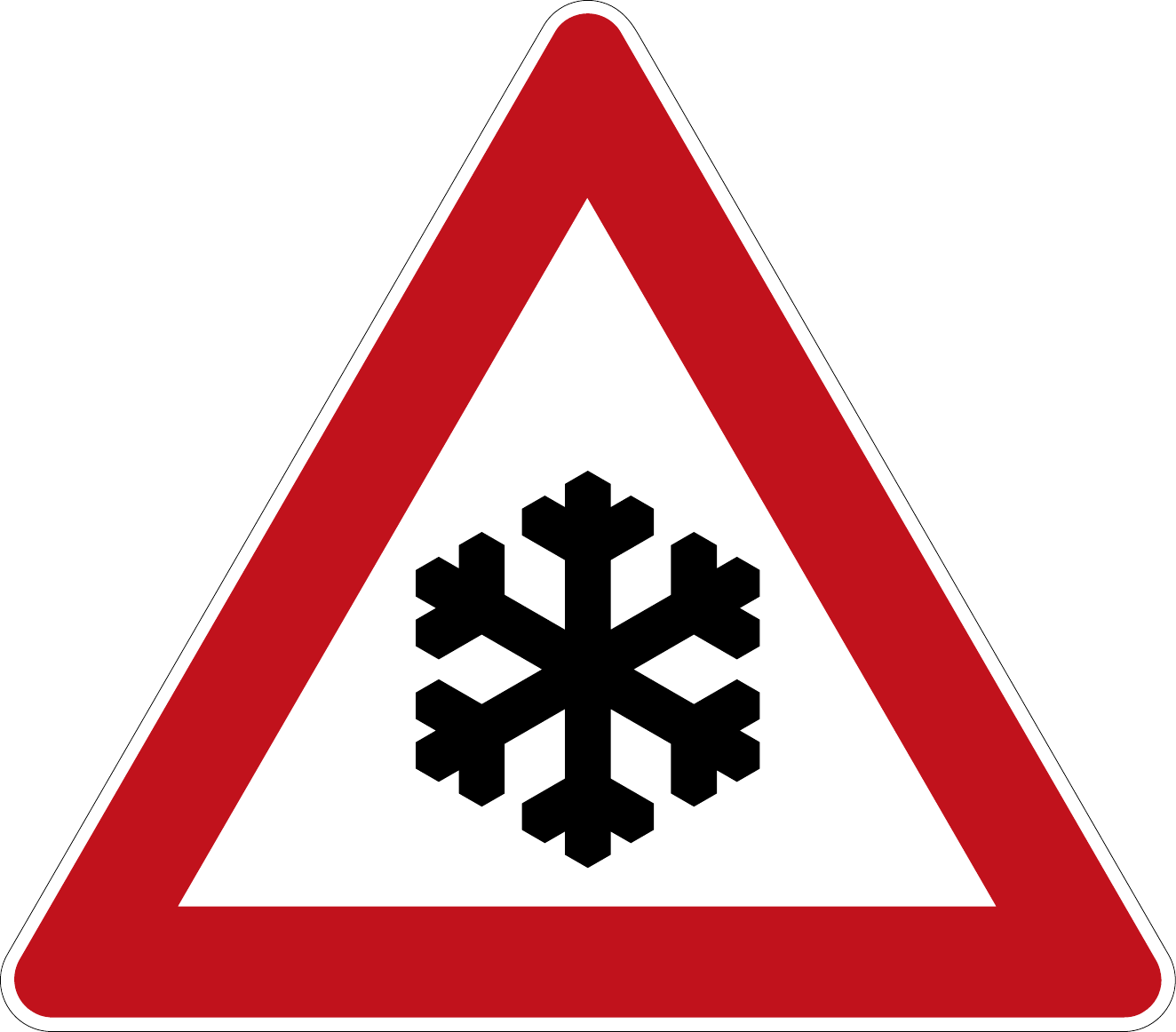}
      \includegraphics[width=.0665\linewidth]{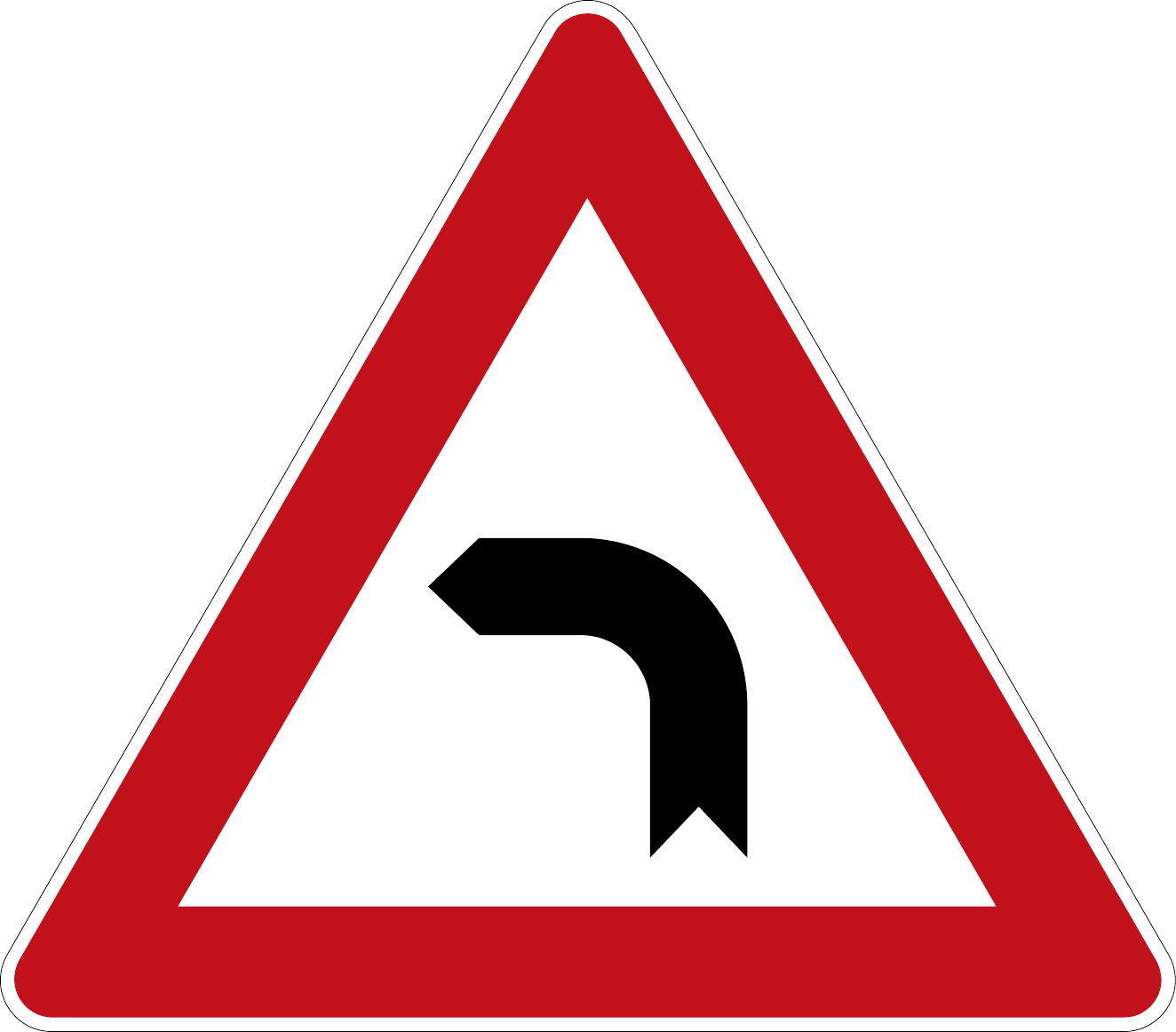}
      \includegraphics[width=.0665\linewidth]{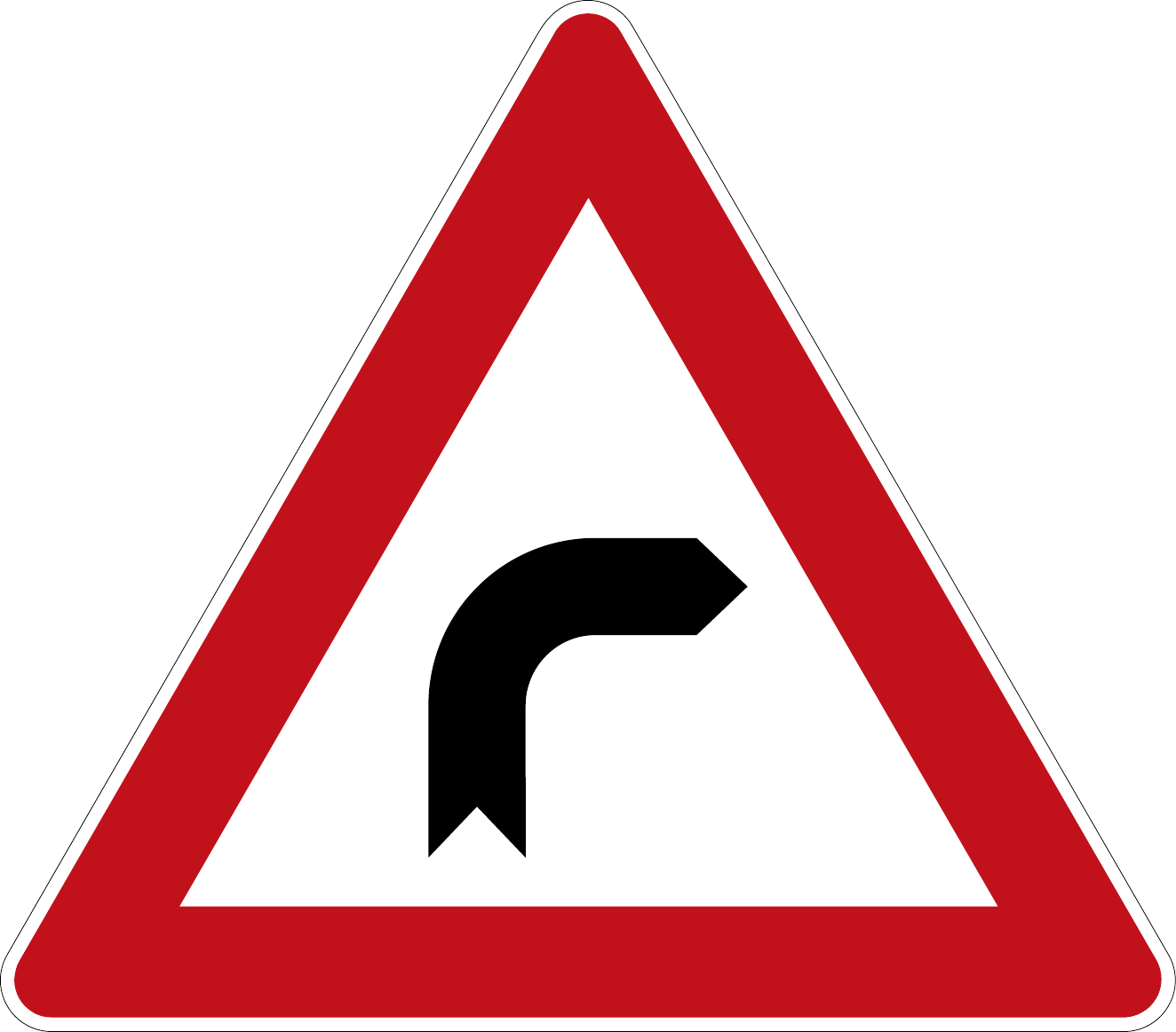}
      \includegraphics[width=.0665\linewidth]{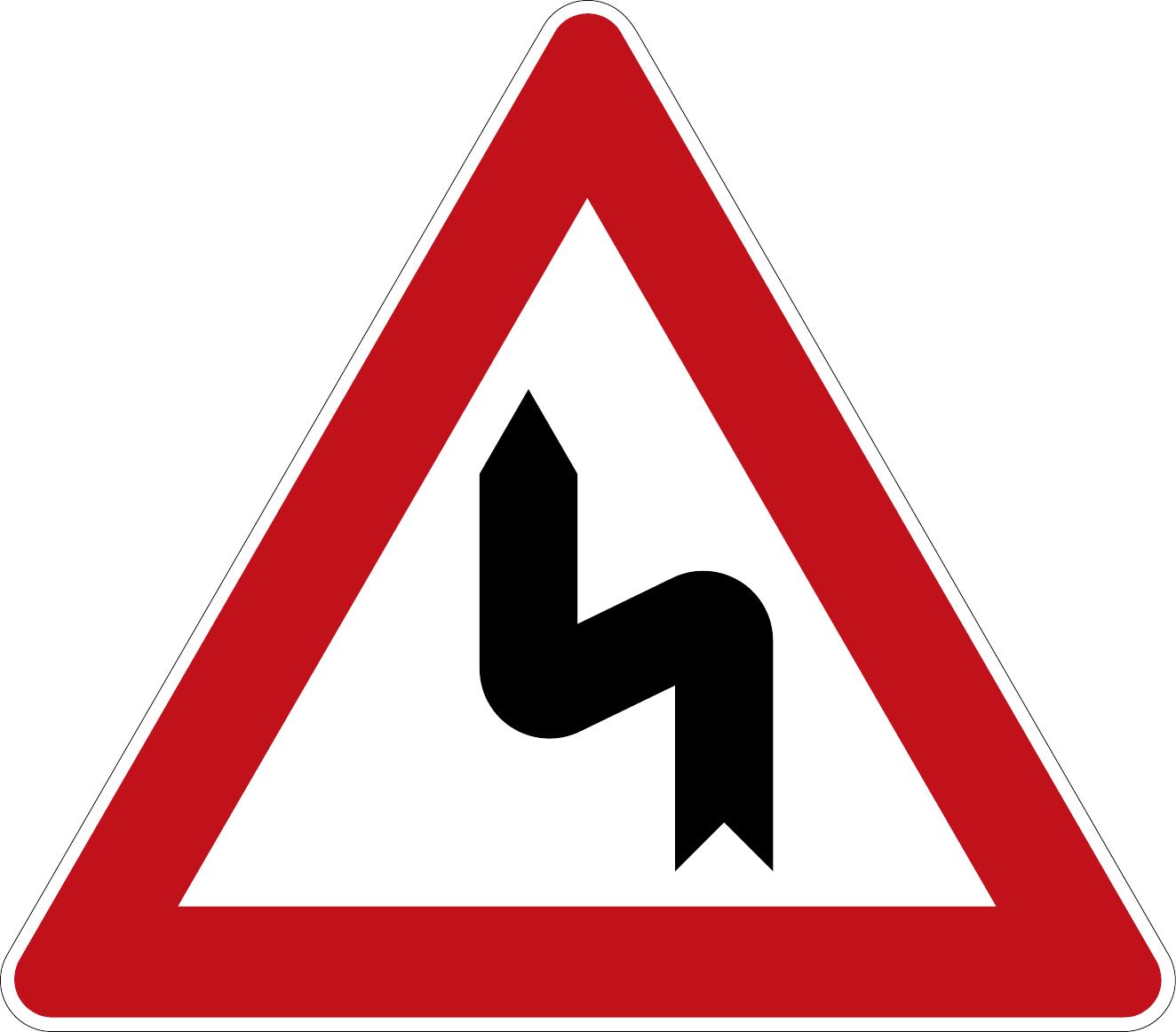}
      \includegraphics[width=.0665\linewidth]{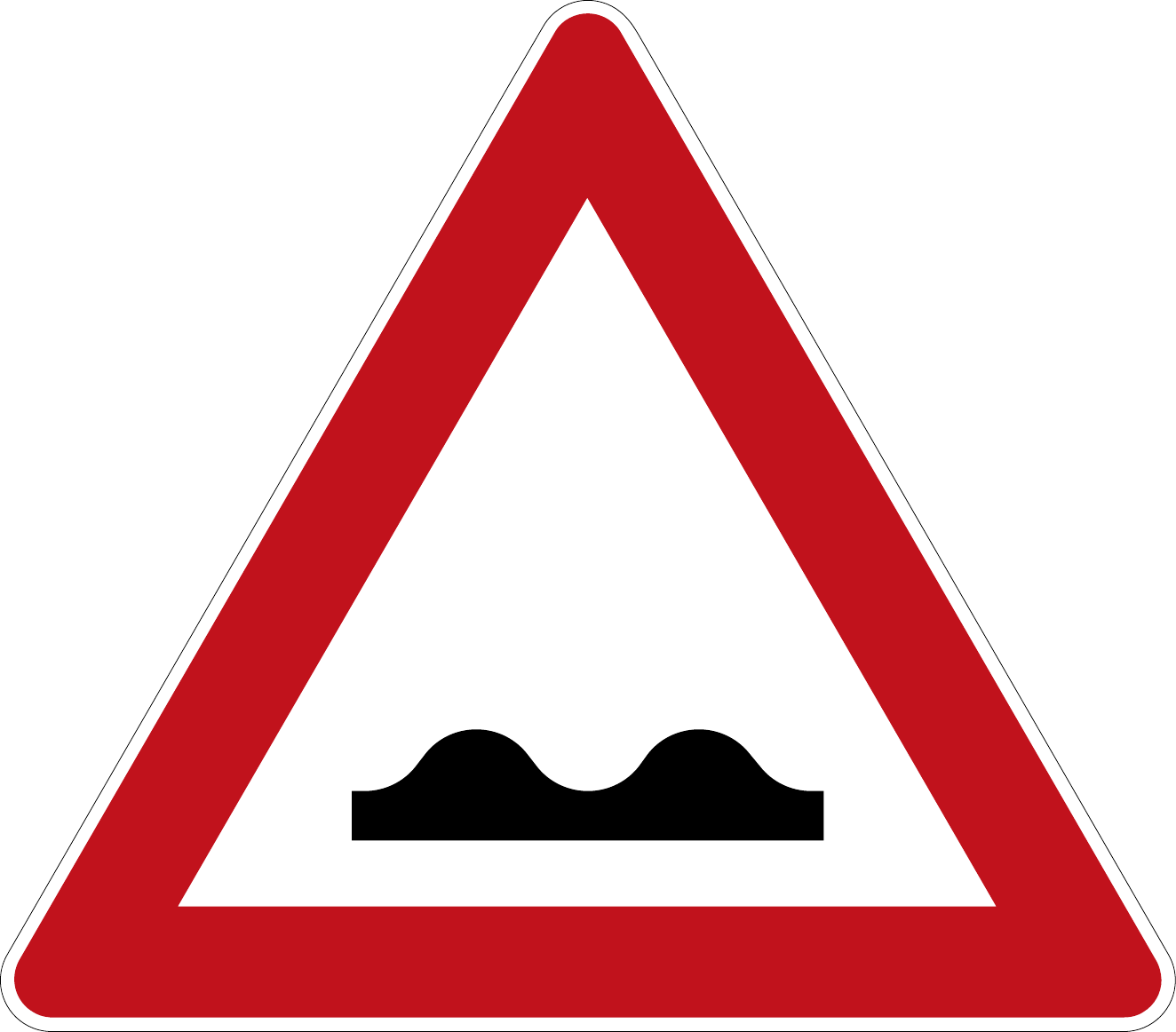}
      \includegraphics[width=.0665\linewidth]{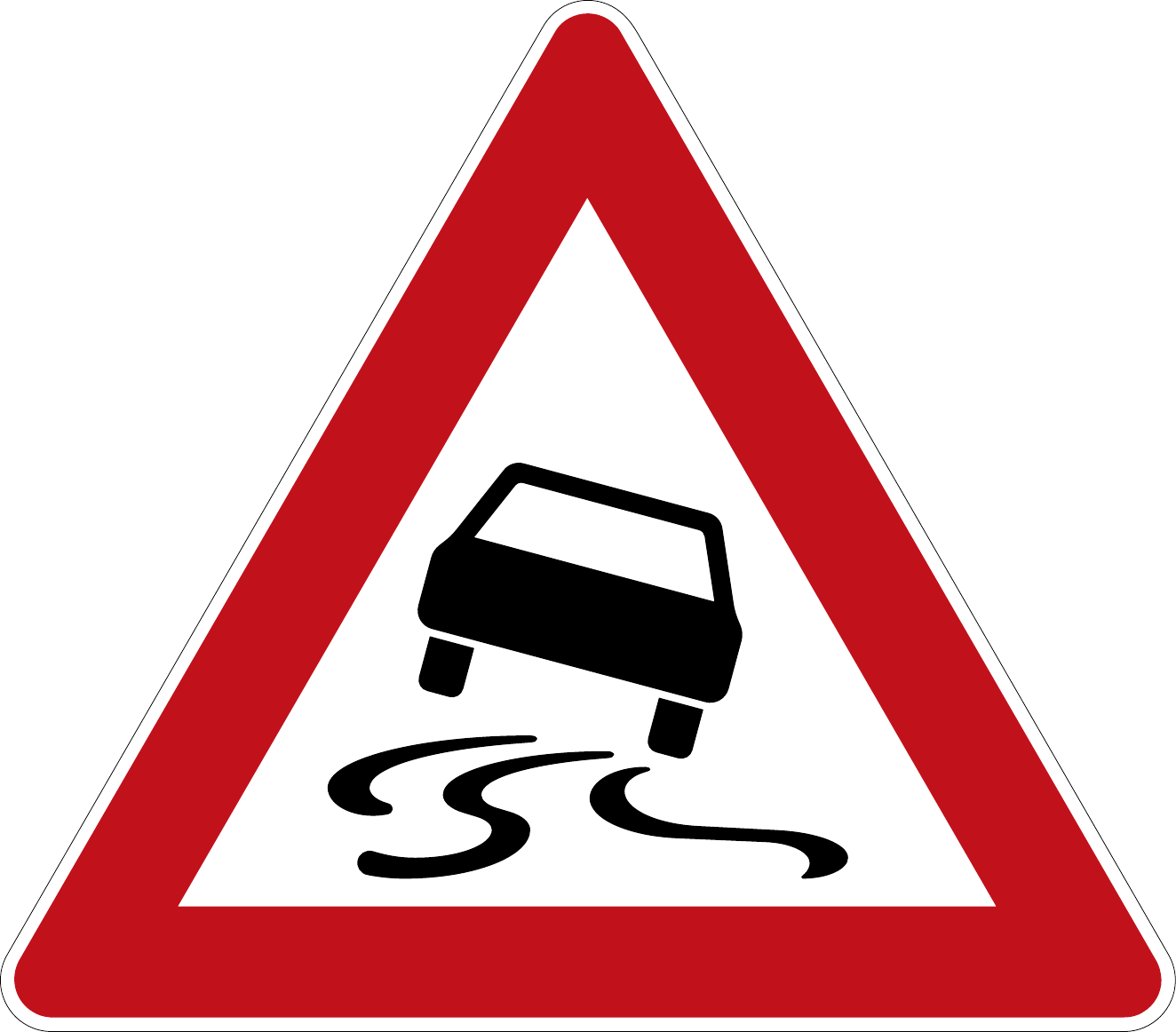}
      \includegraphics[width=.0665\linewidth]{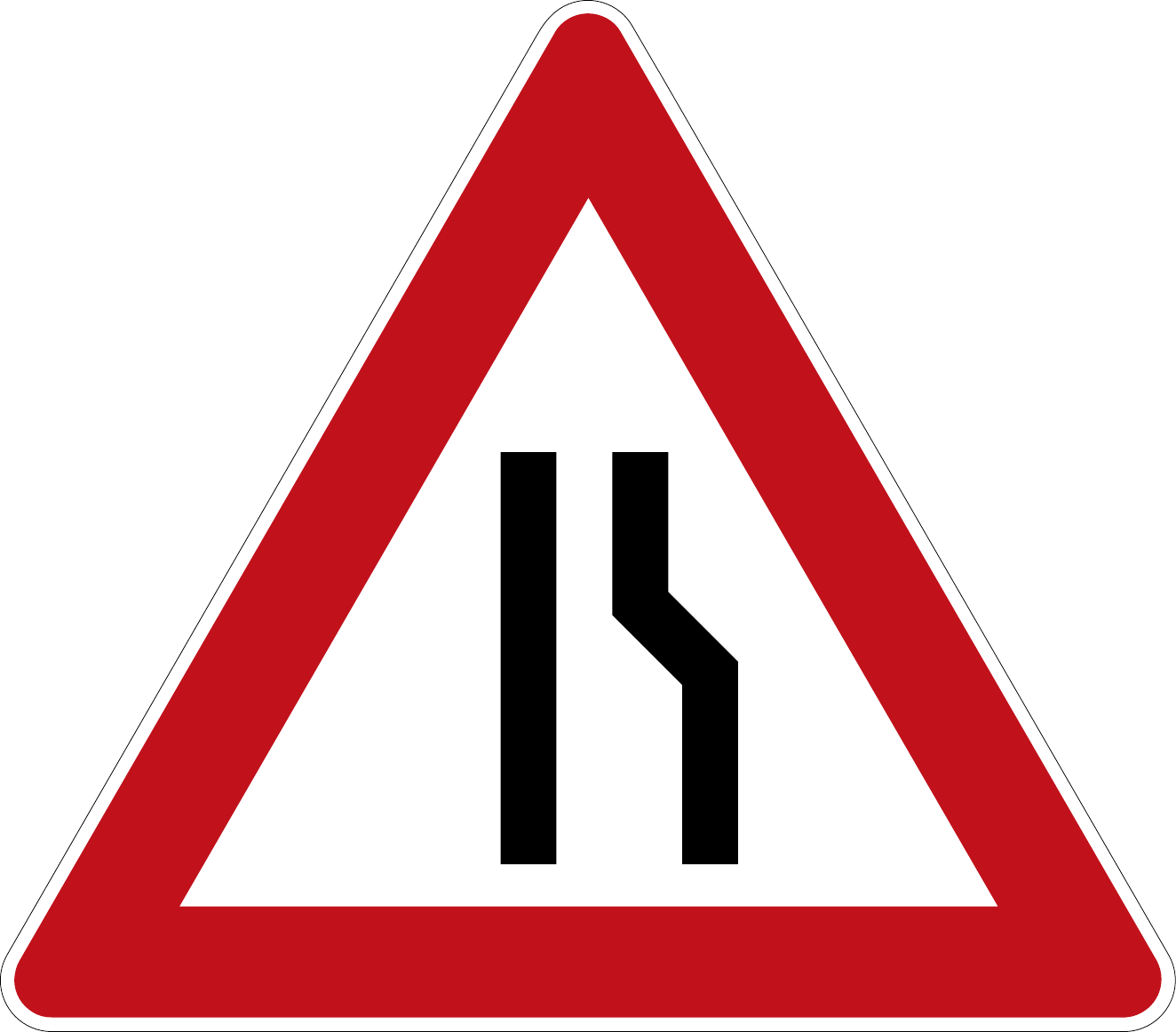}
      \\[2ex]
      \includegraphics[width=.0665\linewidth]{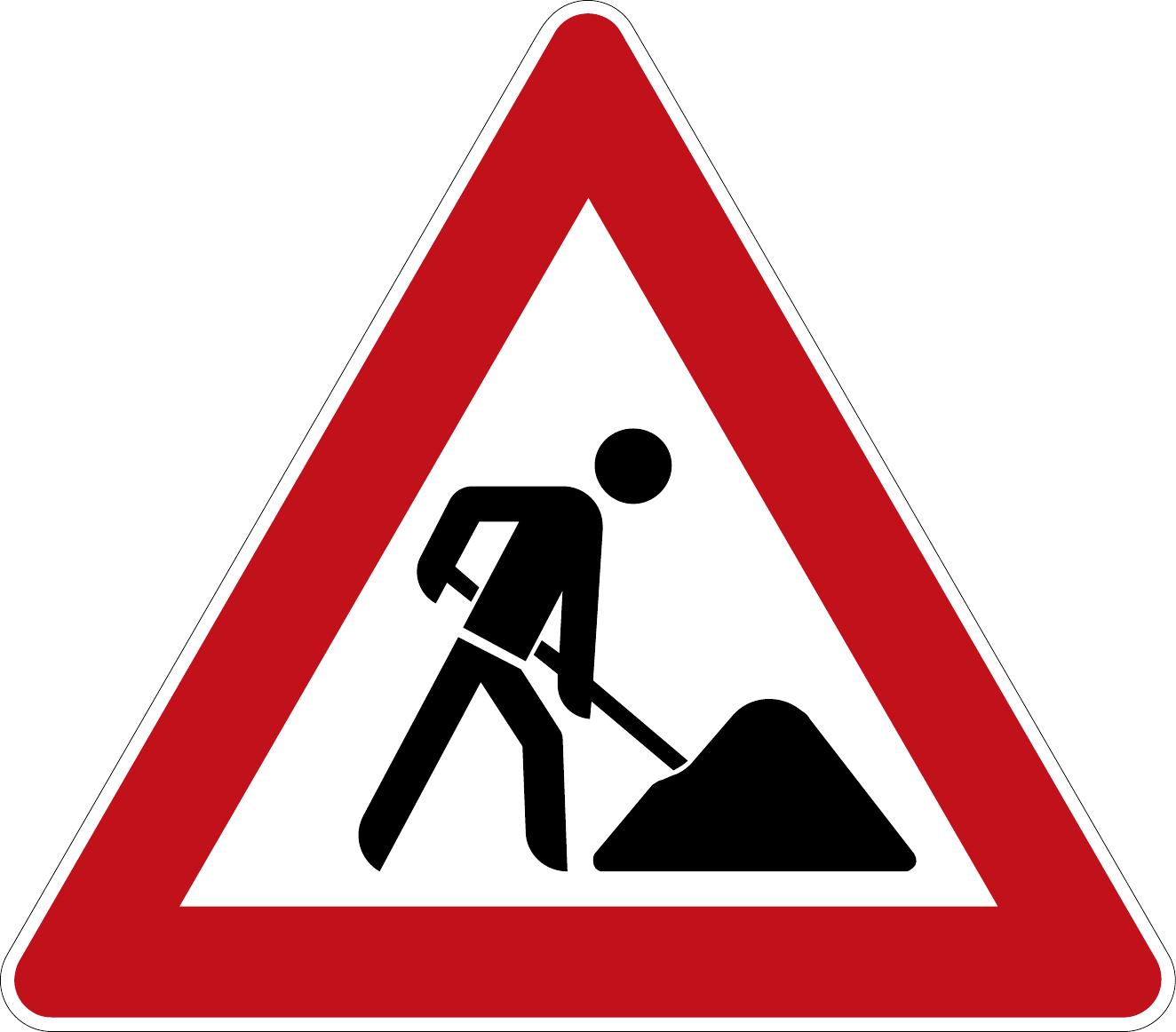}
      \includegraphics[width=.0665\linewidth]{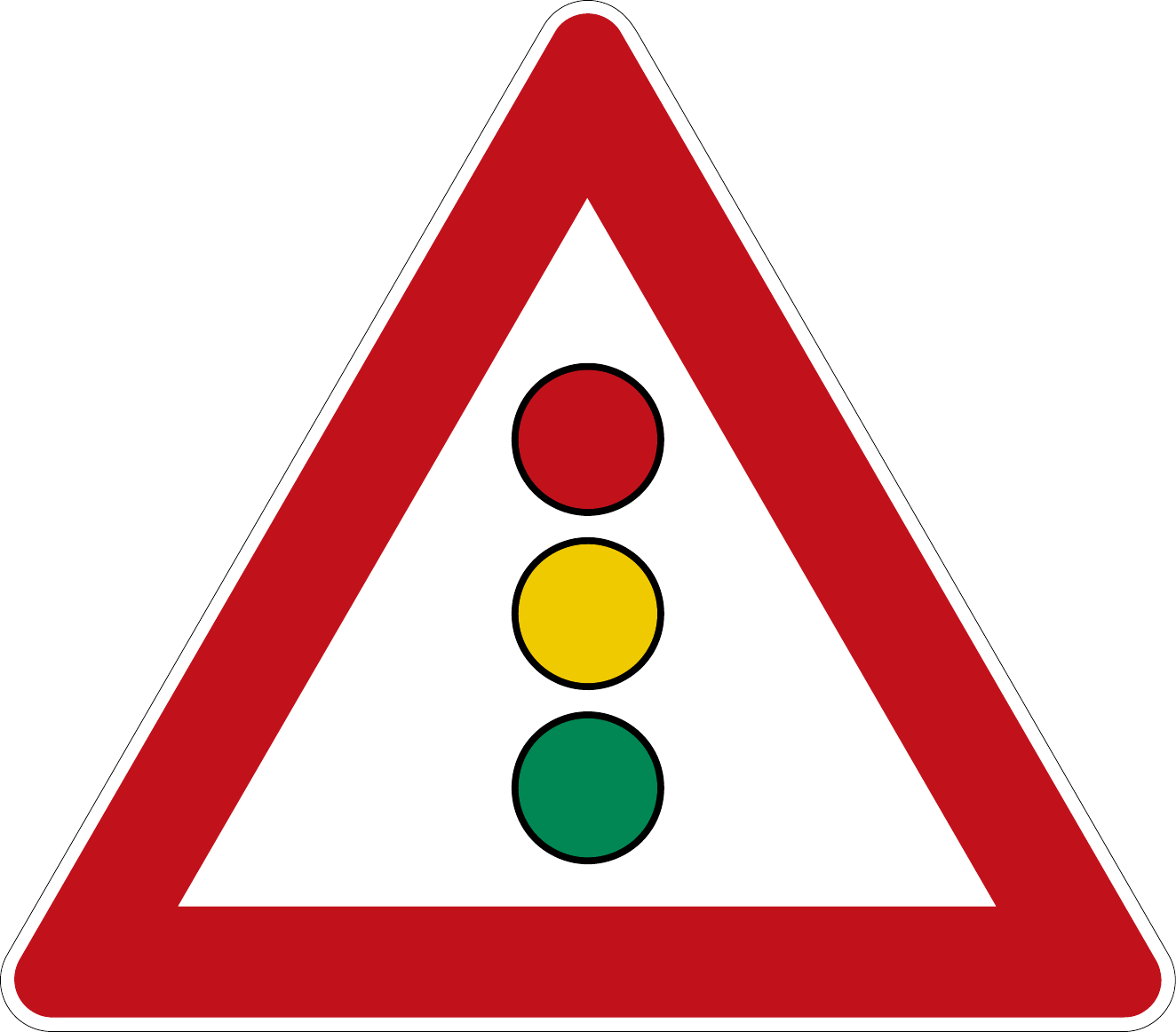}
      \includegraphics[width=.0665\linewidth]{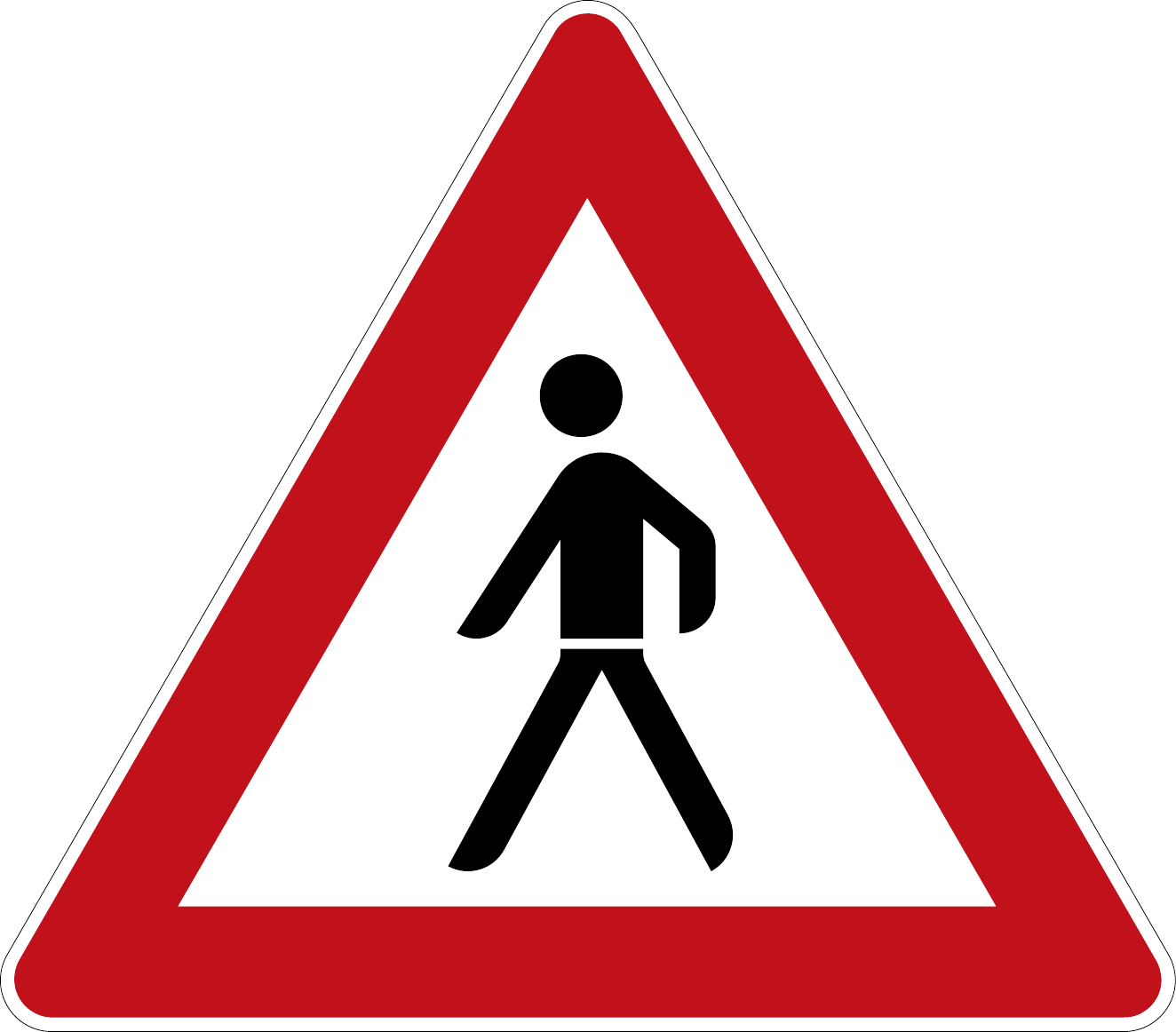}
      \includegraphics[width=.0665\linewidth]{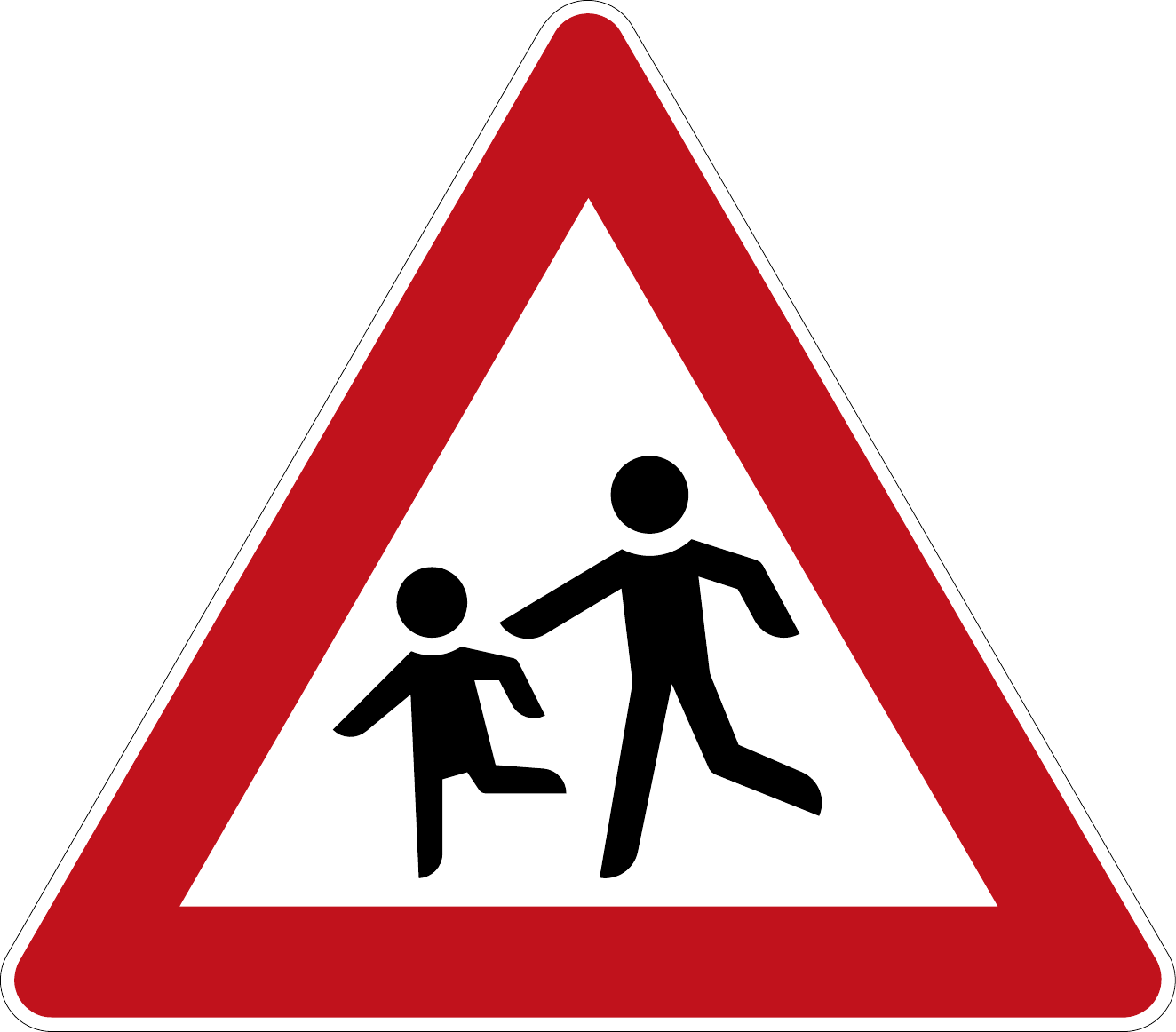}
      \includegraphics[width=.0665\linewidth]{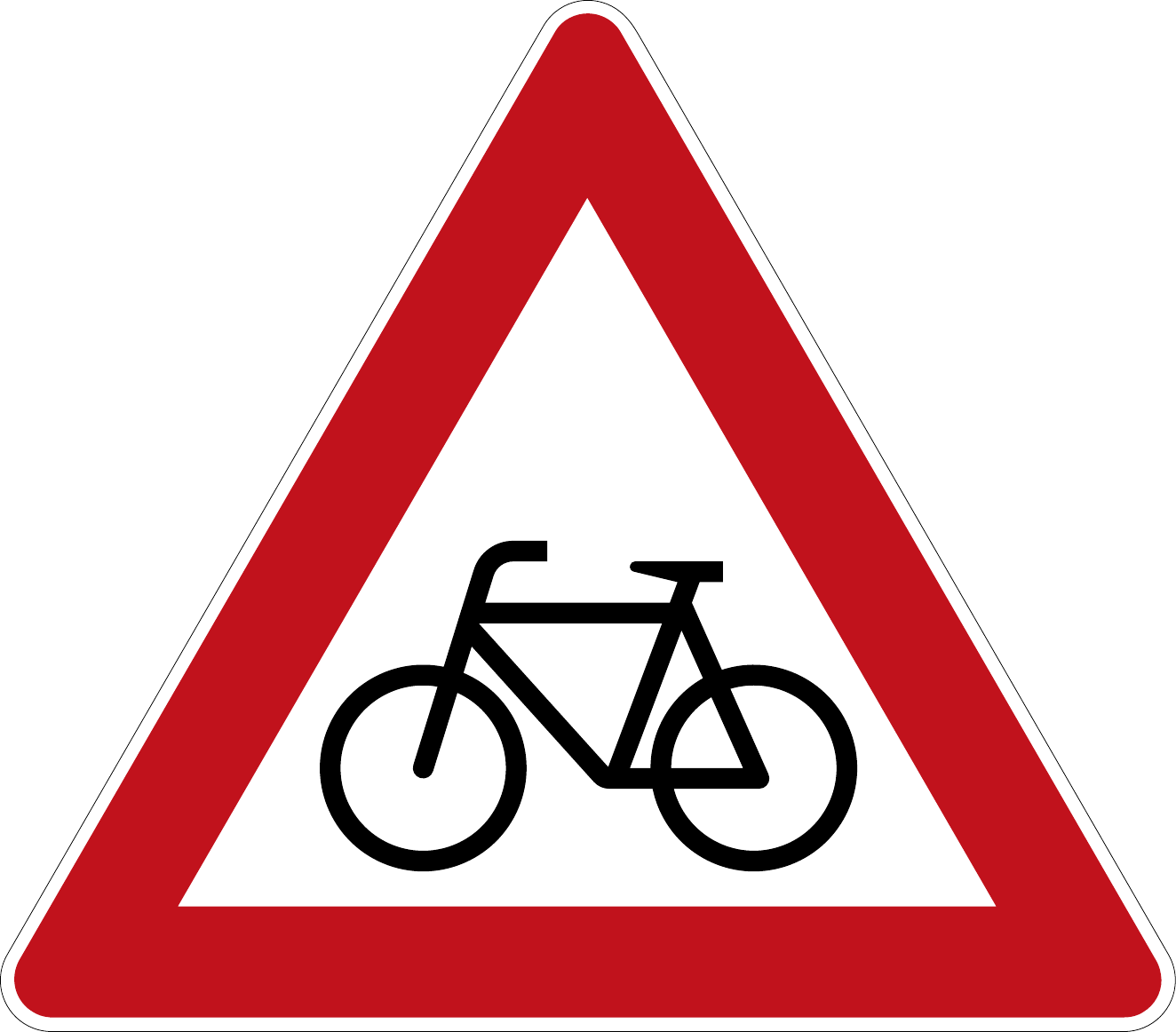}
      \includegraphics[width=.0665\linewidth]{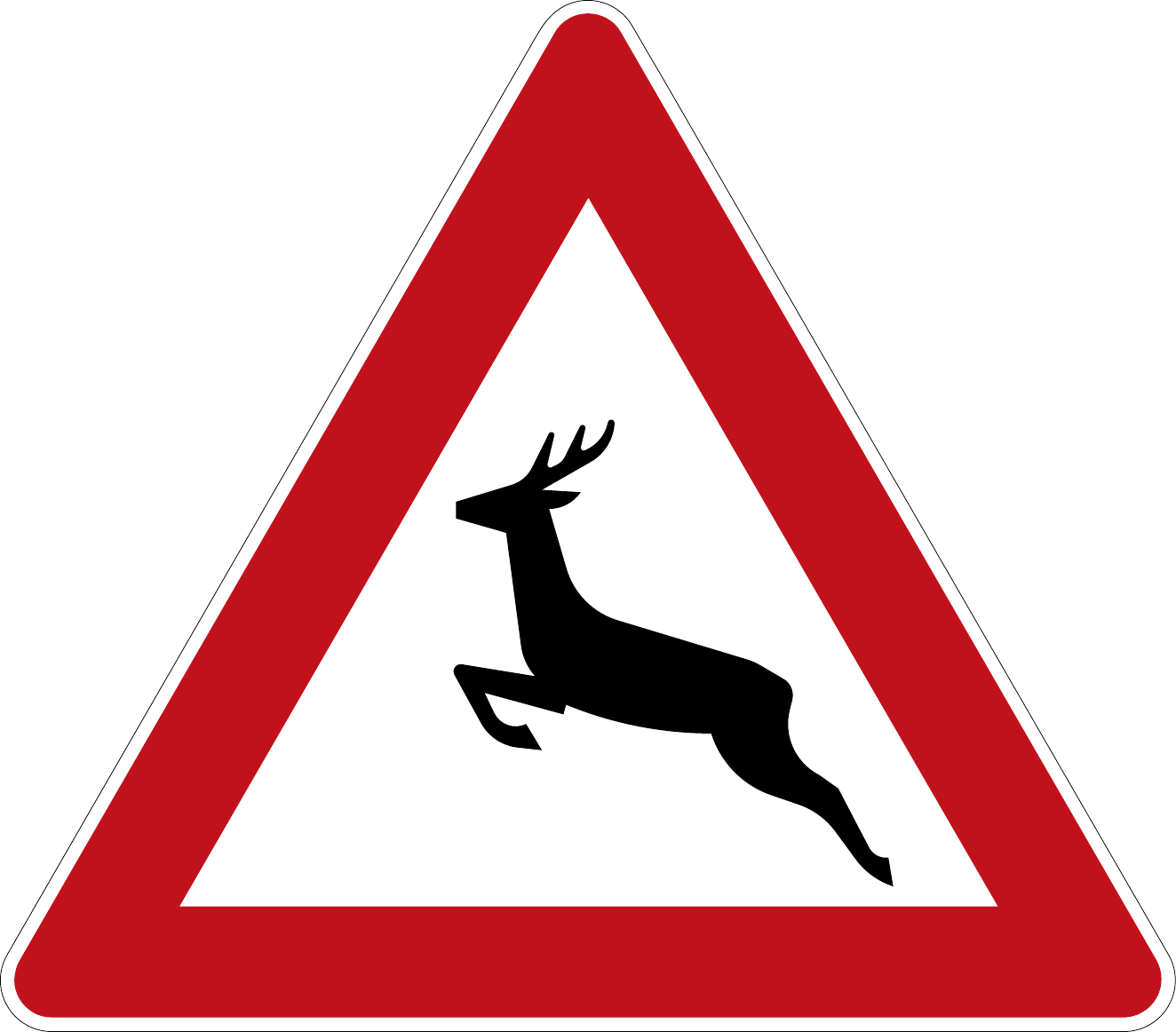}
      \includegraphics[width=.0665\linewidth]{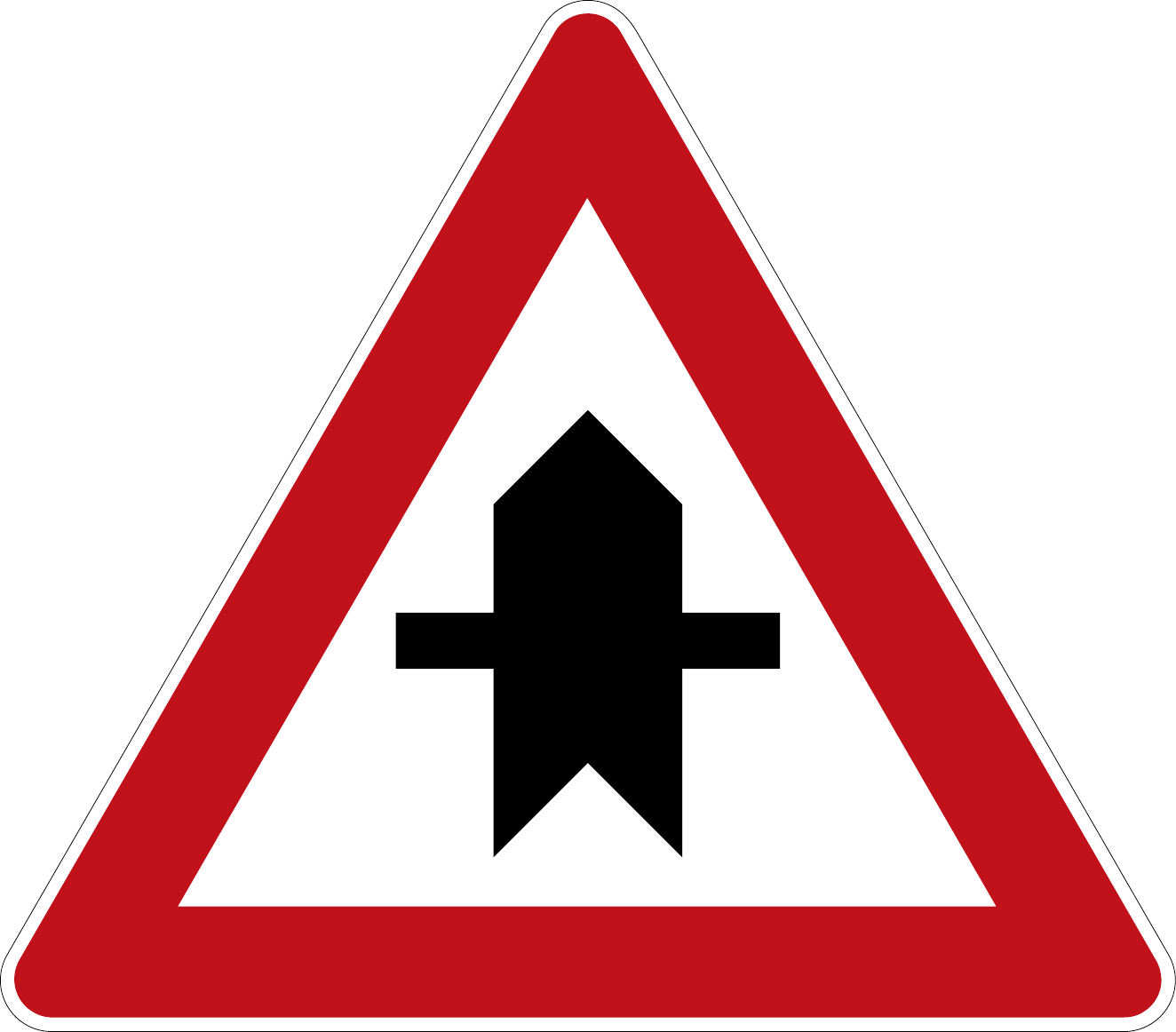}
      \caption{danger signs}
    \end{subfigure}%
    \vskip\baselineskip
    \begin{subfigure}[c]{.25\textwidth}   
      \centering
      \includegraphics[width=.1333\linewidth]{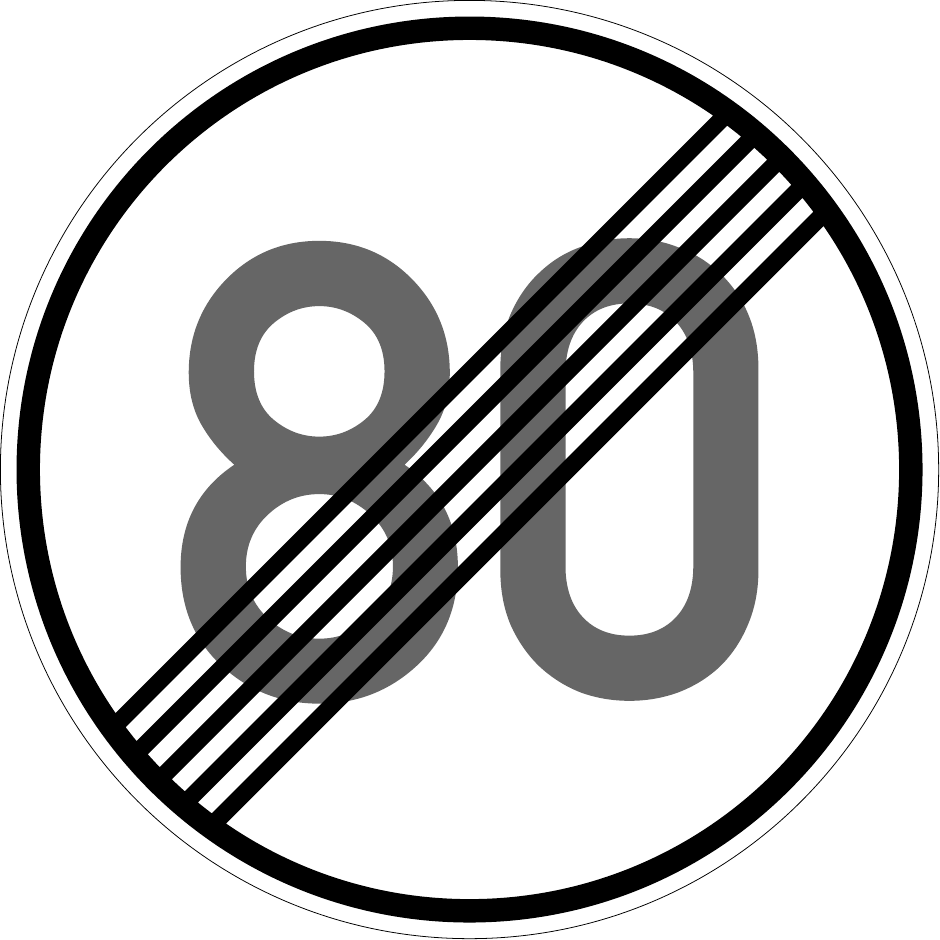}
      \includegraphics[width=.1333\linewidth]{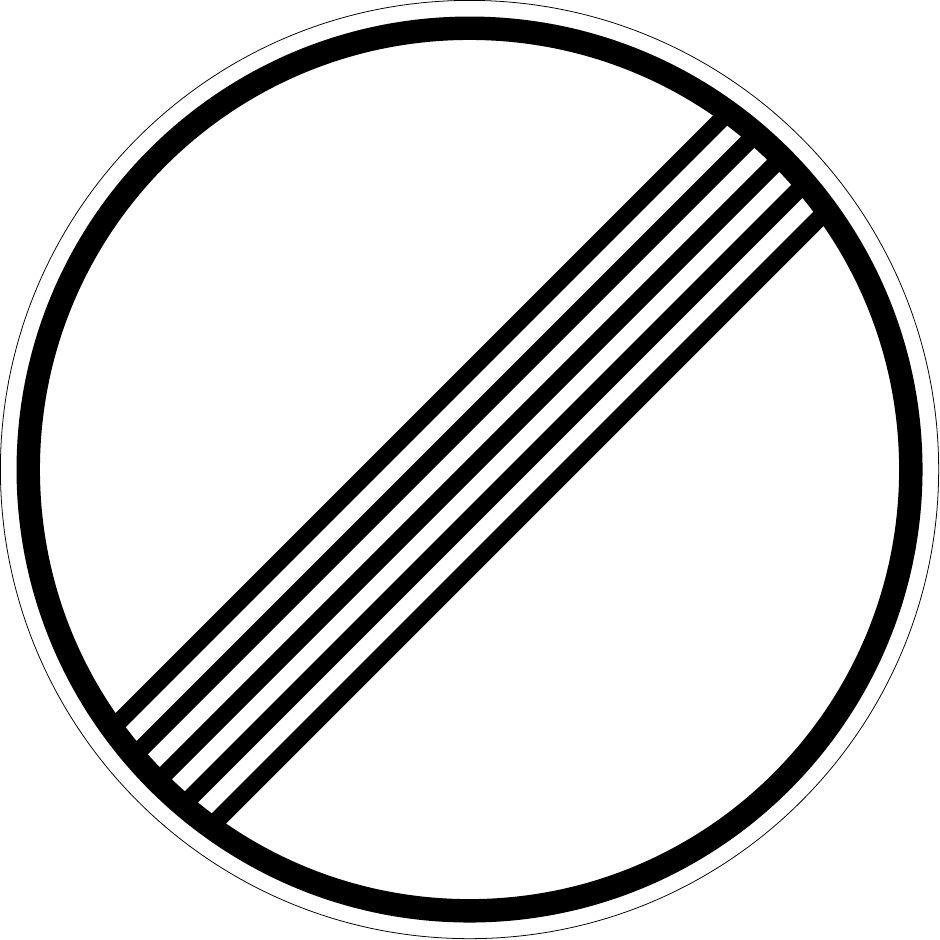}
      \includegraphics[width=.1333\linewidth]{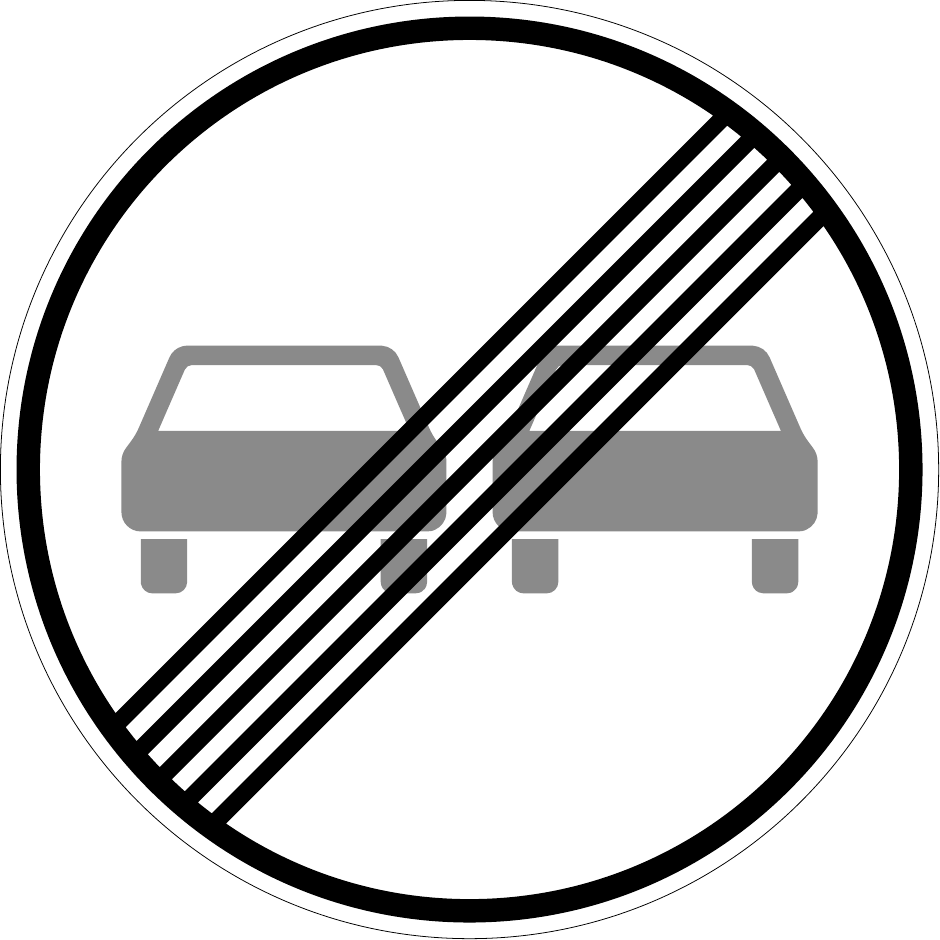}
      \includegraphics[width=.1333\linewidth]{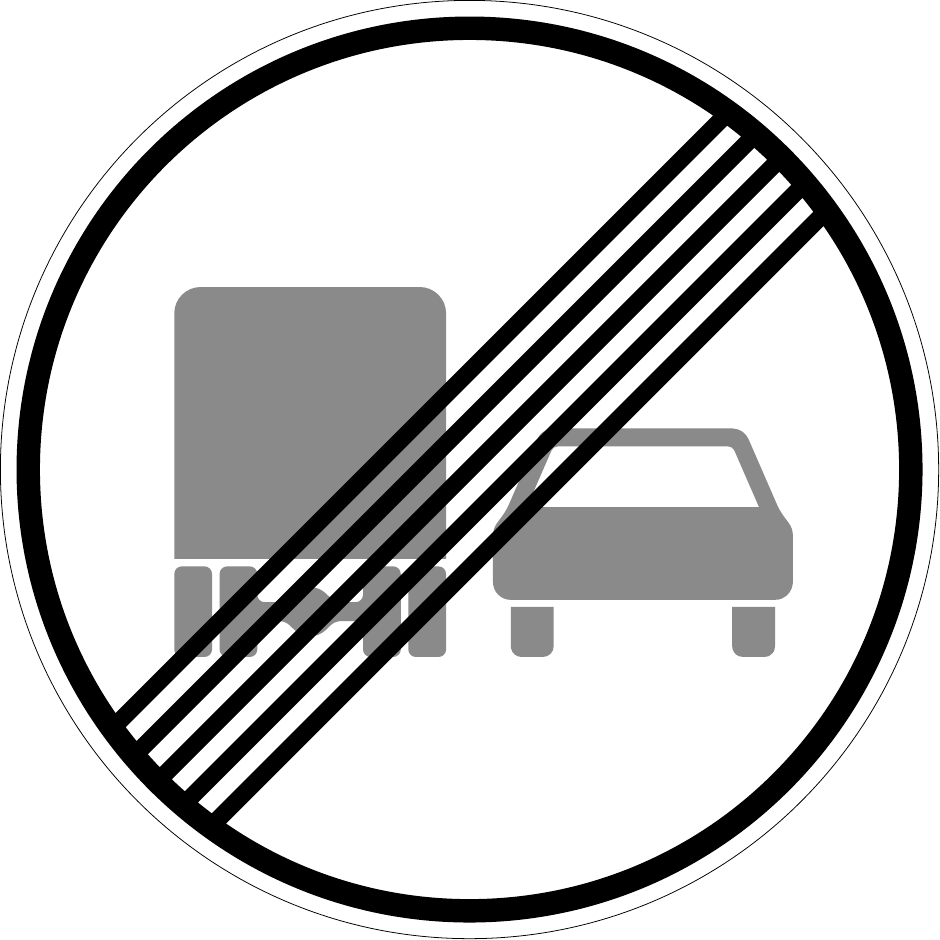}
      \caption{derestriction signs}
    \end{subfigure}%
    \hspace{.2cm}
    \begin{subfigure}[c]{.5\textwidth}
      \centering
      \includegraphics[width=.0665\linewidth]{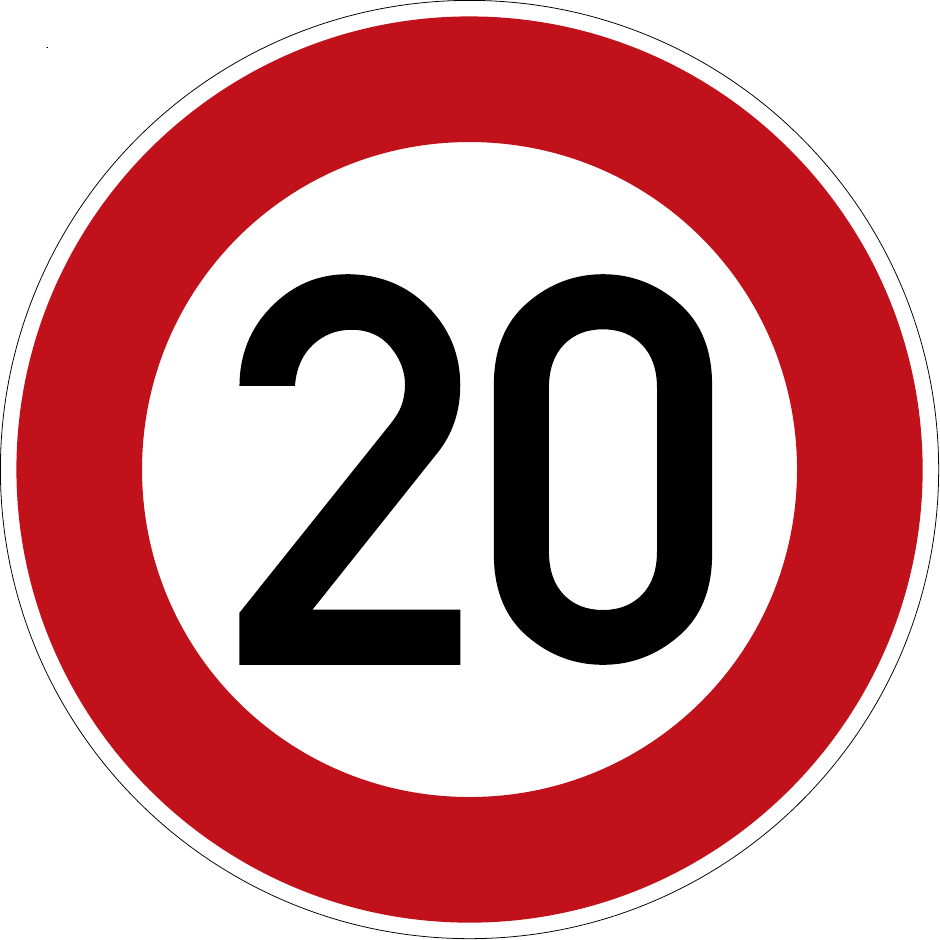}
      \includegraphics[width=.0665\linewidth]{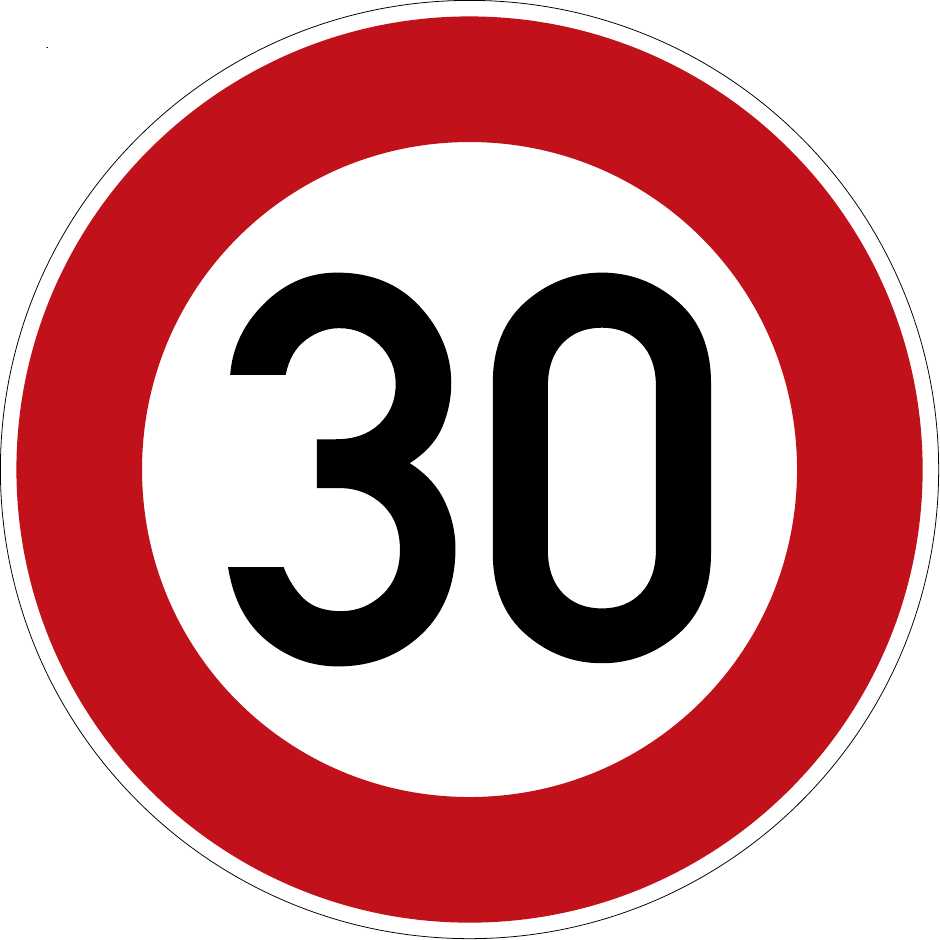}
      \includegraphics[width=.0665\linewidth]{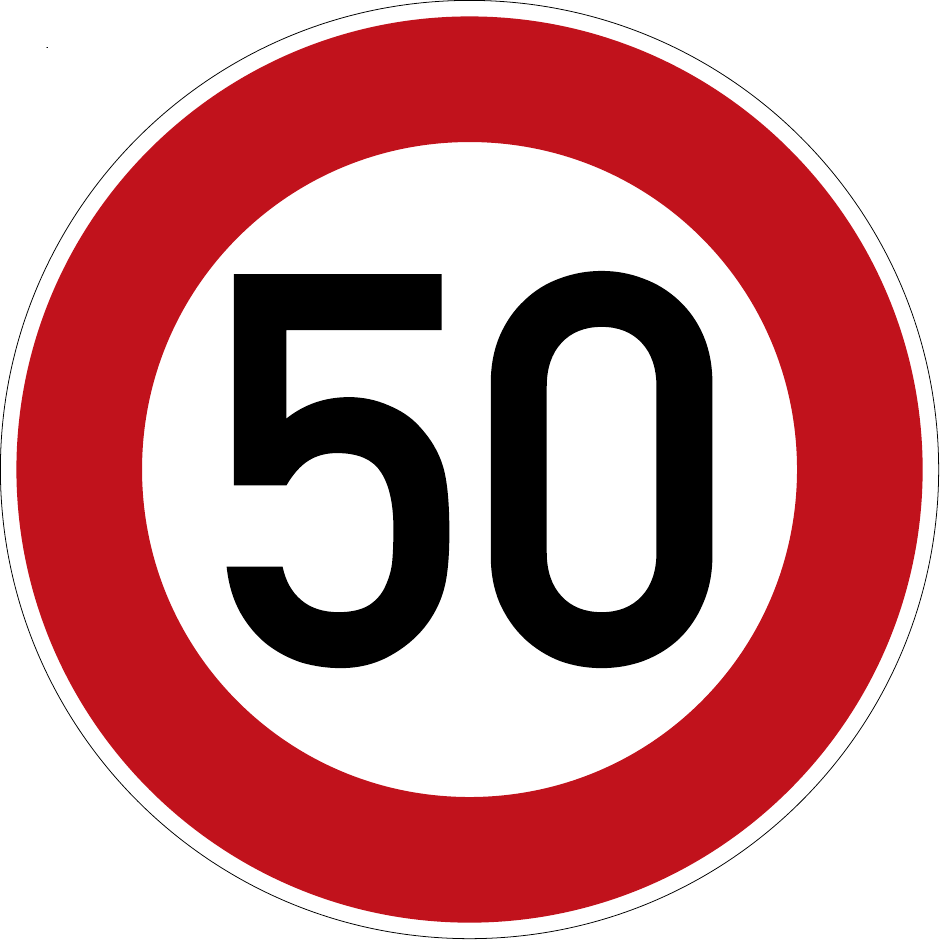}
      \includegraphics[width=.0665\linewidth]{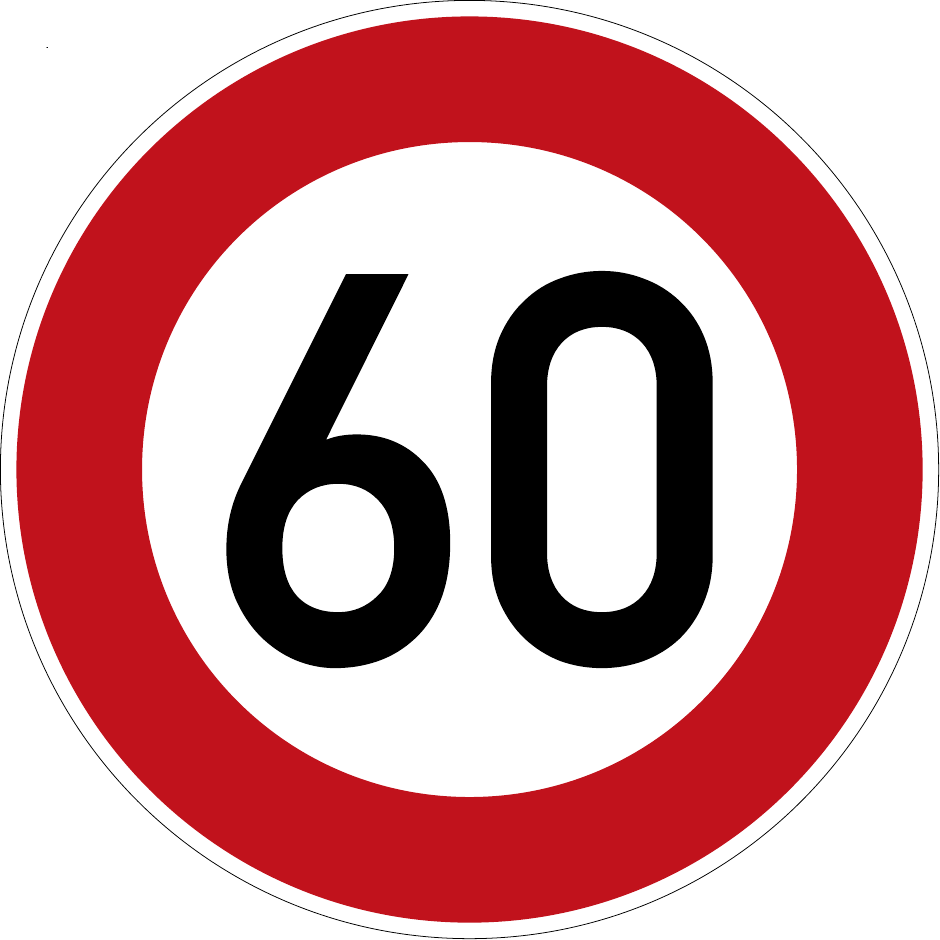}
      \includegraphics[width=.0665\linewidth]{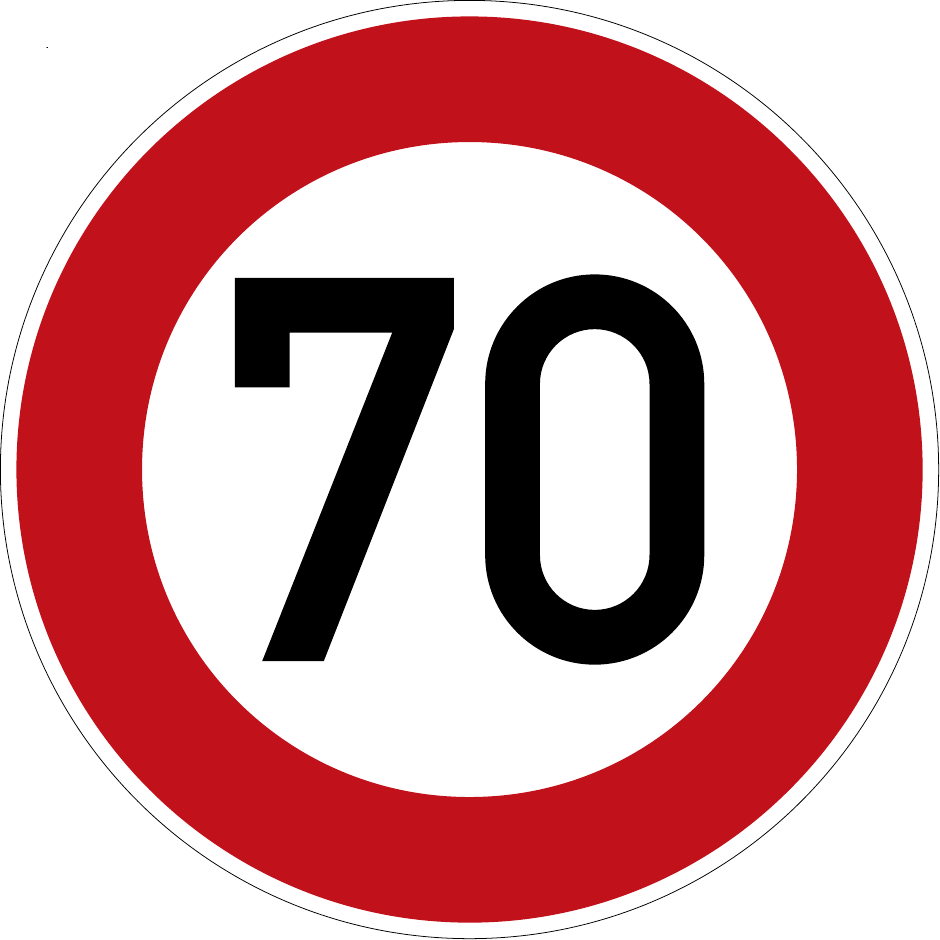}
      \includegraphics[width=.0665\linewidth]{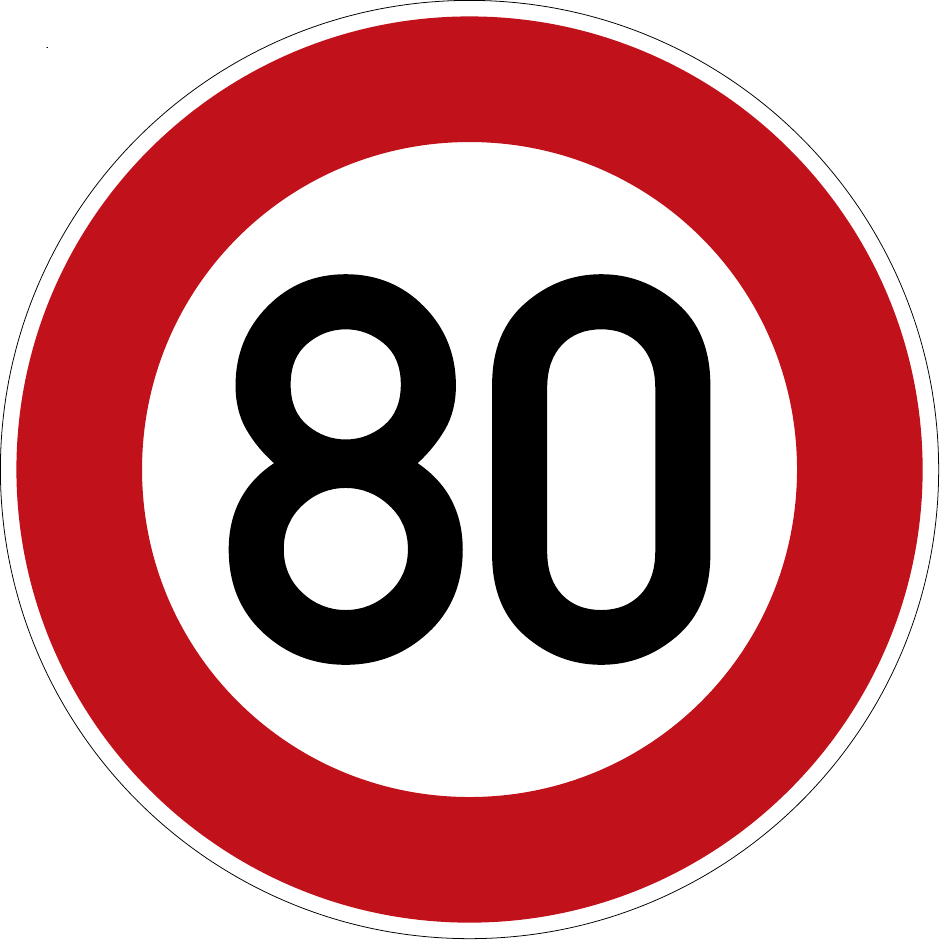}
      \includegraphics[width=.0665\linewidth]{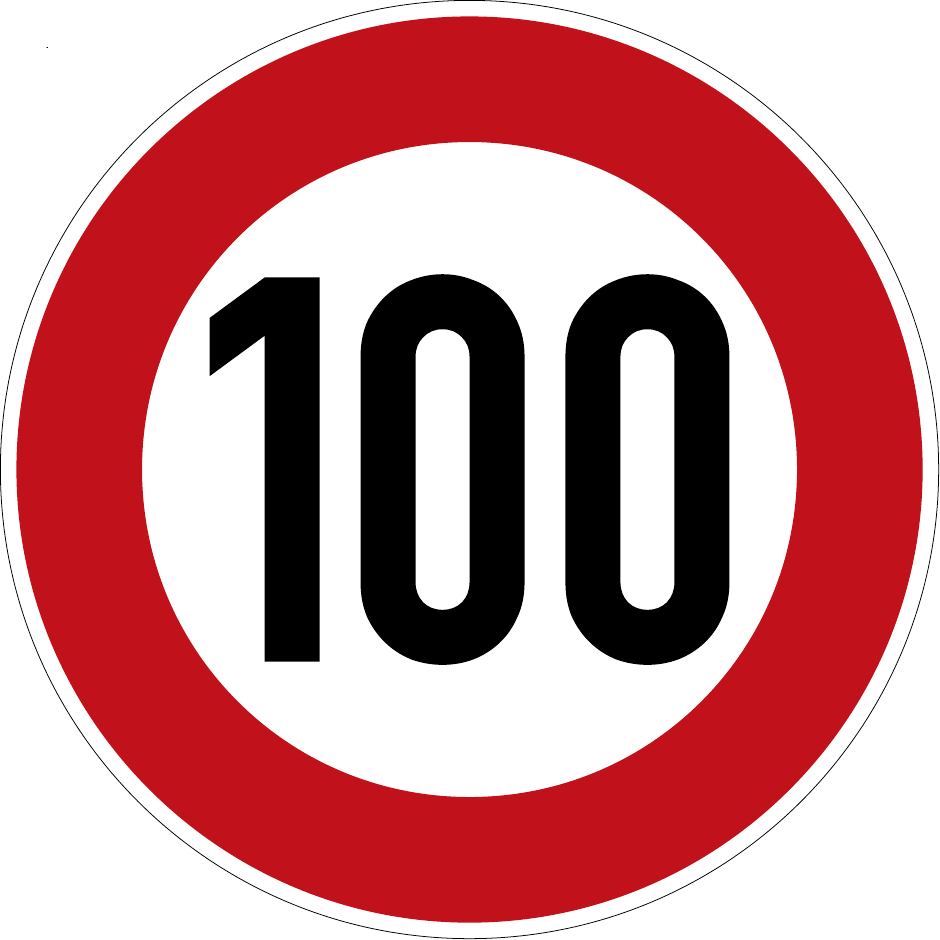}
      \includegraphics[width=.0665\linewidth]{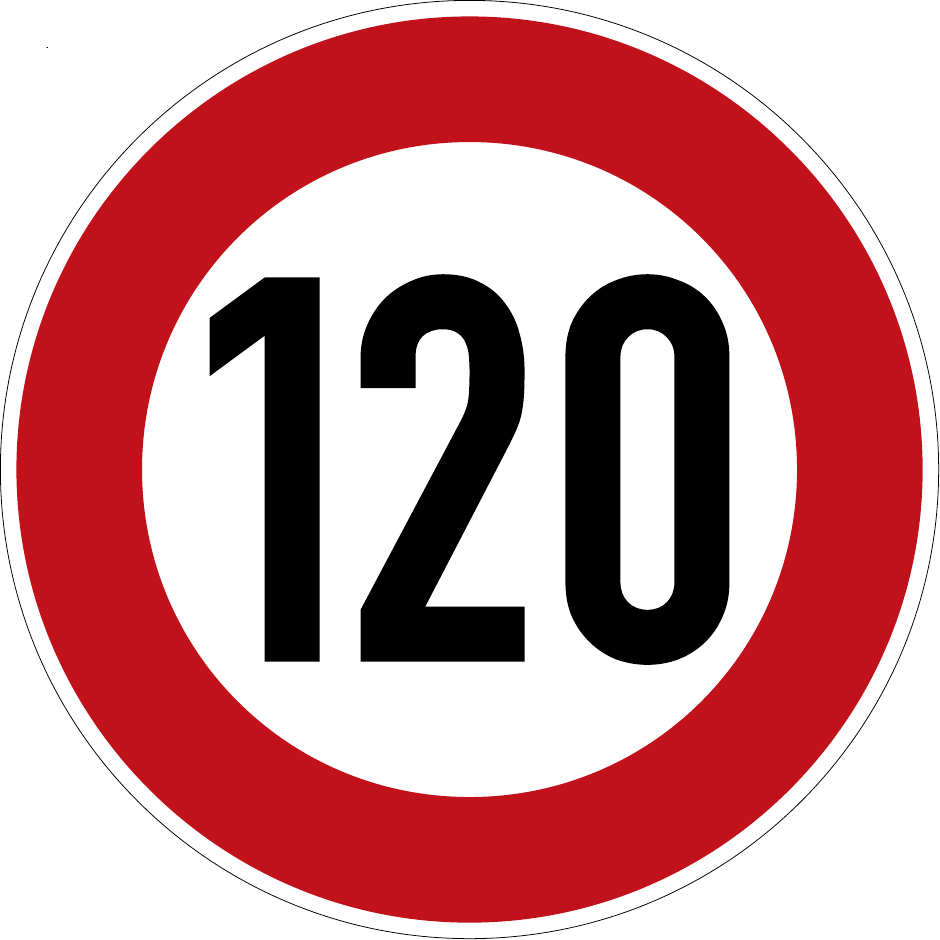}
      \caption{speed limit signs}
    \end{subfigure}%
    \vskip\baselineskip
    \begin{subfigure}[c]{.25\textwidth}   
      \centering
      \includegraphics[width=.1333\linewidth]{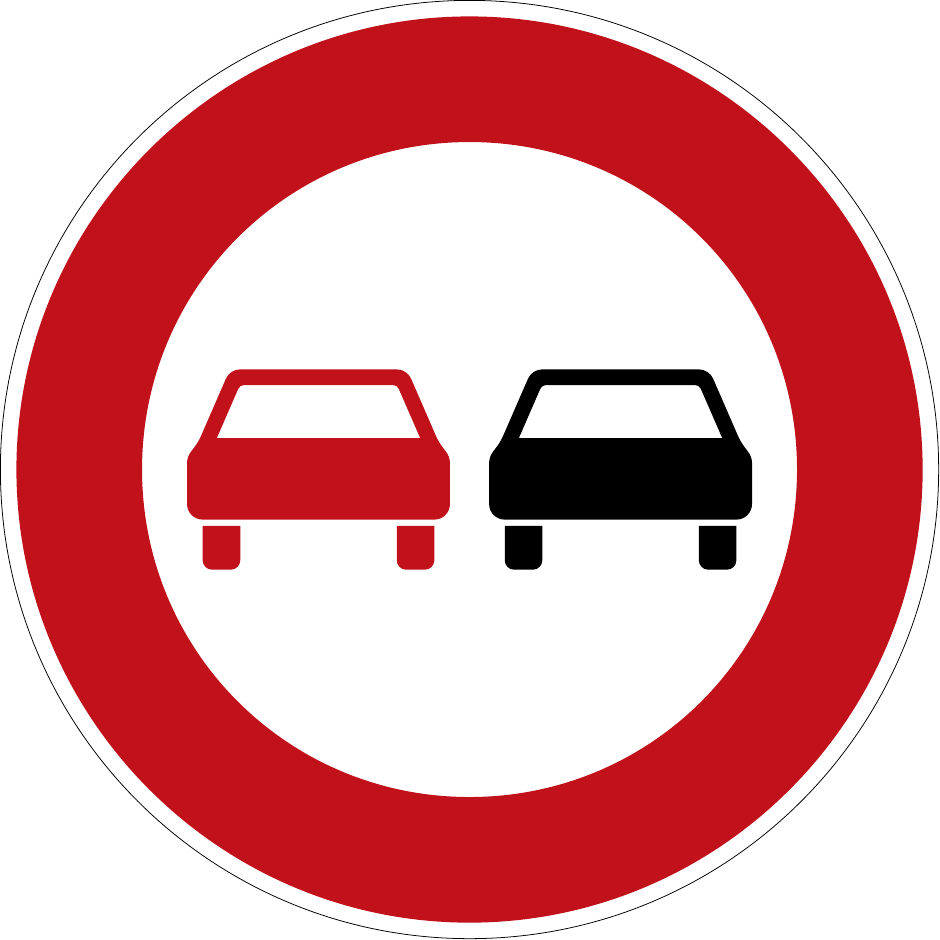}
      \includegraphics[width=.1333\linewidth]{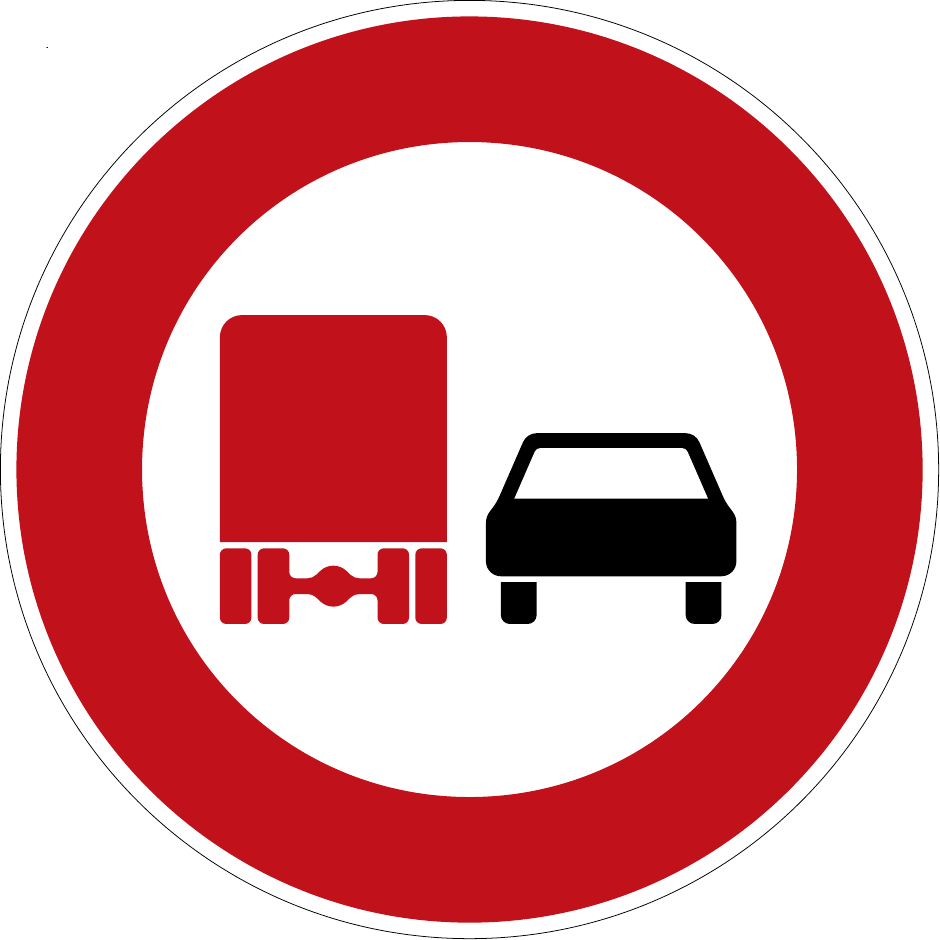}
      \includegraphics[width=.1333\linewidth]{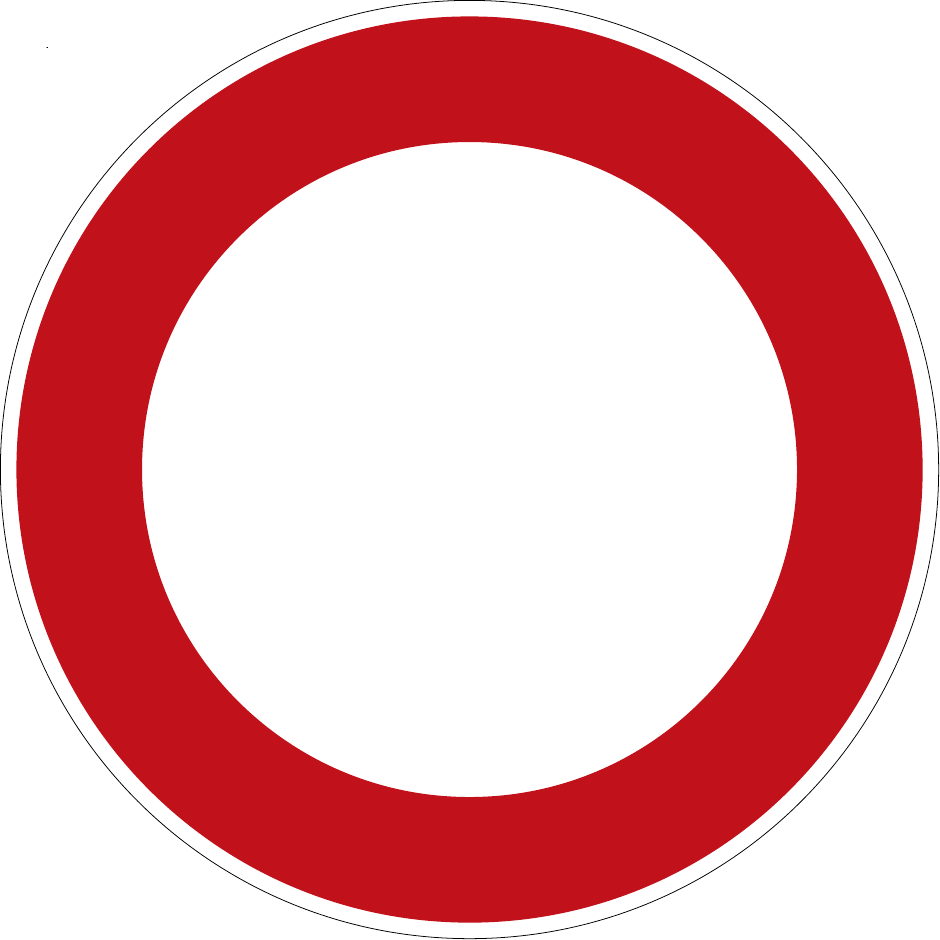}
      \includegraphics[width=.1333\linewidth]{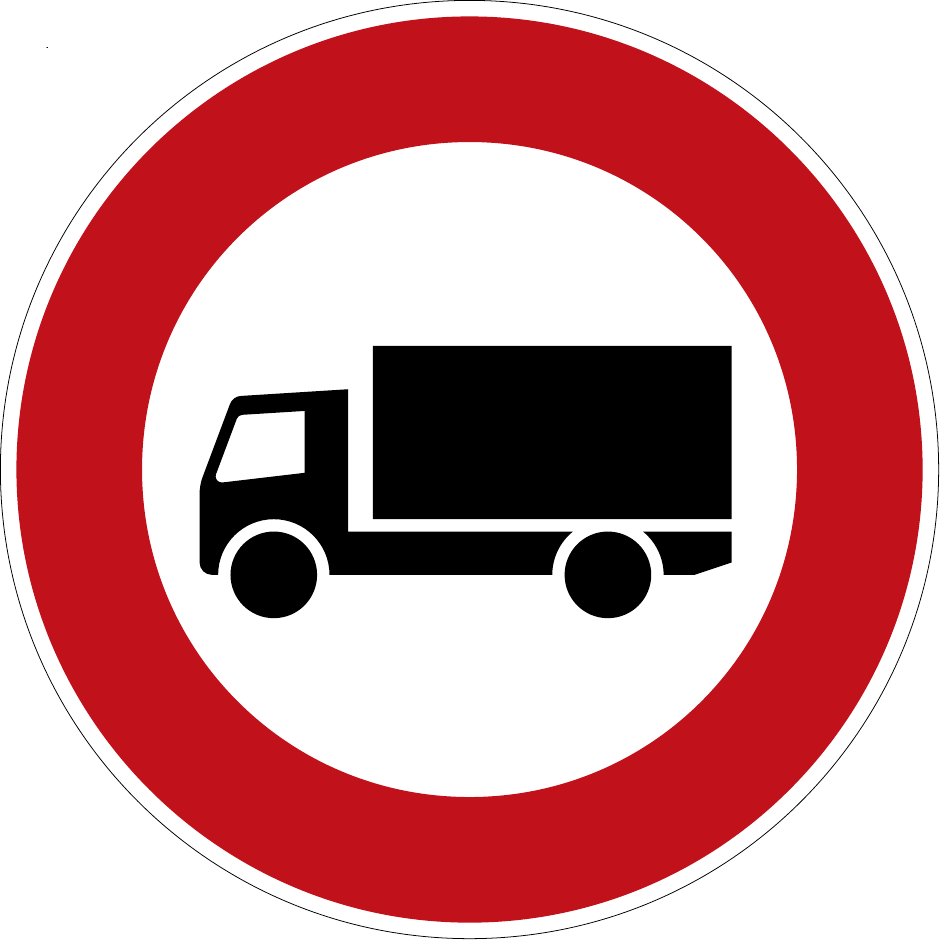}
      \caption{other prohibitory signs}
    \end{subfigure}%
    \hspace{.2cm}
    \begin{subfigure}[c]{.5\textwidth}
      \centering
      \includegraphics[width=.0665\linewidth]{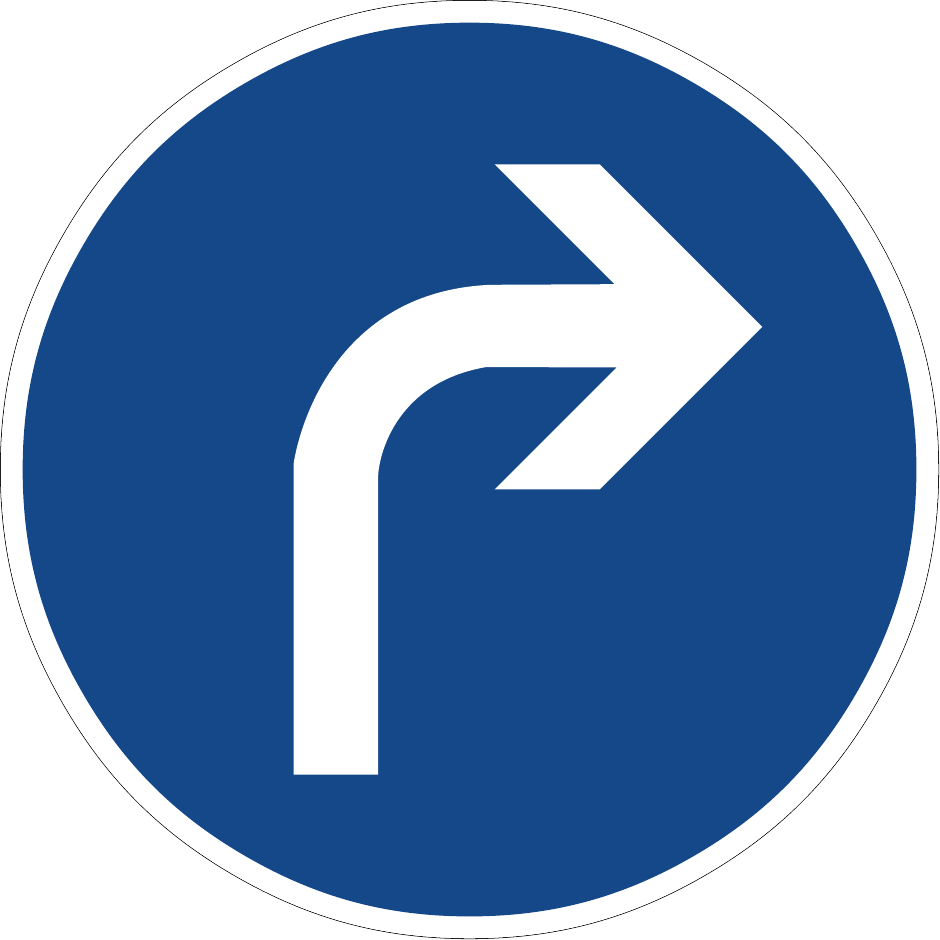}
      \includegraphics[width=.0665\linewidth]{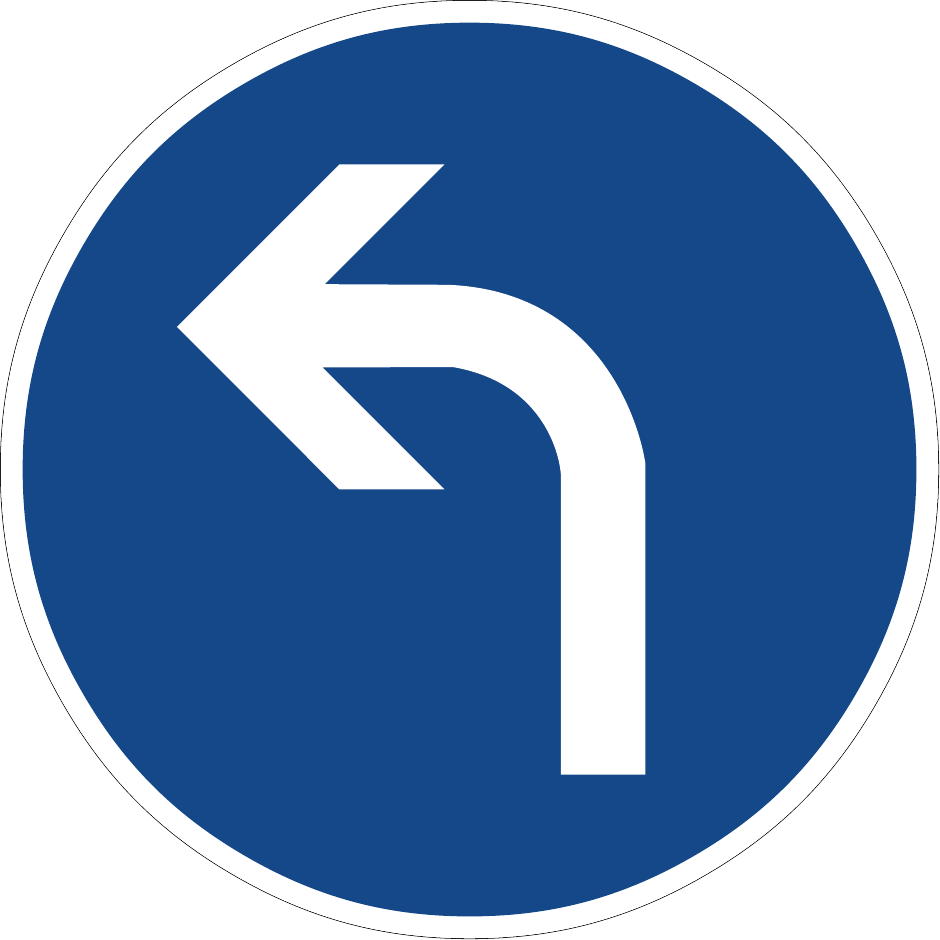}
      \includegraphics[width=.0665\linewidth]{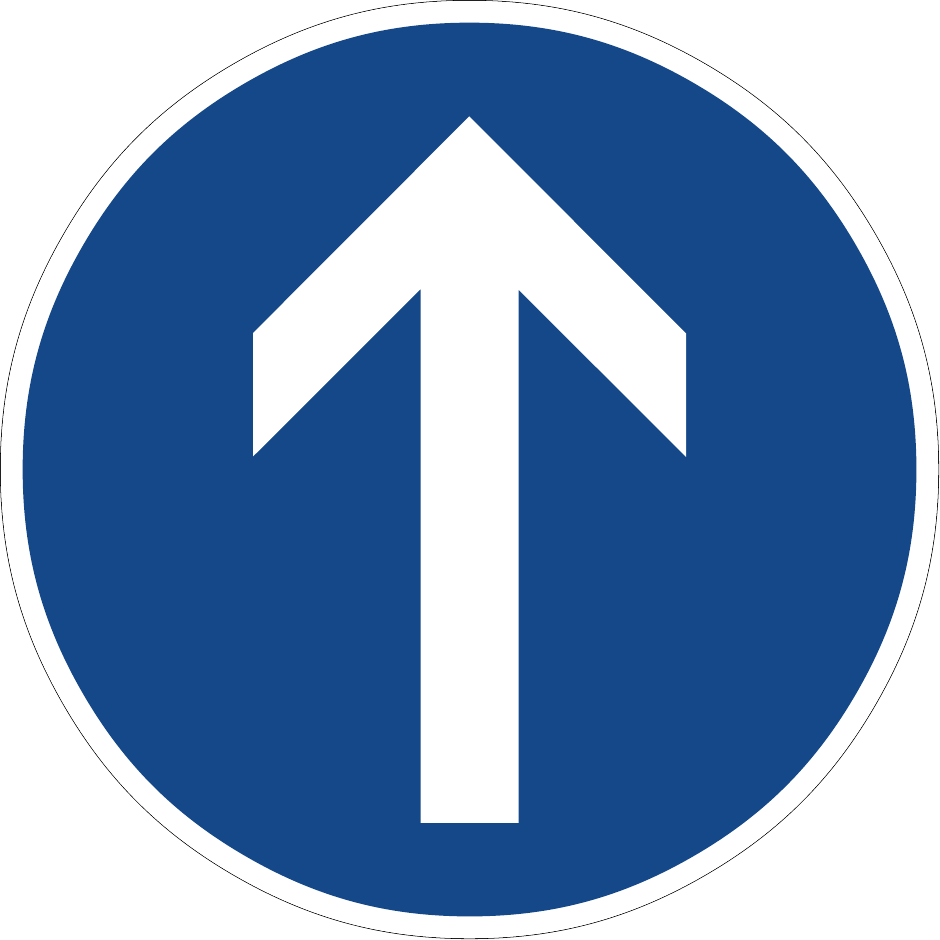}
      \includegraphics[width=.0665\linewidth]{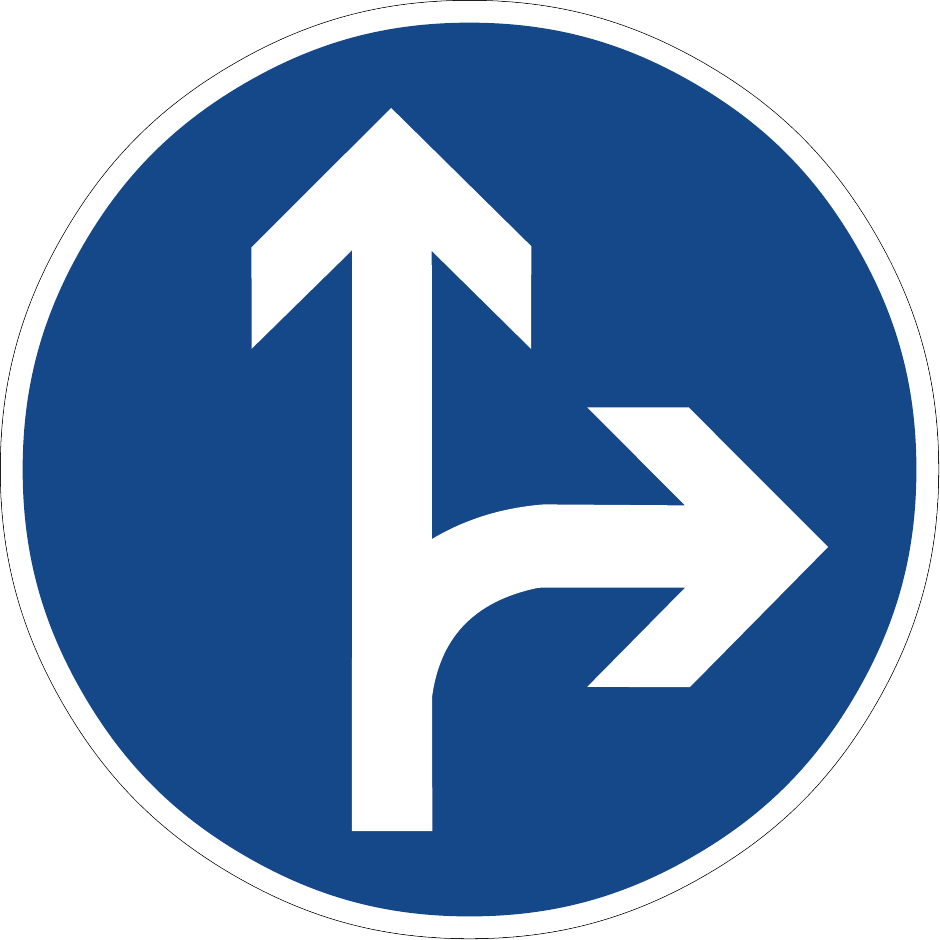}
      \includegraphics[width=.0665\linewidth]{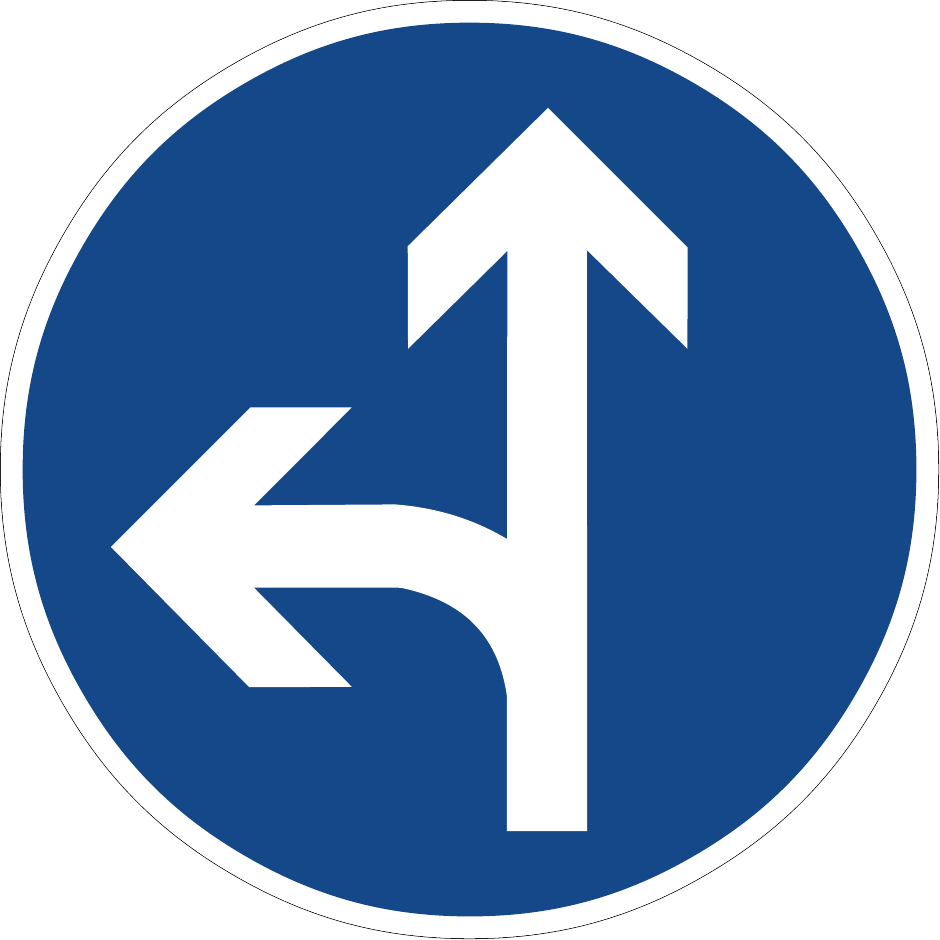}
      \includegraphics[width=.0665\linewidth]{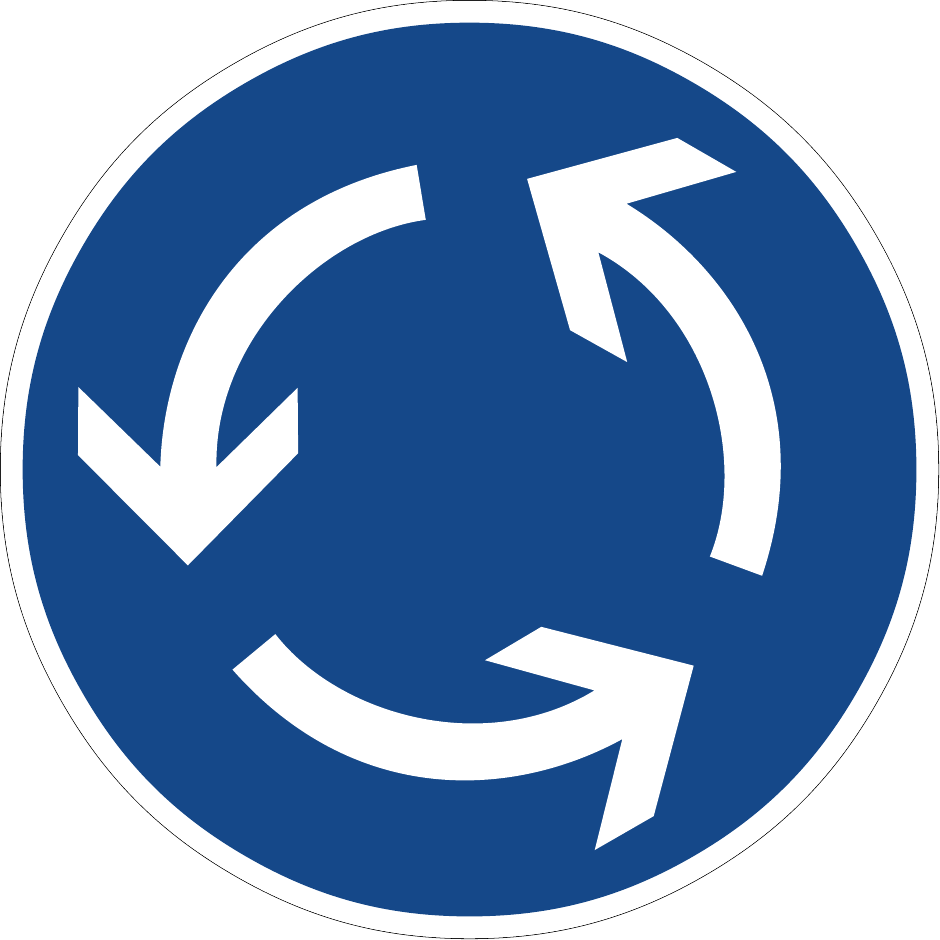}
      \includegraphics[width=.0665\linewidth]{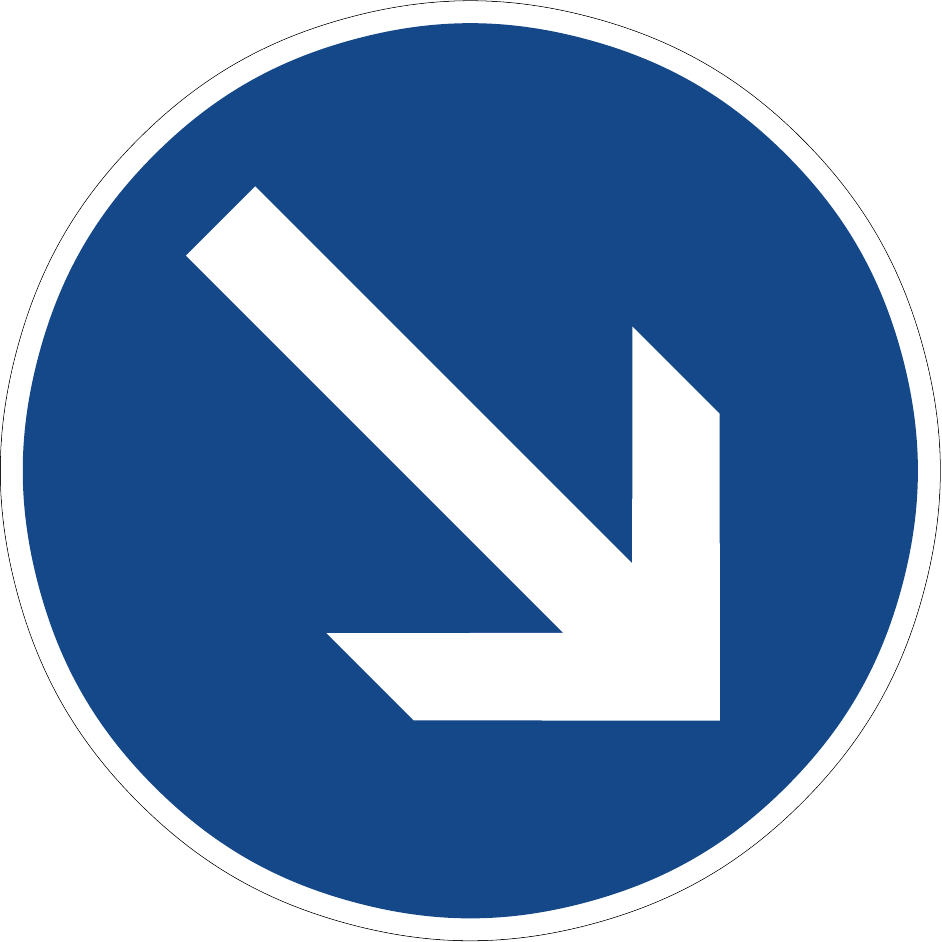}
      \includegraphics[width=.0665\linewidth]{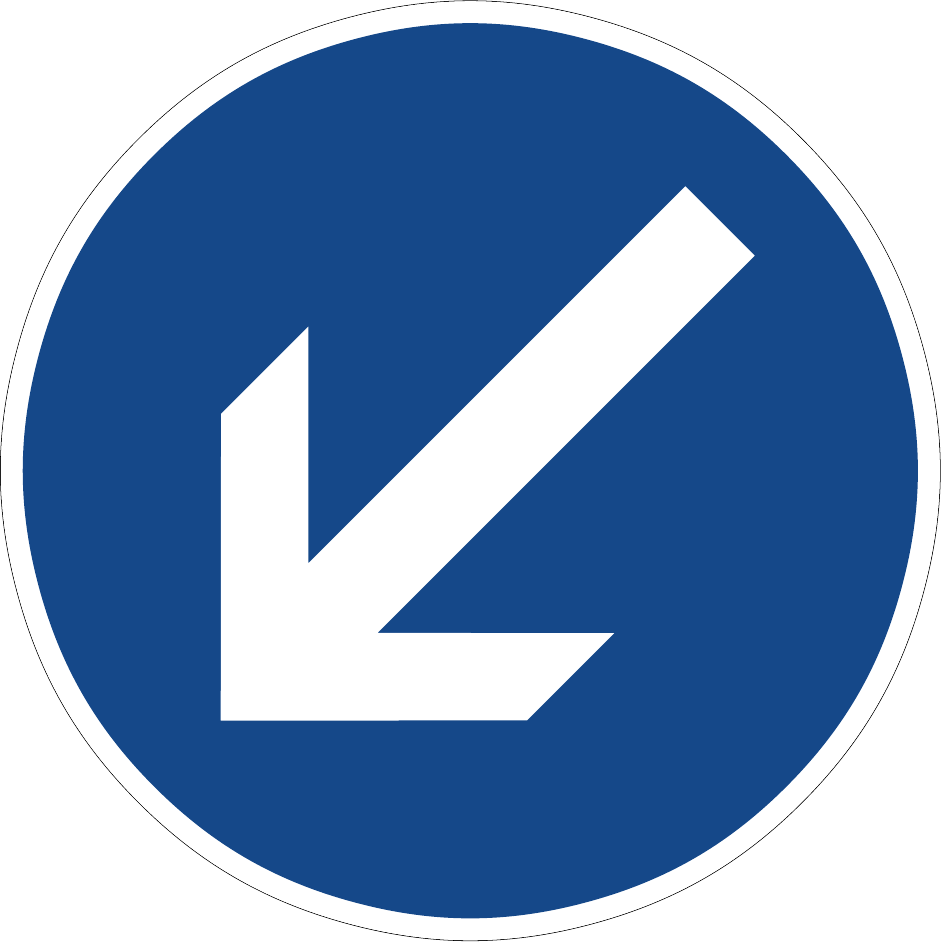}
      \caption{mandatory signs}
    \end{subfigure}%
    \caption{The traffic sign groups for the \textsf{Groups} constraint on GTSRB~\cite{stallkampGermanTrafficSign2011}.} 
  \label{fig:gtsrb_groups}
\end{figure}
  
  In order to compare conjunction and disjunction, we use the \textsf{Groups} constraint introduced in~\cite{fischerDL2TrainingQuerying2019}. This constraint introduces background knowledge into the network to control misclassifications, so that a group of related elements must have a high combined probability, or a low combined probability.
  For example, when the true label is a \qty[per-mode=symbol]{100}{\km\per\hour} speed limit sign, the constraint makes it more preferable to misclassify it as any other speed limit sign, rather than as a completely unrelated sign (such as a stop sign).

  Groups consist of classes of a similar type (e.g. the group of speed signs, the group of danger signs) and are thus used to encode a similarity relation on top of the data.
  We use this constraint on the \ac{gtsrb} dataset with the $6$ groups shown in~\cref{fig:gtsrb_groups}.

  Thus we obtain the group constraint \textsf{Groups} as shown in~\cref{eq:constraint_groups} below\footnote{For brevity, only one conjunct for the group of danger signs is shown, but the constraint we train with also has conjuncts for all groups shown in~\cref{fig:gtsrb_groups}, i.e. unique signs, derestriction signs, speed limit signs, other prohibitory signs, and mandatory signs.}:
  \begin{equation}\label{eq:constraint_groups}
    \mathsf{Groups}(\epsilon, \delta):=\ \forall \vec{x}'\in \epsball{\vec{x}}{\epsilon} \ldotpp \Biggl(\biggl(\sum\limits_{s\in \text{danger signs}}\mathcal{N}(\vec{x}')_s \ \le\ \delta\biggr) \ \vee\ \biggl(\sum\limits_{s\in \text{danger signs}}\mathcal{N}(\vec{x}')_s\ \ge\ 1 - \delta\biggr)\Biggr) \ \wedge\ \ldots
  \end{equation}

  \paragraph{Results}

  \begin{figure}
    \centering
    \begin{tikzpicture}[font=\small]
  \begin{groupplot}[group/results]
    \nextgroupplot[title={Prediction Accuracy (PAcc)},]
\addplot+[mark indices=51, densely dotted] table [y=Test-P-Acc] {r_groups-gtsrb-Baseline.csv};
\addplot+[mark indices=50] table [y=Test-P-Acc] {r_groups-gtsrb-DL2.csv};
\addplot+[mark indices=49] table [y=Test-P-Acc] {r_groups-gtsrb-Goedel.csv};
\addplot+[mark indices=48] table [y=Test-P-Acc] {r_groups-gtsrb-Lukasiewicz.csv};
\addplot+[mark indices=51] table [y=Test-P-Acc] {r_groups-gtsrb-Reichenbach.csv};
\addplot+[mark indices=49] table [y=Test-P-Acc] {r_groups-gtsrb-Yager.csv};

\coordinate (c1) at (rel axis cs:0,1);
    \nextgroupplot[title={Constraint Accuracy (CAcc)},
      yticklabels={},
      xlabel={}
    ]
\addplot+[mark indices=51, densely dotted] table [y=Test-C-Acc] {r_groups-gtsrb-Baseline.csv};
\addplot+[mark indices=50] table [y=Test-C-Acc] {r_groups-gtsrb-DL2.csv};
\addplot+[mark indices=49] table [y=Test-C-Acc] {r_groups-gtsrb-Goedel.csv};
\addplot+[mark indices=48] table [y=Test-C-Acc] {r_groups-gtsrb-Lukasiewicz.csv};
\addplot+[mark indices=51] table [y=Test-C-Acc] {r_groups-gtsrb-Reichenbach.csv};
\addplot+[mark indices=49] table [y=Test-C-Acc] {r_groups-gtsrb-Yager.csv};

\coordinate (c2) at (rel axis cs:0,1);
    \nextgroupplot[title={Constraint Security (CSec)},
      yticklabel pos=right,
      yticklabel style={anchor=east,xshift=2.5em},
      legend to name=full-legend
    ]
\addplot+[mark indices=51, densely dotted] table [y=Test-C-Sec] {r_groups-gtsrb-Baseline.csv};
\addplot+[mark indices=50] table [y=Test-C-Sec] {r_groups-gtsrb-DL2.csv};
\addplot+[mark indices=49] table [y=Test-C-Sec] {r_groups-gtsrb-Goedel.csv};
\addplot+[mark indices=48] table [y=Test-C-Sec] {r_groups-gtsrb-Lukasiewicz.csv};
\addplot+[mark indices=51] table [y=Test-C-Sec] {r_groups-gtsrb-Reichenbach.csv};
\addplot+[mark indices=49] table [y=Test-C-Sec] {r_groups-gtsrb-Yager.csv};
\addlegendentry {Baseline};
\addlegendentry {DL2};
\addlegendentry {Gödel};
\addlegendentry {\L ukasiewicz};
\addlegendentry {Reichenbach};
\addlegendentry {Yager};

\coordinate (c3) at (rel axis cs:1,1);
  \end{groupplot}
  \coordinate (c4) at ($(c1)!.5!(c3)$);
  \node[below, yshift=0.5cm] at (c4 |- current bounding box.south) {\pgfplotslegendfromname{full-legend}};
\end{tikzpicture}
    \caption{Results of training with the $\mathsf{Groups}(\epsilon=\sfrac{16}{255}, \delta=0.02)$ constraint on \ac{gtsrb}.}
    \label{fig:groups_gtsrb}
  \end{figure}
  
  \begin{table}
    \centering
    \caption{Results of training with the $\mathsf{Groups}(\epsilon=\sfrac{16}{255}, \delta=0.02)$ constraint on \ac{gtsrb}. The best result is displayed in boldface.}
    \label{tab:groups_gtsrb}
    \scriptsize
    \begin{tblr}
  {
    colspec={Q[l, mode=text]Q[c, mode=text]Q[c, mode=text]Q[c, mode=text]},
    row{1}={font=\bfseries, mode=text},
  }
    \toprule
      Logic & PAcc & CAcc & CSec \\
    \midrule
Baseline & \qty{98.66}{\percent} & \qty{96.64}{\percent} & \qty{63.61}{\percent} \\
DL2 & \qty{98.27}{\percent} & \qty{98.77}{\percent} & \qty{88.61}{\percent} \\
Gödel & \qty{98.32}{\percent} & \qty{98.85}{\percent} & \qty{93.58}{\percent} \\
\L ukasiewicz & \textbf{\qty{98.05}{\percent}} & \textbf{\qty{98.95}{\percent}} & \textbf{\qty{94.21}{\percent}} \\
Reichenbach & \qty{98.27}{\percent} & \qty{98.87}{\percent} & \qty{92.77}{\percent} \\
Yager & \qty{98.23}{\percent} & \qty{99.15}{\percent} & \qty{93.08}{\percent} \\

    \bottomrule
  \end{tblr}
  \end{table}
  
  As seen in~\cref{fig:groups_gtsrb,tab:groups_gtsrb}, all logics significantly improve constraint security without incurring any noteworthy drop in prediction accuracy.
  The baseline already exhibits a constraint security of \qty{63.61}{\percent}, but all logics are able to improve on this, from \qty{88.61}{\percent} for DL2 to \qty{94.21}{\percent} for the \L ukasiewicz logic, which performs the best overall in this experiment.

  \subsubsection{Comparing implication}
  To compare implication operators, we train with two different constraints.
  The first one is the $\mathsf{EvenOdd}$ constraint shown in~\cref{eq:constraint_even_odd}, meant to introduce additional background knowledge of even and odd numbers into MNIST:
  If the sum of predictions for even digits is high, the sum of predictions for even digits with the true prediction $y$ removed should still be greater than the sum of predictions for odd digits (and vice-versa).
  \begin{equation}\label{eq:constraint_even_odd}
    \begin{aligned}
      \mathsf{EvenOdd}(\epsilon, \delta, \gamma):=\ \forall \vec{x}'\in \epsball{\vec{x}}{\epsilon} \ldotpp &\Biggl(\biggl(\sum\limits_{d\in \text{even}}\mathcal{N}(\vec{x}')_d \ \ge\  \delta\biggr) \implies \biggl(\sum\limits_{d\in \text{even}\setminus\{y\}}\mathcal{N}(\vec{x}')_d \ \ge\  \gamma\sum\limits_{d\in \text{odd}}\mathcal{N}(\vec{x}')_d \biggr)\Biggr)\\
      \ \wedge\ &\Biggl(\biggl(\sum\limits_{d\in \text{odd}}\mathcal{N}(\vec{x}')_d \ \ge\  \delta\biggr) \implies \biggl(\sum\limits_{d\in \text{odd}\setminus\{y\}}\mathcal{N}(\vec{x}')_d \ \ge\  \gamma\sum\limits_{d\in \text{even}}\mathcal{N}(\vec{x}')_d \biggr)\Biggr)
    \end{aligned}      
  \end{equation}
  
  Additionally, we use a constraint based on the class-similarity constraint from DL2~\cite{fischerDL2TrainingQuerying2019} shown in~\cref{eq:constraint_class-similarity_dl2} on CIFAR-10 to introduce background knowledge such as ``a cat is more similar to a dog than to a frog'':
  \begin{equation}\label{eq:constraint_class-similarity_dl2}
    \forall \vec{x}'\in \epsball{\vec{x}}{\epsilon} \ldotpp \biggl( 
    \bigl(\argmax_i \mathcal{N}(\vec{x}')_i=\text{cat}\bigr) \implies \bigl(\mathcal{N}(\vec{x}')_{\text{dog}} \ \ge\  \mathcal{N}(\vec{x}')_{\text{frog}}\bigr) \biggr) \ \wedge\ \ldots
  \end{equation}

  Note that in this formulation, the premise of the implication, i.e. $\argmax_i \mathcal{N}(\vec{x}')_i=\text{cat}$ is either absolutely true or absolutely false.
  In these cases, all fuzzy logics behave the same by definition --- therefore, we use a modified version shown in~\cref{eq:constraint_class-similarity} for our \textsf{ClassSimilarity} constraint\footnote{We only show one conjunct in~\cref{eq:constraint_class-similarity} for brevity, but in addition to ``cat is more similar to dog than to frog'', the constraint we train with also contains ``airplane is more similar to ship than to dog'', ``automobile is more similar to truck than to cat'', ``bird is more similar to airplane than to dog'', ``deer is more similar to horse than to truck'', ``dog is more similar to cat than to bird'', ``frog is more similar to ship than to truck'', ``horse is more similar to deer than to airplane'', ``ship is more similar to airplane than to deer'', and ``truck is more similar to automobile than to airplane'' --- one for each of the ten classes.}.
  By replacing the premise with $\mathcal{N}(\vec{x}')_{\text{cat}}\ge\delta$, the premise will now always take on a fuzzy truth value, which allows the comparison of  the different implication operators.
  \begin{equation}\label{eq:constraint_class-similarity}
    \mathsf{ClassSimilarity}(\epsilon,\delta):=\ \forall \vec{x}'\in \epsball{\vec{x}}{\epsilon} \ldotpp \biggl(\bigl(\mathcal{N}(\vec{x}')_{\text{cat}}\ \ge\ \delta\bigr) \implies \bigl(\mathcal{N}(\vec{x}')_{\text{dog}}\ \ge\ \mathcal{N}(\vec{x}')_{\text{frog}}\bigr) \biggr) \ \wedge\ \ldots
  \end{equation}

  For both constraints, networks were initially trained without the logical constraint for $10$ epochs to ensure they could produce accurate predictions, before introducing  the logical loss to guide the models towards satisfying the constraint.

  \paragraph{Results}
  
  \begin{figure}
    \begin{subfigure}{\textwidth}
      \centering
      \begin{tikzpicture}[font=\small]
  \begin{groupplot}[group/results]
    \nextgroupplot[title={Prediction Accuracy (PAcc)},]
\addplot+[mark indices=51, densely dotted] table [y=Test-P-Acc] {r_even-odd-mnist-Baseline.csv};
\addplot+[mark indices=51] table [y=Test-P-Acc] {r_even-odd-mnist-DL2.csv};
\addplot+[mark indices=51] table [y=Test-P-Acc] {r_even-odd-mnist-Goedel.csv};
\addplot+[mark indices=47] table [y=Test-P-Acc] {r_even-odd-mnist-Goguen.csv};
\addplot+[mark indices=51] table [y=Test-P-Acc] {r_even-odd-mnist-KleeneDienes.csv};
\addplot+[mark indices=51] table [y=Test-P-Acc] {r_even-odd-mnist-Lukasiewicz.csv};
\addplot+[mark indices=51] table [y=Test-P-Acc] {r_even-odd-mnist-Reichenbach.csv};
\addplot+[mark indices=51] table [y=Test-P-Acc] {r_even-odd-mnist-ReichenbachSigmoidal.csv};
\addplot+[mark indices=51] table [y=Test-P-Acc] {r_even-odd-mnist-Yager.csv};

\coordinate (c1) at (rel axis cs:0,1);
    \nextgroupplot[title={Constraint Accuracy (CAcc)},
      yticklabels={},
      xlabel={}
    ]
\addplot+[mark indices=51, densely dotted] table [y=Test-C-Acc] {r_even-odd-mnist-Baseline.csv};
\addplot+[mark indices=51] table [y=Test-C-Acc] {r_even-odd-mnist-DL2.csv};
\addplot+[mark indices=51] table [y=Test-C-Acc] {r_even-odd-mnist-Goedel.csv};
\addplot+[mark indices=47] table [y=Test-C-Acc] {r_even-odd-mnist-Goguen.csv};
\addplot+[mark indices=51] table [y=Test-C-Acc] {r_even-odd-mnist-KleeneDienes.csv};
\addplot+[mark indices=51] table [y=Test-C-Acc] {r_even-odd-mnist-Lukasiewicz.csv};
\addplot+[mark indices=51] table [y=Test-C-Acc] {r_even-odd-mnist-Reichenbach.csv};
\addplot+[mark indices=51] table [y=Test-C-Acc] {r_even-odd-mnist-ReichenbachSigmoidal.csv};
\addplot+[mark indices=51] table [y=Test-C-Acc] {r_even-odd-mnist-Yager.csv};

\coordinate (c2) at (rel axis cs:0,1);
    \nextgroupplot[title={Constraint Security (CSec)},
      yticklabel pos=right,
      yticklabel style={anchor=east,xshift=2.5em},
      legend to name=full-legend
    ]
\addplot+[mark indices=51, densely dotted] table [y=Test-C-Sec] {r_even-odd-mnist-Baseline.csv};
\addplot+[mark indices=51] table [y=Test-C-Sec] {r_even-odd-mnist-DL2.csv};
\addplot+[mark indices=51] table [y=Test-C-Sec] {r_even-odd-mnist-Goedel.csv};
\addplot+[mark indices=47] table [y=Test-C-Sec] {r_even-odd-mnist-Goguen.csv};
\addplot+[mark indices=51] table [y=Test-C-Sec] {r_even-odd-mnist-KleeneDienes.csv};
\addplot+[mark indices=51] table [y=Test-C-Sec] {r_even-odd-mnist-Lukasiewicz.csv};
\addplot+[mark indices=51] table [y=Test-C-Sec] {r_even-odd-mnist-Reichenbach.csv};
\addplot+[mark indices=51] table [y=Test-C-Sec] {r_even-odd-mnist-ReichenbachSigmoidal.csv};
\addplot+[mark indices=51] table [y=Test-C-Sec] {r_even-odd-mnist-Yager.csv};
\addlegendentry {Baseline};
\addlegendentry {DL2};
\addlegendentry {Gödel};
\addlegendentry {Goguen};
\addlegendentry {Kleene-Dienes};
\addlegendentry {\L ukasiewicz};
\addlegendentry {Reichenbach};
\addlegendentry {sig. Reichenbach};
\addlegendentry {Yager};

\coordinate (c3) at (rel axis cs:1,1);
  \end{groupplot}
  \coordinate (c4) at ($(c1)!.5!(c3)$);
  \node[below, yshift=0.5cm] at (c4 |- current bounding box.south) {\pgfplotslegendfromname{full-legend}};
\end{tikzpicture}
      \caption{Results of training with the $\mathsf{EvenOdd}(\epsilon=0.4,\delta=0.6,\gamma=3)$ constraint on MNIST.}
      \label{fig:even-odd-mnist}
    \end{subfigure}
    
    \bigskip

    \begin{subfigure}{\textwidth}
      \centering
      \begin{tikzpicture}[font=\small]
  \begin{groupplot}[group/results]
    \nextgroupplot[title={Prediction Accuracy (PAcc)},]
\addplot+[mark indices=49, densely dotted] table [y=Test-P-Acc] {r_class-similarity-cifar10-Baseline.csv};
\addplot+[mark indices=49] table [y=Test-P-Acc] {r_class-similarity-cifar10-DL2.csv};
\addplot+[mark indices=49] table [y=Test-P-Acc] {r_class-similarity-cifar10-Goedel.csv};
\addplot+[mark indices=51] table [y=Test-P-Acc] {r_class-similarity-cifar10-Goguen.csv};
\addplot+[mark indices=47] table [y=Test-P-Acc] {r_class-similarity-cifar10-KleeneDienes.csv};
\addplot+[mark indices=47] table [y=Test-P-Acc] {r_class-similarity-cifar10-Lukasiewicz.csv};
\addplot+[mark indices=48] table [y=Test-P-Acc] {r_class-similarity-cifar10-Reichenbach.csv};
\addplot+[mark indices=47] table [y=Test-P-Acc] {r_class-similarity-cifar10-ReichenbachSigmoidal.csv};
\addplot+[mark indices=47] table [y=Test-P-Acc] {r_class-similarity-cifar10-Yager.csv};

\coordinate (c1) at (rel axis cs:0,1);
    \nextgroupplot[title={Constraint Accuracy (CAcc)},
      yticklabels={},
      xlabel={}
    ]
\addplot+[mark indices=49, densely dotted] table [y=Test-C-Acc] {r_class-similarity-cifar10-Baseline.csv};
\addplot+[mark indices=49] table [y=Test-C-Acc] {r_class-similarity-cifar10-DL2.csv};
\addplot+[mark indices=49] table [y=Test-C-Acc] {r_class-similarity-cifar10-Goedel.csv};
\addplot+[mark indices=51] table [y=Test-C-Acc] {r_class-similarity-cifar10-Goguen.csv};
\addplot+[mark indices=47] table [y=Test-C-Acc] {r_class-similarity-cifar10-KleeneDienes.csv};
\addplot+[mark indices=47] table [y=Test-C-Acc] {r_class-similarity-cifar10-Lukasiewicz.csv};
\addplot+[mark indices=48] table [y=Test-C-Acc] {r_class-similarity-cifar10-Reichenbach.csv};
\addplot+[mark indices=47] table [y=Test-C-Acc] {r_class-similarity-cifar10-ReichenbachSigmoidal.csv};
\addplot+[mark indices=47] table [y=Test-C-Acc] {r_class-similarity-cifar10-Yager.csv};

\coordinate (c2) at (rel axis cs:0,1);
    \nextgroupplot[title={Constraint Security (CSec)},
      yticklabel pos=right,
      yticklabel style={anchor=east,xshift=2.5em},
      legend to name=full-legend
    ]
\addplot+[mark indices=49, densely dotted] table [y=Test-C-Sec] {r_class-similarity-cifar10-Baseline.csv};
\addplot+[mark indices=49] table [y=Test-C-Sec] {r_class-similarity-cifar10-DL2.csv};
\addplot+[mark indices=49] table [y=Test-C-Sec] {r_class-similarity-cifar10-Goedel.csv};
\addplot+[mark indices=51] table [y=Test-C-Sec] {r_class-similarity-cifar10-Goguen.csv};
\addplot+[mark indices=47] table [y=Test-C-Sec] {r_class-similarity-cifar10-KleeneDienes.csv};
\addplot+[mark indices=47] table [y=Test-C-Sec] {r_class-similarity-cifar10-Lukasiewicz.csv};
\addplot+[mark indices=48] table [y=Test-C-Sec] {r_class-similarity-cifar10-Reichenbach.csv};
\addplot+[mark indices=47] table [y=Test-C-Sec] {r_class-similarity-cifar10-ReichenbachSigmoidal.csv};
\addplot+[mark indices=47] table [y=Test-C-Sec] {r_class-similarity-cifar10-Yager.csv};
\addlegendentry {Baseline};
\addlegendentry {DL2};
\addlegendentry {Gödel};
\addlegendentry {Goguen};
\addlegendentry {Kleene-Dienes};
\addlegendentry {\L ukasiewicz};
\addlegendentry {Reichenbach};
\addlegendentry {sig. Reichenbach};
\addlegendentry {Yager};

\coordinate (c3) at (rel axis cs:1,1);
  \end{groupplot}
  \coordinate (c4) at ($(c1)!.5!(c3)$);
  \node[below, yshift=0.5cm] at (c4 |- current bounding box.south) {\pgfplotslegendfromname{full-legend}};
\end{tikzpicture}
      \caption{Results of training with the $\mathsf{ClassSimilarity}(\epsilon=\sfrac{24}{255}, \delta=0.1)$ constraint on CIFAR-10.}
      \label{fig:class-similarity_cifar10}
    \end{subfigure}
    \caption{Results of training with the $\mathsf{EvenOdd}$ constraint on MNIST in~\cref{fig:even-odd-mnist} and of training with the $\mathsf{ClassSimilarity}$ on CIFAR-10 in~\cref{fig:class-similarity_cifar10}.}
    \label{fig:even-odd-class-similarity}
  \end{figure}

  \begin{table}
    \centering
    \caption{Results of training with the $\mathsf{EvenOdd}$ constraint on MNIST in~\cref{tab:even-odd-mnist} and of training with the $\mathsf{ClassSimilarity}$ on CIFAR-10 in~\cref{tab:class-similarity_cifar10}.
    The best result is displayed in boldface in each table.}
    \label{tab:even-odd-class-similarity}
    \begin{subtable}{.45\textwidth}
      \centering
      \caption{$\mathsf{EvenOdd}(\epsilon=0.4,\delta=0.6,\gamma=3)$ on MNIST.}
      \label{tab:even-odd-mnist}
      \scriptsize
      \begin{tblr}
  {
    colspec={Q[l, mode=text]Q[c, mode=text]Q[c, mode=text]Q[c, mode=text]},
    row{1}={font=\bfseries, mode=text},
  }
    \toprule
      Logic & PAcc & CAcc & CSec \\
    \midrule
Baseline & \qty{91.43}{\percent} & \qty{29.08}{\percent} & \qty{25.18}{\percent} \\
DL2 & \qty{91.50}{\percent} & \qty{97.80}{\percent} & \qty{72.39}{\percent} \\
Gödel & \qty{91.66}{\percent} & \qty{99.48}{\percent} & \qty{89.47}{\percent} \\
Goguen & \qty{91.00}{\percent} & \qty{61.98}{\percent} & \qty{48.27}{\percent} \\
Kleene-Dienes & \qty{90.41}{\percent} & \qty{79.41}{\percent} & \qty{77.59}{\percent} \\
\L ukasiewicz & \qty{90.83}{\percent} & \qty{100.00}{\percent} & \qty{99.93}{\percent} \\
Reichenbach & \textbf{\qty{91.10}{\percent}} & \textbf{\qty{100.00}{\percent}} & \textbf{\qty{99.94}{\percent}} \\
sig. Reichenbach & \qty{90.97}{\percent} & \qty{100.00}{\percent} & \qty{99.72}{\percent} \\
Yager & \qty{90.70}{\percent} & \qty{100.00}{\percent} & \qty{99.57}{\percent} \\

    \bottomrule
  \end{tblr}
    \end{subtable}%
    \hfil
    \begin{subtable}{.45\textwidth}
      \centering
      \caption{$\mathsf{ClassSimilarity}(\epsilon=\sfrac{24}{255}, \delta=0.1)$ on CIFAR-10.}
      \label{tab:class-similarity_cifar10}
      \scriptsize
      \begin{tblr}
  {
    colspec={Q[l, mode=text]Q[c, mode=text]Q[c, mode=text]Q[c, mode=text]},
    row{1}={font=\bfseries, mode=text},
  }
    \toprule
      Logic & PAcc & CAcc & CSec \\
    \midrule
Baseline & \qty{62.83}{\percent} & \qty{64.96}{\percent} & \qty{36.96}{\percent} \\
DL2 & \qty{65.54}{\percent} & \qty{78.23}{\percent} & \qty{61.69}{\percent} \\
Gödel & \qty{58.94}{\percent} & \qty{65.08}{\percent} & \qty{86.08}{\percent} \\
Goguen & \textbf{\qty{52.50}{\percent}} & \textbf{\qty{97.65}{\percent}} & \textbf{\qty{79.65}{\percent}} \\
Kleene-Dienes & \qty{13.93}{\percent} & \qty{99.38}{\percent} & \qty{21.79}{\percent} \\
\L ukasiewicz & \qty{18.66}{\percent} & \qty{99.97}{\percent} & \qty{97.07}{\percent} \\
Reichenbach & \qty{13.06}{\percent} & \qty{99.93}{\percent} & \qty{13.25}{\percent} \\
sig. Reichenbach & \qty{16.05}{\percent} & \qty{99.21}{\percent} & \qty{9.54}{\percent} \\
Yager & \qty{16.59}{\percent} & \qty{99.47}{\percent} & \qty{24.60}{\percent} \\

    \bottomrule
  \end{tblr}
    \end{subtable}
  \end{table}

  In the following, it is important to not just look at prediction accuracy, constraint accuracy, and constraint security, but to also take into account the percentage the constraints are satisfied vacuously (i.e. where the premise of the implication is false).
  
  In the case of the $\mathsf{EvenOdd}$ constraint (assuming a reasonably balanced dataset, as is the case for MNIST) we would expect each conjunct of the constraint to be vacuously satisfied roughly half of the time.
  On the baseline model, the first conjunct is in fact vacuously satisfied \qty{62.75}{\percent} of the time, and the second \qty{55.54}{\percent}.
  When training with DL2, the first conjunct is vacuously satisfied \qty{16.97}{\percent} of the time, and the second \qty{90.52}{\percent}.
  For the Gödel logic, we observe \qty{94.45}{\percent} and \qty{94.96}{\percent}, and \qty{70.45}{\percent} and \qty{59.58}{\percent} for the Goguen logic.
  The Kleene-Dienes logic seems more balanced: the first conjunct is vacuously satisfied \qty{52.34}{\percent} of the time, and the second \qty{55.10}{\percent}.
  Remarkably, for the \L ukasiewicz, Reichenbach, sigmoidal Reichenbach, and Yager logic, the first conjunct is never satisfied vacuously, while the second is always satisfied vacuously, which in turn means that the network learns to make the sum of predictions of even numbers never exceed the threshold $\delta$, and to always make the sum of predictions of odd numbers exceed the threshold.

  In this experiment, there is no noticeable effect on prediction accuracy, as the premise of the implication does not refer to only one class prediction, but rather a sum of multiple predictions.
  Even when that sum is low, the prediction of the true class can still be the highest overall, which does not affect prediction accuracy negatively.
  
  However, with the $\mathsf{ClassSimilarity}$ constraint on CIFAR-10, where the premise of the implication of each conjunct refers to only one class prediction, a significant drop in prediction accuracy can be observed for the Kleene-Dienes, \L ukasiewicz, Reichenbach, sigmoidal Reichenbach, and Yager logics.
  This means, for example, the network learns to satisfy the implication $\bigl(\mathcal{N}(\vec{x}')_{\text{cat}}\ \ge\ \delta\bigr) \implies \bigl(\mathcal{N}(\vec{x}')_{\text{dog}}\ \ge\ \mathcal{N}(\vec{x}')_{\text{frog}}\bigr)$ by making the prediction for the class cat low, even if it is the correct class---this then leads to incorrect predictions and drastically reduced prediction accuracy.
  
  A slightly less severe drop in prediction accuracy occurs for the Gödel and Goguen logic, while the DL2 logic even leads to slightly improved prediction accuracy over the baseline.

  The phenomenon of neural networks having a tendency to vacuously satisfy constraints is known and has been called \emph{shortcut satisfaction} by \citet{liLearningLogicalConstraints2022}, and \emph{implication bias} by \citet{heReducedImplicationbiasLogic2023}.

  \subsection{Formally verifying neural networks trained with differentiable logics}\label{subsec:verification}
  Drawing from our motivation of using differentiable logics to obtain correct-by-construction neural networks, we also investigate whether training with logical constraints can indeed lead to networks that verifiably satisfy constraints more after training.

  \subsubsection{The Vehicle language}
  We use the expressive functional language Vehicle~\cite{daggittVehicleInterfacingNeural2022,daggittVehicleTutorialNeural2023,daggittVehicle2024} for stating neural network specifications.
  From these specifications, Vehicle can generate loss functions for training\footnote{In this work, we use Vehicle specifically as a frontend for the formal verifier, and not to generate loss functions for training.} with TensorFlow~\cite{abadiTensorFlowSystemLargescale2016}, compile verifier-specific queries for neural network verifiers and run the verifiers.
  Currently, Vehicle supports the Marabou~\cite{katzMarabouFrameworkVerification2019,wuMarabou20Versatile2024} neural network verifier, a sound and complete neural network verifier utilising \ac{smt} and \ac{milp} solving as well as various abstract interpretation and bounds tightening algorithms.
  After running the verifier, Vehicle can also integrate the resulting proofs with the interactive theorem prover Agda.

  \subsubsection{Limitations of neural network verification}
  Neural network verification has been shown to be NP-hard~\cite{katzReluplexEfficientSMT2017}, and many verification algorithms do not scale well to big network architectures.
  For this reason, to make formal verification tractable, we used a very lightweight \ac{cnn} with two convolutional layers ($4$ and $8$ filters) with ReLU activations followed by two fully connected layers ($16$ and $10$ units, ReLU in the hidden layer).

  The \textsf{StandardRobustness} constraint shown in~\cref{eq:constraint_standard_robustness} cannot be verified due to limitations of the tools used (as the constraint is essentially a 2-safety property~\cite{clarksonHyperproperties2010}, involving two network executions).
  Additionally, the \textsf{StrongClassificationRobustness} constraint shown in~\cref{eq:robustness_constraint_dl2}, which compares the predicted probability of the true class against some threshold probability, requires the $\mathrm{softmax}$ function, which is currently not supported by Marabou for \ac{smt} solving.
  Hence, we use the modified version shown in~\cref{eq:strong_classification_robustness_constraint} in training and verification.
  \begin{equation}\label{eq:strong_classification_robustness_constraint}
      \mathsf{StrongClassificationRobustness}(\epsilon,\delta):=\ \forall \vec{x}'\in \epsball{\vec{x}}{\epsilon} \ldotpp \mathrm{logit}(\vec{x}')_y\ge \delta.
  \end{equation}
  This constraint does not refer to the predicted probabilities, but rather operates on the raw logits the network outputs.
  This constraint can be verified by Marabou, but it does have the drawback of being less interpretable, as raw logits are unbounded real numbers.
 
  \subsubsection{Experimental setup \& results}

  \begin{table}
      \small
      \centering
\begin{talltblr}[
    note{a}={Marabou was run with a per-image timeout of \qty{30}{\second}. The table reports the verified fraction of queries that did not time out.},
    caption={Verified constraint satisfaction of the $\mathsf{StrongClassificationRobustness}(\epsilon=0.4, \delta=3.0)$ constraint for values of $\epsilon\in\{0.1,0.2,0.3,0.35,0.4\}$ on $500$ randomly chosen images of the MNIST test set.},
    label={tab:verification}
]
  {
    colspec={lQ[c, m, mode=text]Q[c, m, mode=text]Q[c, m, mode=text]Q[c, m, mode=text]Q[c, m, mode=text]Q[c, m, mode=text]Q[c, m, mode=text]Q[c, m, mode=text]},
    row{1}={font=\bfseries, mode=text},
  }
    \toprule
      \SetCell[r=2,c=1]{l} Logic & \SetCell[r=2,c=1]{c} {Prediction\\Accuracy} & \SetCell[r=2,c=1]{c} {Constraint\\Accuracy} & \SetCell[r=2,c=1]{c} {Constraint\\Security} & \SetCell[r=1,c=5]{c} Verified Constraint Satisfaction \\
      \cline{5-9}
      & & & & $\epsilon=0.1$ & $\epsilon=0.2$ & $\epsilon=0.3$ & $\epsilon=0.35$ & $\epsilon=0.4$\\
    \midrule
Baseline & \qty{96.50}{\percent} & \qty{90.88}{\percent} & \qty{79.68}{\percent} & {\qty{49.87}{\percent} \\ $(\sfrac{195}{391})$\TblrNote{a}} & {\qty{0.68}{\percent} \\ $(\sfrac{3}{444})$\TblrNote{a}} & {\qty{0}{\percent} \\ $(\sfrac{0}{500})$} & {\qty{0}{\percent} \\ $(\sfrac{0}{500})$} & {\qty{0}{\percent} \\ $(\sfrac{0}{500})$} \\ \addlinespace
DL2 & \qty{93.07}{\percent} & \qty{100}{\percent} & \qty{100}{\percent} & {\qty{99.80}{\percent} \\ $(\sfrac{495}{496})$\TblrNote{a}} & {\qty{92.98}{\percent} \\ $(\sfrac{384}{413})$\TblrNote{a}} & {\qty{55.29}{\percent} \\ $(\sfrac{183}{331})$\TblrNote{a}} & {\qty{33.91}{\percent} \\ $(\sfrac{117}{345})$\TblrNote{a}} & {\qty{20.51}{\percent} \\ $(\sfrac{73}{356})$\TblrNote{a}} \\ \addlinespace
Fuzzy Logic & \qty{94.87}{\percent} & \qty{100}{\percent} & \qty{100}{\percent} & {\qty{99.80}{\percent} \\ $(\sfrac{498}{499})$\TblrNote{a}} & {\qty{92.70}{\percent} \\ $(\sfrac{368}{397})$\TblrNote{a}} & {\qty{52.16}{\percent} \\ $(\sfrac{157}{301})$\TblrNote{a}} & {\qty{27.18}{\percent} \\ $(\sfrac{84}{309})$\TblrNote{a}} & {\qty{9.22}{\percent} \\ $(\sfrac{27}{293})$\TblrNote{a}} \\
    \bottomrule
  \end{talltblr}
  \end{table}

  The verification experiment was conducted on a system with an Intel i9 13900KF CPU with \qty{64}{\giga\byte} of RAM and \qty{64}{\giga\byte} of swap memory.
  We trained the neural network without differentiable logics (baseline), as well as with DL2 and fuzzy logic to satisfy $\mathsf{StrongClassificationRobustness}(\epsilon=0.4,\delta=3.0)$.
  We used Marabou to check the verified constraint satisfaction for values $\epsilon\in\{0.1,0.2,0.3,0.35,0.4\}$ on $500$ of the $\num{10000}$ images of the MNIST test set.
  Marabou was run with a timeout of \qty{30}{\second} per image.
  The whole verification experiment took \qty{18}{\hour} \qty{53}{\minute} to complete.
  
  The results are presented in~\cref{tab:verification}.
  Two key insights stand out:
  \begin{inparaenum}[(i)]
    \item training with logical constraints (regardless of the logic translation used) leads to significantly improved verified constraint satisfaction compared to the baseline, and
    \item the verified constraint satisfaction at $\epsilon=0.4$ (the exact value used during training) is notably lower than the constraint security achieved during training.
  \end{inparaenum}

  As explained earlier, the discrepancy between high constraint security and significantly reduced verified constraint satisfaction is due to the fact that constraint security is both a measure of how robust the network is against counterexamples, but at the same time influenced by how successful the oracle is in finding counterexamples during training.
  Constraint security is high if the oracle cannot find a counterexample, which is not only the case when the network is robust, but also when the attack parameters are suboptimal.
  Additionally, even assuming a perfect oracle that always succeeds in finding counterexamples, minimising an upper bound on the worst-case loss during training does not provide any formal guarantees, as noted by~\citet{urbanReviewFormalMethods2021}.

  The high baseline constraint security can be explained by the fact that the constraint~\cref{eq:strong_classification_robustness_constraint} only requires the prediction of the true label is greater than the threshold, and can be trivially satisfied if the predictions for \emph{all} labels are greater than the threshold.
  An improved version of this constraint would need to check if all other predictions except for the true label are less than the threshold; however, this would increase the number of verification queries that need to be run for each image from $1$ to $10$.

  Lastly, this verification experiment was only run for a small fraction (\qty{5}{\percent}) of the MNIST test set---however, even verifying all $\num{10000}$ images would not provide any guarantees for unseen data outside of that dataset.\footnote{A problem inherent to formal verification of neural networks: verification algorithms are general enough to verify properties for specific regions of the input space, but specifying those is not feasible for high-dimensional spaces like images or videos, which in practice limits verification to point-wise verification.}
  
  \section{Discussion and Future Work}
  \label{sec:conclusion}

  The theoretic comparison in~\cref{sec:theory} provided different ways to interpret logic relaxations.
  \begin{itemize}
      \item The shadow-lifting property of a conjunction is desirable for learning, as it allows for gradual satisfaction.
      The investigation in \cref{sec:shadow-lifting} showed that no fuzzy logic conjunction apart from the Reichenbach one satisfies shadow-lifting.
      The DL2 conjunction satisfies both shadow-lifting and has behaviour that is well-suited for learning, having strong derivatives everywhere.
      The Reichenbach conjunction, despite satisfying shadow-lifting, does not exhibit behaviour suitable for learning, as its derivatives are low when they should be high to enable learning in the first place.
      The Yager conjunction vanishes on a considerable part of its domain, but its derivatives are well-suited for learning.
      \item Investigating implication with respect to whether the logics follow Modus Ponens, or, more importantly, Modus Tollens reasoning, showed that only the Reichenbach logic closely follows both Modus Ponens and Modus Tollens~\cref{sec:modus-ponens-modus-tollens}.
      Other logics (such as the Yager logic, which follows only Modus Tollens reasoning, and DL2, which only follows Modus Ponens reasoning and only for a limited domain) only allow for either Modus Ponens or Modus Tollens reasoning, (wheras others do not allow for either) suggesting that training with these logics would not lead to useful updates during training.
      \item In \cref{sec:consistency}, we investigated fuzzy logics from a logic perspective as opposed to examining their behaviour in the training process.
      The \L ukasiewicz, sigmoidal Reichenbach, and Goguen logics were found to be the most consistent (i.e., they preserve truth the most), with the Gödel logic being the least consistent.
  \end{itemize}

  Our experimental evaluation of differentiable logics in training in~\cref{subsec:training_experiment} complements the theoretical one, and shows most importantly, that training with any logic will generally lead to improved constraint security, at a slight expense of prediction accuracy.
  This phenomenon is commonly known and was first reported by \citet{tsiprasRobustnessMayBe2018}.

  When comparing conjunction operators, theoretical investigations suggested that the shadow-lifting property of a logic would have an important impact on the learning process, which our experiment (see~\cref{fig:groups_gtsrb,tab:groups_gtsrb}) did not confirm: the only shadow-lifting conjunction did not perform best, and the differences between different logics were not significant.

  The theoretical investigations into a logic's ability to preserve truth from~\cref{sec:consistency} suggest that the \L ukasiewicz logic is the most consistent, and the Gödel logic the least.
  Our experiments comparing implication (see~\cref{fig:even-odd-mnist,tab:even-odd-mnist,fig:class-similarity_cifar10,tab:class-similarity_cifar10}) confirm that neural networks tend to vacuously satisfy implications, as reported by \citet{liLearningLogicalConstraints2022,heReducedImplicationbiasLogic2023}.

  We conclude that in general, theoretical properties such as shadow-lifting, the ability to perform Modus Ponens or Modus Tollens reasoning, and consistency, do not have as much of an impact as the operators having strong derivatives.

  Using Auto-PGD and GradNorm in training to ensure logics can perform optimally, the differences in performance between logics are more clear, but still depend on the specific task at hand.

  The verification experiment in~\cref{subsec:verification} has shown that training with logical constraints leads to improved verified constraint satisfaction, although there is a significant gap between constraint satisfaction in training and verified constraint satisfaction.
  Combining training with logical constraints (to find counterexamples outside of the training set) with formal verification of the neural network after training (limited to point-wise verification) seems like a promising approach towards safe-by-construction neural networks.
  
  \subsection{Lessons learnt}
  The large number of hyperparameters, including standard \ac{ml} hyperparameters (such as batch size, optimiser, learning rate, and network architecture) and hyperparameters introduced by the counterexample finding approach (such as iterations, step size, number of restarts) and the asymmetry hyperparameter introduced by GradNorm make it difficult (if not impossible) to provide a definite answer for the question of what logical relaxation to train to satisfy arbitrary constraints on arbitrary datasets on an arbitrary network in the future.

  Instead of trying to find a single best logic that works well in all possible use cases, a more fruitful research direction would be to explore what kinds of logical constraints require logics that exhibit certain geometric properties (such as shadow-lifting, or Modus Ponens / Modus Tollens reasoning).
  
  \subsection{Future work}
  Our experiments have shown that learning with differentiable logics can generally improve how much a \ac{ml} model satisfies a constraint.
  Imposing logical constraints on the training process in this manner could be a step in the direction of verified ML, allowing the use of continuous-learning in self-improving ML-enabled autonomous systems.
  It has to be noted that in contrast to formal verifiers, training with logical loss does not formally guarantee properties to hold in all possible cases.
  We highlight a few more areas for future work in the following.

  \paragraph{Reusing logical constraints during inference}
  Any logical constraints imposed on the learning process are unavailable during inference.
  This is due to the differentiable logics acting as a regulariser during training.
  The trained model can therefore not make use of the logical constraints to check its predictions, for example to attach confidence scores to its predictions.
  \citet{giunchigliaDeepLearningLogical2022a} provide a survey of learning with logical constraints and also investigate ways to guarantee satisfaction of logical constraints after training, by means of constraining the output.
  
  \paragraph{Probabilistic logics}
  Despite expressing satisfaction of formulas on $[0,1]$, fuzzy logics are inherently not probabilistic, having been designed instead for reasoning in the presence of vagueness\footnote{Which has consequences for the semantics: for example, in probability theory, given $P=Q=0.5$, the probability of ``$P$ and $Q$'' is $\operatorname{Pr}[P\cap Q]=0.25$, whereas in fuzzy logic, given $p=q=0.5$, the value of $p\wedge q$ is $0.5$ using the Gödel t-norm $\min\{x,y\}$, or $0$, using the \L ukasiewicz t-norm $\max\{0,x+y-1\}$.}.
  We point to DeepProbLog~\cite{manhaeveDeepProbLogNeuralProbabilistic2018} as one example for a probabilistic logic for use with deep learning.
  In the context of neural networks, which often output probabilities, it could be more natural to reason about probabilities instead of vagueness, especially for constraints that include probabilities~\cite{farrellExploringRequirementsSoftware2023}.
  
  \paragraph{Properties \& expressivity}
  A common problem with verifying ML is the lack of specifications, as noted by \citet{seshiaFormalSpecificationDeep2018,leuckerFormalVerificationNeural2020,farrellExploringRequirementsSoftware2023}.
  Most properties in the literature are limited to robustness against slight perturbations, although differentiable logics can not only relate network inputs and outputs, but could also refer to the inner workings of the neural network, such as weights and activation states.
  A related area is to investigate whether learning with logical loss can be used to show that desired properties continue to hold when retraining the network.

  Lastly, logical constraints are often encoded as propositional logic constraints.
  For example, the ROAD-R \cite{giunchigliaROADRAutonomousDriving2023} benchmark provides videos annotated with propositional logic constraints encoding background knowledge, but more expressive logical constraints are planned for future work.
  Looking at differentiable logics beyond propositional logic is a direction worth exploring; \citet{varnaiRobustnessMetricsLearning2020a} provide a differentiable logic for \ac{stl}, and \citet{leungBackpropagationSignalTemporal2021,xieEmbeddingSymbolicTemporal2021,xuDonPourCereal2022} have translations for \ac{ltl}.

  \paragraph{Certified training}
  As seen in~\cref{subsec:verification} and reported by~\citet{urbanReviewFormalMethods2021}, training with differentiable logics using PGD to minimise a lower bound on the worst-case loss does not yield formal guarantees.
  Instead of finding a worst perturbation, we will continue our work by investigating approaches based on certified training such as~\citet{wongProvableDefensesAdversarial2018,wongScalingProvableAdversarial2018,mirmanDifferentiableAbstractInterpretation2018a,raghunathanCertifiedDefensesAdversarial2018}.

  \section*{Acknowledgements}
  This publication has emanated from research conducted with the financial support of Taighde Éireann – Research Ireland under Grant number 20/FFP-P/8853.

  We thank Rozenn Dahyot for generously providing the compute resources used in this work.
  We are also grateful to Matthew L. Daggitt for his prompt support and usability improvements to the Vehicle tool.
  Finally, we appreciate the reviewers for their helpful comments and feedback, which helped improve the clarity of the paper.
  
  \bibliographystyle{elsarticle-num-names} 
  \bibliography{a_references.bib}
\end{document}